\documentclass[a4paper,11pt]{article}
\pdfoutput=1
\usepackage{jheppub}
\usepackage{amsmath,amssymb}
\usepackage{bbold}
\usepackage{graphicx}
\usepackage{float}
\usepackage{color}
\usepackage [utf8]{inputenc}
\usepackage{subfigure}
\usepackage{euscript}
\usepackage{placeins}
\usepackage{diagbox}
\usepackage[thinlines]{easytable}
\usepackage{ulem}
\usepackage{xcolor}
\usepackage{rotating}

\renewcommand{\d}{\textrm{d}}

\def\OO{\mathcal{O}}

\def\bx{{\boldsymbol x}}

\def\gsqa{g_{3\mathrm{d}}^2a}
\def\gE{g_{3\mathrm{d}}}
\def\gsq{g_{3\mathrm{d}}^2}
\def\gthree{g_{3\mathrm{d}}^3}
\def\gfour{g_{3\mathrm{d}}^4}
\def\gsix{g_{3\mathrm{d}}^6}

\def\mDsq{m_{\mathrm{D}}^2}

\def\Tr{\mathrm{Tr}}

\def\nf{n_\mathrm{f}}

\def\d{\mathrm{d}}

\def\bmu{\bar{\mu}}
\def\bmi{\bar{\mu}_i}
\def\bmj{\bar{\mu}_j}
\def\bmk{\bar{\mu}_k}
\def\bmm{\bar{\mu}_m}
\def\bmn{\bar{\mu}_n}
\def\bml{\bar{\mu}_l}
\def\bmf{\bar{\mu}_f}
\def\gammaE{\gamma_\mathrm{E}}
\def\chitd{\chi_{3 \mathrm{d}}}
\def\LMSb{\Lambda_{\overline{\mathrm{MS}}}}
\def\zR{z_\mathrm{R}}
\def\zI{z_\mathrm{I}}
\def\Tc{T_\mathrm{c}}

\begin{document}


\title{High order quark number susceptibilities in hot QCD from lattice EQCD}

\author[a]{Kari Rummukainen,}
\author[a,b]{Niels Schlusser}
\affiliation[a]{Department of Physics \& Helsinki Institute of Physics, P.O. Box 64, FI-00014 University of Helsinki}
\affiliation[b]{Biozentrum, University of Basel, 4056 Basel, Switzerland}
\emailAdd{kari.rummukainen@helsinki.fi,  niels.schlusser@helsinki.fi}

\abstract{
Building on the experience of \cite{Hietanen:2008tv}, we develop a formalism to construct operators for higher derivatives of the pressure in hot QCD with respect to the quark chemical potential $\mu$. We provide formulae for the operators up to the sixth derivative, and obtain continuum-extrapolated results from lattice EQCD at zero and finite $\mu$ and at six different pairs of temperature $T$ and number of massless quark flavors $\nf$. Our data is benchmarked against full-QCD lattice and perturbative results, allowing to judge the quality of the perturbative series expansion in EQCD and the dimensional reduction procedure as a whole.
}
\maketitle

\section{Introduction}
\label{sec:intro}
It has been discovered at RHIC \cite{Adler:2003kt,Adams:2004bi} and LHC \cite{ALICE:2011ab,Chatrchyan:2013nka,Aad:2014fla,ALICE:2016kpq} that the Quark-Gluon Plasma (QGP) behaves effectively as an almost perfect fluid. Therefore, it seems natural to attempt modeling the QGP's dynamics with hydrodynamic equations \cite{Elze:1989un,Gale:2013da}. Beyond equations of continuity for all conserved charges and an equation of motion for the velocity field, a thermodynamic equation of state (EOS) is required.
While experimental measurements of the equation of state are underway, our work will focus on the theoretical side. A broad variety of methods has been applied to study the equation of state of the QGP, i.e.\ the pressure $p$ as a function of the temperature $T$ and the quark chemical potential $\mu$, for instance holographic methods \cite{CasalderreySolana:2011us}, functional methods \cite{Fu:2021oaw,Herbst:2013ufa}, perturbation theory \cite{Kajantie:2002wa,Ipp:2006ij}, and lattice simulations \cite{Borsanyi:2013bia,Bazavov:2014pvz}. Our approach will concentrate on the two latter, methodologically more conservative approaches.

Perturbation theory relies on the system being weakly coupled, meaning that the temperature must be much higher than the renormalization scale $\Lambda$. In particular, the temperature must be far above the pseudocritical temperature $\Tc$ of the predicted crossover between the hadronic phase of nuclear matter and the QGP phase at $\mu=0$.
On the other hand, lattice computations tackle the path integral numerically. For convergence it is important that the measure of the path integral is strictly positive. A non-vanishing quark chemical potential, however, manifests in an imaginary part of the action \cite{Hasenfratz:1983ba}. The measure of the path integral becomes not strictly positive, and one runs into the famous sign problem of QCD which makes direct simulations at $\mu \neq 0$ in practice impossible. Numerous ways have been developed to circumvent that problem, one of which is the Taylor expansion of the pressure in terms of the quark chemical potential $\mu$ around $\mu=0$ \cite{Gavai:2001fr,Allton:2002zi,Mondal:2021jxk}. The Taylor coefficients itself are computed as increasingly complicated correlation functions at $\mu = 0$. Currently, there are numerical results up to the sixth order in $\mu$ \cite{Bazavov:2017dus}. From the ratio of the coefficients, one can also estimate the radius of convergence of the Taylor series and therefore infer a lower bound on the location of the conjectured tricritical point on the $\mu$-axis of the QCD phase diagram. 

At temperatures of a few times the crossover temperature $\Tc$, QCD thermodynamics does not yet behave fully perturbatively. The reason for that are the gluon Matsubara zero modes which contribute a factor of the inverse strong coupling constant $1/g$ to each closed loop and therefore modify the $g^2$-suppression per loop to $g^2/g=g$ \cite{Linde:1980ts}. Fortunately, these problematic modes can be separately treated in a three-dimensional effective theory, 'electrostatic QCD' (EQCD) \cite{Appelquist:1981vg,Nadkarni:1982kb,Braaten:1995cm}.
Beyond the advantage of allowing for a separate treatment and resummation of the Matsubara zero modes, EQCD also has the upside of a much milder sign problem. In fact, an analytic continuation of the quark chemical potential is possible, enabling us to solve EQCD directly on the lattice at moderate non-vanishing $\mu$.
 
The goal of this work is twofold: 
First, we compute Taylor coefficients from EQCD at vanishing chemical potential in order to compare to both lattice results at temperatures slightly above $\Tc$ and perturbative predictions at very high temperature. The higher cumulants have proven to be an especially well-suited quantity to judge the lower end of the range of validity of EQCD in the past \cite{Hietanen:2008tv}. As even higher cumulants are expected to be even less dominated by ultraviolet effects, this should also hold for the higher cumulants. This part also serves as a crosscheck of our work.
Second, we give numerical values of the quark number susceptibility up to sixth order directly simulated at nonvanishing chemical potential $\mu$, a region that is still inaccessible to traditional lattice simulations.

The paper is organized as follows: We give the formulae for the dimensional reduction procedure from full QCD to EQCD in Sec.~\ref{sec:thermo_from_EQCD} and specify our scenarios of interest. A recipe for the derivation of operators for the higher cumulants and how to analytically continue them to make them accessible to lattice simulations is given in Sec.~\ref{sec:higher_cumulants}. Sec.~\ref{sec:latt_implement} contains the details of our lattice implementation, in particular how the operators presented in the previous section translate to the lattice. Numerical results, for second to sixth derivatives with respect to the quark chemical potential $\mu$ are given in Sec.~\ref{sec:results}, and compared to predictions from perturbation theory and four-dimensional lattice simulations. We conclude with Sec.~\ref{sec:concl}. 
Appendices contain a table with our simulation parameters, analytic continuation of the derivative expressions beyond the third order, and comprehensive tables with values for all derivatives at all temperatures.

\section{Thermodynamics from Electrostatic Quantum Chromodynamics}
\label{sec:thermo_from_EQCD}
At high temperatures thermal four-dimensional QCD reduces to electrostatic QCD, which only treats the gluonic Matsubara zero mode dynamically and hides all other modes -- gluonic and fermionic -- in effective field theory parameters. The action of EQCD reads
\begin{align}	\label{eq:EQCD_cont_action}
S_{\mathrm{EQCD}} &= \int \! \d^3x 
\bigg( \frac{1}{2 \gsq} \Tr \, F^{ij} F^{ij} + \Tr \, D^i \Phi D^i \Phi \notag \\
& \hspace{53pt} + \mDsq \Tr \, \Phi^2 + i \gamma \, \Tr \, \Phi^3 + \lambda \, (\Tr \, \Phi^2)^2 \bigg) \, ,
\end{align}
with the now dimensionful gauge coupling $\gsq$.
Since all derivatives in time direction vanish, gauge freedom does not prevent the former $A^0$ field from acquiring a screening mass $\mDsq$. Therefore, the zero component of the four-dimensional gauge field $A^0$ turns into SU(3) adjoint representation scalar field $\Phi = \phi^a T^a$, where $T^a = \lambda^a/2$, $a=1\ldots 8$ are the Gell-Mann matrices. The gluon self-interaction generates a four-point interaction proportional to $\lambda$. The term cubic in $\Phi$ explicitly violates the otherwise valid mirror symmetry $\Phi \to - \Phi$. It can be directly linked to a nonvanishing quark chemical potential $\mu$ in four dimensions.

Due to its three dimensional nature, EQCD is a super-renormalizable theory, meaning that all correlation functions can be rendered finite at all orders with a finite number of counterterms. In particular, only the screening mass $\mDsq$ in \ref{eq:EQCD_cont_action} receives counterterms (at 1 and 2 loop order), which therefore is the only scale-dependent parameter. Consequently, one can use the dimensionful gauge coupling $\gsq$ to set the scale and proceed with the dimensionless ratios
\begin{equation}
y = \frac{\mDsq}{\gfour} \bigg\vert_{\Lambda = \gsq} \, , \qquad x = \frac{\lambda}{\gsq} \, , \qquad z = \frac{\gamma}{\gthree} \, .
\end{equation}

\subsection{EFT parameter matching}
The values of the EQCD parameters can be rigorously derived from a perturbative matching computation \cite{Kajantie:1997tt,Hart:2000ha}. The starting point is the 1 loop running coupling of full QCD
\begin{equation}	
g^2 = \; \frac{24 \pi^2}{33 - 2 \nf} \frac{T}{\ln \left( \frac{\overline{\Lambda}_\mathrm{g}}{\Lambda_{\overline{\mathrm{MS}}}} \right)} \, .
\label{eq:match_gsq} 
\end{equation}
Consulting the overview over the state of the matching in Sec.~6.3.1 of \cite{Ghiglieri:2020dpq}, we find that the precision bottleneck is the $z$-matching, which is only available at $\OO(g^3)$. Therefore, we can express the explicit dependence of the EFT parameters on $g^2$ through $x$, and truncate the expansion of the other parameters in powers of $x$ at $\OO(x)$, since parametrically $x \sim g^4/g^2 = g^2$.
\begin{align}
x =& \; \frac{9 - \nf}{33 - 2 \nf} \frac{1}{\ln \left( \frac{\hat{\mu} \mu_\mathrm{T}}{\Lambda_{\overline{\mathrm{MS}}}} \right)} + \OO(x) \label{eq:match_x} \\
y =& \; \frac{(9 - \nf) (6 + \nf)}{144 \pi^2 x} 
+ \sum_{i=1}^{\nf} \frac{3 (9 - \nf)}{144 \pi^2 x} \bmi^2  \notag \\
&+ \frac{486 - 33 \nf - 11 \nf^2 - 2 \nf^3}{96 \pi^2 (9 - \nf)} 
+ \sum_{i=1}^{\nf} \frac{3 (7 + \nf)}{96 \pi^2} \bmi^2 + \OO(x) \notag \\
\equiv& \; y_0 + y_2 \sum_{i=0}^{\nf} \bmi^2\label{eq:match_y}  \\
z =& \, \sum_{i=1}^{\nf} \frac{\bmi}{3 \pi} + \OO(x) \label{eq:match_z} \, .
\end{align}
Here $\bar\mu_i = \mu_i/T$, and $\mu_i$ is the quark chemical potential for flavor $i$.
Relevant scales for the matching are 
\begin{equation}
\hat{\mu} = \exp \left( \frac{\ln 4 \; (36 \nf - 4 \nf^2) - 162 + 30 \nf + 4 \nf^2}{(66 - 4 \nf)(9 - \nf)} \right) \, ,
\end{equation}
and 
\begin{equation}
\overline{\Lambda}_\mathrm{g} = 4 \pi T \exp \left( \frac{-3 + 4 \nf \ln 4}{66 - 4 \nf} - \gammaE \right) \, .
\end{equation}
For the (perturbative) scale-setting, we use $\LMSb = 341 \, \mathrm{MeV}$ determined in \cite{Bruno:2017gxd} for three-flavor QCD. We investigate EQCD corresponding to the six scenarios of full QCD in Tab.~\ref{tab:matching}. 
\begin{table}[ptbh] 
\centering {\small
\begin{tabular}{|c|c|c|c|c|c|}	
\hline 
$T \; \left[ \mathrm{MeV} \right]$ & $\nf$ & $g^2$ & $x$ & $y_0$ & $y_2$  \\
\hline
$277$ & $3$ & $4.389885$ & $0.1143767$ & $0.3733574$ & $0.126563$ \\
$400$ & $3$ & $3.708094$ & $0.0961857$ & $0.4361834$ & $0.147505$ \\
$600$ & $3$ & $3.165581$ & $0.0818254$ & $0.5055096$ & $0.170614$ \\
$2500$ & $4$ & $2.188389$ & $0.0446669$ & $0.7981834$ & $0.243255$ \\
$25000$ & $5$ & $1.517625$ & $0.0237276$ & $1.2509489$ & $0.346349$ \\
$100000$ & $5$ & $1.260187$ & $0.0199537$ & $1.4977310$ & $0.413653$ \\
\hline
\end{tabular} }
\caption{
  Our six scenarios of interest. 
}
\label{tab:matching}
\end{table}
An analytic continuation $i z \to z$ will finally allow us to get rid of the prefactor of $i$ in front of the cubic term in \eqref{eq:EQCD_cont_action}. This procedure was found to be highly superior to the more naive continuation of $\bmu$ \cite{Hietanen:2008tv}, since we only continue the linear contributions in $\bmu$, which appear in $z$, and leave the quadratic, i.e.\ real, contributions by $\bmu$ to $y$ (see \eqref{eq:match_y}) untouched. For the time being, especially for the derivation of the investigated operators, we keep the cubic term in its original complex form, and perform the analytic continuation of action and operators just before discretizing the theory on the lattice.
For each of the scenarios in Table \ref{tab:matching}, we simulate at six different values of the (then analytically continued) cubic coupling
\begin{equation}
z = \sum_{i=1}^{\nf} \frac{\mu_i}{3 \pi^2 T} = 0.0, \, 0.025, \, 0.05, \, 0.1, \, 0.15, \, 0.2 \, .
\end{equation}

\subsection{Matching the pressure to full QCD}
Our goal is to study different susceptibilities, i.e.\ derivatives of the pressure, in full QCD using simulations of EQCD. To this end, we do not only have to match the parameters in the EQCD action to full QCD, but also the pressure itself.
The philosophy for this is quite simple; we take the perturbative pressure of QCD,  subtract its perturbative EQCD counterpart and substitute it with the lattice EQCD result.
The remaining perturbative part is UV-dominated and can be reliably evaluated in perturbation theory:
\begin{align}
p_\mathrm{QCD} &= \left( p_\mathrm{QCD} - p_\mathrm{EQCD} \right) + p_\mathrm{EQCD} \notag \\
&\approx \left( p^\mathrm{PT}_\mathrm{QCD} - p^\mathrm{PT}_\mathrm{EQCD} \right) + p^\mathrm{latt}_\mathrm{EQCD} \; .
\end{align}
Depending on which derivative we want to investigate, we take derivatives of the above expression with respect to a given number of $\mu_i$.
The four-dimensional pressure was computed perturbatively up to $\OO(g^5)$ in ref.~\cite{Ipp:2006ij}. This computation also contains the three-dimensional pressure, which is divergent by itself, though. These divergences cancel against divergences in the hard sector in the four-dimensional case. In the purely three-dimensional case, however, we need to introduce a divergent cosmological constant into the EQCD action that cures these divergences.

\section{Higher-order cumulants in EQCD}
\label{sec:higher_cumulants}
Beyond the second order derivatives studied in \cite{Hietanen:2008tv}, one can also investigate higher derivatives of the pressure with respect to different quark chemical potentials, sometimes also called cumulants. Since we only consider massless quarks, they are degenerate and we do not need to specify the name of the quark, e.g.\
\begin{equation}
\chi^{ij} = \chi^{ud} = \chi^{us} = \chi^{ds} \, ,
\end{equation}
and conversely for higher derivatives.
Even though a brute force derivation of the corresponding correlation functions in EQCD is possible, it can be dramatically simplified with a little additional work.

\subsection{Derivation of higher order cumulants}
We calculate derivatives of the pressure, i.e.\
\begin{equation}
p(T,\bmu) = - f(T,\bmu) = \lim_{V \to \infty} \frac{1}{V} \ln \mathcal{Z}(T,\bmu) \, ,
\end{equation}
with the free energy density $f$ and the volume $V$.
Recalling that only fully connected correlation functions contribute to the free energy -- and consequently the pressure -- helps simplifying the calculation drastically. Subtractions of partly connected correlation functions can therefore be dropped in the following, and re-instated in the end. This also means that we only have to keep track of all derivatives acting directly on $\mathcal{Z}$, since the derivatives acting on the $1/\mathcal{Z}$-normalization of correlation functions generate the partly connected subtractions.
Moreover, we need to take a closer look at the ($\bmu$-dependent) part of the EQCD action
\begin{equation}
S_\mathrm{EQCD}^{\bmu} = \int \! \d^3x \; y_2 \sum_f \bmf^2 \Tr \, \Phi^2(\bx) + i \frac{1}{3 \pi} \sum_f \bmf \Tr \, \Phi^3(\bx) \, .
\end{equation}
Derivatives can only act on $\mathcal{Z}$ in two different ways: either as a single derivative on the exponential function or as a double derivative\footnote{We simplified the notation as $1/V \sum_{\bx} \Tr \, \Phi^2(x) \to \Phi^2$, and analogously for $\Phi^3$.}
\begin{align}	
- \frac{\partial S_\mathrm{EQCD}}{\partial \bmi} &= - 2 y_2 \bmi \Phi^2 - i \frac{\Phi^3}{3 \pi} \label{eq:single_der_action} \\
- \frac{\partial^2 S_\mathrm{EQCD}}{\partial \bmi \partial \bmj} &= - 2 y_2 \delta_{ij} \Phi^2 \label{eq:double_der_action} \, .
\end{align}
Note the minus sign due to the weight of the exponential in the Euclidean path-integral being $\exp(- S_\mathrm{EQCD})$.
Calculating ever-higher derivatives of the pressure thus amounts to combining the two above expressions in all possible ways maintaining the symmetry in the indices. Similar to the subtractions, it is sufficient to keep only specific combinations of the indices and re-instate the symmetry of the derivative in the end, bearing in mind that derivatives acting on continuous functions commute. Since the dimensional reduction to EQCD only works away from any phase transition, i.e.\ a non-analyticity in the free energy $y$, anyway, it is safe to assume that derivatives with respect to the quark chemical potential $\bmu$ acting on the pressure commute.
As a first pedagogical example, we re-derive the quadratic quark number susceptibility, as used in \cite{Hietanen:2008tv}. Since it is a second derivative, \eqref{eq:single_der_action} can only contribute as 
\begin{equation}
\left( -\frac{\partial S_\mathrm{EQCD}}{\partial \bmi} \right)
\left( -\frac{\partial S_\mathrm{EQCD}}{\partial \bmj} \right) \, 
\end{equation}
and \eqref{eq:double_der_action} as 
\begin{equation}
\left( -\frac{\partial^2 S_\mathrm{EQCD}}{\partial \bmi \partial \bmj} \right) \, ,
\end{equation}
where the symmetrization in the indices $i$ and $j$ is already intact and does not yet need to be enforced at this stage.
We combine the two above terms to
\begin{align}	\label{eq:quadratic_sus}
\chi_{ij} &= 4 y_2^2 \bmi \bmj K_{2,1} + i \frac{2 y_2}{3 \pi} (\bmi + \bmj) K_{2,2} - \frac{1}{(3 \pi)^2} K_{2,3} - 2 \delta_{ij} y_2 K_{1,1}  
\end{align}
with the help of the condensates
\begin{align*}
K_{1,1} &\equiv \langle \Phi^2 \rangle \\
K_{2,1} &\equiv V \gsix \langle \left( \Phi^2 - \langle \Phi^2 \rangle \right)^2 \rangle \\
K_{2,2} &\equiv V \gsix \langle \left( \Phi^2 - \langle \Phi^2 \rangle \right) \left( \Phi^3 - \langle \Phi^3 \rangle \right) \rangle \\
K_{2,3} &\equiv V \gsix \langle \left( \Phi^3 - \langle \Phi^3 \rangle \right)^2 \rangle \, . 
\end{align*} 
Let us caution the reader once more that the subtractions of the partly connected contributions have to be re-instated in the end. Keeping track of all the subtractions throughout the derivation does not provide any further insights about the underlying mechanisms.

As a first non-trivial example, we present the derivation of the expression for the cubic derivative of the pressure with respect to the chemical potential. Once more, this can only receive two possible contributions:
\begin{equation}
\left( -\frac{\partial S_\mathrm{EQCD}}{\partial \bmi} \right)
\left( -\frac{\partial S_\mathrm{EQCD}}{\partial \bmj} \right)
\left( -\frac{\partial S_\mathrm{EQCD}}{\partial \bmk} \right)
\end{equation}
and 
\begin{equation}
\left( -\frac{\partial S_\mathrm{EQCD}}{\partial \bmi} \right)
\left( -\frac{\partial^2 S_\mathrm{EQCD}}{\partial \bmj \partial \bmk} \right) + \dots \, ,
\end{equation}
where the dots will denote the symmetrizations in the respective indices in the following.
We can combine these two possibilities to 
\begin{align}	\label{eq:cubic_sus}
\chi_{ijk} =& 
	- 8 y_2^3 \bmi \bmj \bmk K_{3,1}
	- i \frac{4 y_2^2}{3 \pi} 
	\left( \bmi \bmj + \bmj \bmk + \bmk \bmi \right) K_{3,2} \notag \\
	&+ \frac{2 y_2}{9 \pi^2} \left( \bmi + \bmj + \bmk \right) K_{3,3}
	+ i \frac{1}{(3 \pi)^3} K_{3,4} \notag \\
	&+ 4 y_2^2 \left( \delta_{ij} \bmk + \delta_{jk} \bmi + \delta_{ki} \bmj \right) K_{2,1}
	+ i \frac{2 y_2}{3 \pi} K_{2,2} \, ,
\end{align}
where we defined the condensates
\begin{align*}
K_{3,1} &\equiv V^2 \gE^{12} \Big\langle 
	\left( \Phi^2 - \langle \Phi^2 \rangle \right)^3 \Big\rangle\\
K_{3,2} &\equiv V^2 \gE^{12} \Big\langle 
	\left( \Phi^2 - \langle \Phi^2 \rangle \right)^2 
	\left( \Phi^3 - \langle \Phi^3 \rangle \right) \Big\rangle \\
K_{3,3} &\equiv V^2 \gE^{12} \Big\langle 
	\left( \Phi^2 - \langle \Phi^2 \rangle \right) 
	\left( \Phi^3 - \langle \Phi^3 \rangle \right)^2 \Big\rangle \\
K_{3,4} &\equiv V^2 \gE^{12} \Big\langle 
	\left( \Phi^3 - \langle \Phi^3 \rangle \right)^3 \Big\rangle \, .
\end{align*}
Considering \eqref{eq:cubic_sus}, we observe that by setting $\bmu=0$, only $K_{2,2}$ and $K_{3,4}$ survive. These two condensates, in turn, are odd in $\Phi^3$, therefore they should be consistent with $0$ numerically at vanishing $z$.\footnote{We note that 
the EQCD action \ref{eq:EQCD_cont_action} can break the symmetry $\Phi \leftrightarrow -\Phi$ spontaneously, leading to non-zero $\langle \Phi^3\rangle$ \cite{Kajantie:1998yc}.  However, the dimensional reduction of hot QCD maps to the symmetric phase of EQCD with $\langle\Phi\rangle =0$ if $z=0$.}
Consequently, \eqref{eq:cubic_sus} reflects the expectation that an odd derivative in $\bmu$ should vanish at vanishing $z$.

\subsection{Analytic continuation}
We still cannot compute $K_{i,j}$ -observables defined above directly on the lattice due to the imaginary part of the action. This can be cured by analytic continuation of $z$. It was found in \cite{Hietanen:2008tv} that this way of analytic continuation is highly superior to a more straight-forward continuation in the full quark chemical potential $\mu$, since the leading contribution of $\mu$ is quadratically through $y$. An analytic continuation in $z$ has the advantage of leaving $y$ unchanged and only modifying the imaginary part of the action.
As for the second derivatives, the condensates contributing to the higher order derivatives also only mildly vary with $iz$ and therefore can be expanded in powers of $iz$. Depending on how many powers of $\langle \Phi^3 - \langle \Phi^3 \rangle \rangle$ a condensate contains it contains either exclusively even or exclusively odd powers of $iz$. 
Consequently, we apply the following procedure for the analytic continuation of the $n$-th derivative:
\begin{enumerate}
\item We expand the condensates with the highest sum of powers of $\Phi^2$ and $\Phi^3$ to first non-trivial order.
\item The coefficients of condensates with the highest power sum also re-occur in condensates with a lower power sum. We expand the lower condensates to consistent order in $iz$, i.e.\ the order where the highest power sum coefficients become relevant.
\item We analytically continue $iz \to z$.
\item We re-phrase the expanded expressions in terms of the respective condensates at imaginary chemical potential $\zI$. As a matter of choice, we set $\zR = \zI$ in order to make the factors of $\zI / \zR$ unity.
\end{enumerate} 
As a first instructive example, we start with the analytic continuation of \eqref{eq:quadratic_sus}, as already done in \cite{Hietanen:2008tv}. We start with expanding 
\begin{align}
K_{2,1} &= V \gsix \big\langle \left( \Phi^2 - \langle \Phi^2 \rangle \right)^2 \big\rangle \approx a_{21,0} \notag \\
K_{2,2} &= V \gsix \big\langle \left( \Phi^2 - \langle \Phi^2 \rangle \right) \left( \Phi^3 - \langle \Phi^3 \rangle \right) \big\rangle \approx  i a_{22,1} z \notag \\
K_{2,3} &= V \gsix \big\langle \left( \Phi^3 - \langle \Phi^3 \rangle \right)^2 \big\rangle \approx a_{23,0} \, ,
\end{align}
where coeffcients $a_{ij,k}$ are assumed to be independent of $z$.
In the next step, we notice that $K_{2,2}$ and $K_{1,1}$ are related by
\begin{equation}
K_{2,2} = \frac{\partial K_{1,1}}{\partial (iz)} \, ,
\end{equation}
so in order to obtain a nontrivial contribution by $K_{2,2}$ which also contributes everywhere else consistently, we expand
\begin{equation}
K_{1,1} = \big\langle \Phi^2 \big\rangle \approx a_{11,0} + a_{11,2} z^2
\end{equation}
and identify
\begin{equation}
a_{22,1} = -2 a_{11,2} \, .
\end{equation}
When performing the analytic continuation, we therefore end up with 
\begin{align}	\label{eq:ac_quadratic_sus}
\left( \chitd^{ij} \right)^{\mathrm{ac}} =& \; 4 y_2^2 \bmi \bmj K_{2,1}(\zI) + \frac{2 y_2}{3 \pi} \frac{\zR}{\zI} (\bmi + \bmj) K_{2,2}(\zI) - \frac{1}{(3 \pi)^2} K_{2,3}(\zI) \notag \\
&- 2 \delta_{ij} y_2 \left( K_{1,1}(\zI) + \frac{\zR^2}{\zI} K_{2,2}(\zI) \right) \, ,
\end{align}
where all condensates $K$ now have $\zI$ as an argument as opposed to $\zR$ before. As a matter of choice, we decide to set $\zR=\zI$ to force all ratios of $\zR/\zI$ to unity.

Applying the same procedure to the third derivative in \eqref{eq:cubic_sus}, we obtain the condensates
\begin{align*}
K_{2,1} &= V \gsix \Big\langle 
	\left( \Phi^2 - \langle \Phi^2 \rangle \right)^2 \Big\rangle
\approx a_{21,0} + a_{21,2} z^2 \\
K_{2,2} &= V \gsix \Big\langle 
	\left( \Phi^2 - \langle \Phi^2 \rangle \right)
	\left( \Phi^3 - \langle \Phi^3 \rangle \right) \Big\rangle
\approx - 2 i z a_{11,2} \\
K_{3,1} &\equiv V^2 \gE^{12} \Big\langle 
	\left( \Phi^2 - \langle \Phi^2 \rangle \right)^3 \Big\rangle 
\approx a_{31,0} \\
K_{3,2} &\equiv V^2 \gE^{12} \Big\langle 
	\left( \Phi^2 - \langle \Phi^2 \rangle \right)^2 
	\left( \Phi^3 - \langle \Phi^3 \rangle \right) \Big\rangle
= \frac{\partial K_{2,1}}{\partial (iz)}
\approx - 2 i z a_{21,2} \\
K_{3,3} &\equiv V^2 \gE^{12} \Big\langle 
	\left( \Phi^2 - \langle \Phi^2 \rangle \right) 
	\left( \Phi^3 - \langle \Phi^3 \rangle \right)^2 \Big\rangle \\
&= \frac{\partial K_{2,2}}{\partial (iz)} = \frac{\partial^2 K_{1,1}}{\partial^2 (iz)}
\approx - 2 a_{11,2} \\
K_{3,4} &\equiv V^2 \gE^{12} \Big\langle 
	\left( \Phi^3 - \langle \Phi^3 \rangle \right)^3 \Big\rangle
= \frac{\partial K_{2,3}}{\partial (iz)} 
\approx - 2 i z a_{23,2} \, .
\end{align*} 
After analytic continuation we obtain
\begin{align}	\label{eq:ac_cubic_sus}
\left( \chitd^{ijk} \right)^{ac} =& 
	- 8 y_2^3 \bmi \bmj \bmk K_{3,1}(z_I) \notag \\
&- 	\frac{4 y_2^2}{3 \pi} 
	\left( \bmi \bmj + \bmj \bmk + \bmk \bmi \right) 
	\frac{z_R}{z_I} K_{3,2} (z_I) \notag \\
&+ 	\frac{2 y_2}{9 \pi^2} \left( \bmi + \bmj + \bmk \right) 
	K_{3,3}(z_I)
+ 	\frac{1}{(3 \pi)^3} \frac{z_R}{z_I} K_{3,4}(z_I) \notag \\
&+ 	4 y_2^2 
	\left( \delta_{ij} \bmk + \delta_{jk} \bmi + \delta_{ki} \bmj \right) 
	\left( K_{2,1}(z_I) + \frac{z_R^2}{z_I} K_{3,2}(z_I) \right) \notag \\
&+ \frac{2 y_2}{3 \pi} \frac{z_R}{z_I} K_{2,2}(z_I) \, .
\end{align}
The procedure outlined in this section can be easily generalized to even higher orders. We present the analytically continued expressions for the quartic, quintic, and sextic derivatives in Appendix \ref{app:express_higher_der} for the sake of readability.

\section{Lattice implementation}
\label{sec:latt_implement}
Discretizing \eqref{eq:EQCD_cont_action} on a spatial grid with finite lattice spacing $a$ yields
\begin{align}	\label{eq:EQCD_latt_action}
S_{\mathrm{EQCD},L} =& \; \beta \sum_{x, i>j} \left( 1 - \frac{1}{3} \Box_{x,ij} \right) \notag \\
&+ 2 \sum_{x,i} \Tr \, \left( \Phi_\mathrm{L}^2(x) - \Phi_\mathrm{L}(x) U_i(x) \Phi_\mathrm{L}(x + a \hat{i}) U^\dagger_i(x) \right) \notag \\
&+ \sum_x Z_4 (x + \delta x) \left( \Tr \, \Phi_\mathrm{L}^2(x) \right)^2 + Z_2 (y + \delta y) \Tr \, \Phi^2_\mathrm{L} (x) + Z_3 z \Tr \, \Phi^3_\mathrm{L}(x) \\
\Box_{x,ij} \equiv& \; U_i(x) U_j(x + a \hat{i}) U^\dagger_i(x + a \hat{j}) U^\dagger_j(x) \, ,
\end{align}
with the spatial link variables $U_i(x)$ connecting lattice sites $x$ and $x + a \hat{i}$, and the rescaled lattice version of the scalar field $\Phi_\mathrm{L}$. Both $x$ and $y$ receive multiplicative and additive renormalization. Since EQCD is a super-renormalizable field theory, only a finite number of divergent counterterms suffice to render all amplitudes of the theory finite. In particular, only the scalar mass term $y$ receives a divergent contribution, which can be eliminated analytically by a calculation of $\delta y$ in lattice perturbation theory \cite{Moore:1997np}. Beyond the elimination of divergent bits, it is also possible to pursue the lattice improvement procedure further in order to bring our lattice implementation of EQCD even closer to the continuum. While the elimination of $\OO(a)$ errors in the EFT parameters of EQCD was recently completed \cite{Moore:2019lua} for $z=0$, the $\OO(a)$ behavior of most of the condensates that we measure is not yet investigated. Moreover, the emergence of the $z$-term in the EQCD action can also modify the known $\OO(a)$ behavior of the EFT parameters. Therefore, we use a partly-$\OO(a)$-improved set of the parameters \cite{Moore:1997np}, meaning that $\beta$, $Z_4$, $\delta x$, and $Z_2$ are accurate to $\OO(a)$, while we set $Z_3=1$ and $\delta y$ is only utilized to $\OO(\ln a)$, meaning that both $z$ and $y$ still give rise to $\OO(a)$-correction on the Lagrangian level. A compendium of the lattice renormalization constants can be found in Appendix~A of \cite{DOnofrio:2014mld}. Note that $z$ does not acquire additive renormalization due to the $\Phi^3 \to - \Phi^3$-symmetry of the Lagrangian. 
 
\begin{figure}[htbp!] 
\centering 
	\begin{tabular}{ccc}
		\includegraphics[scale=0.55]{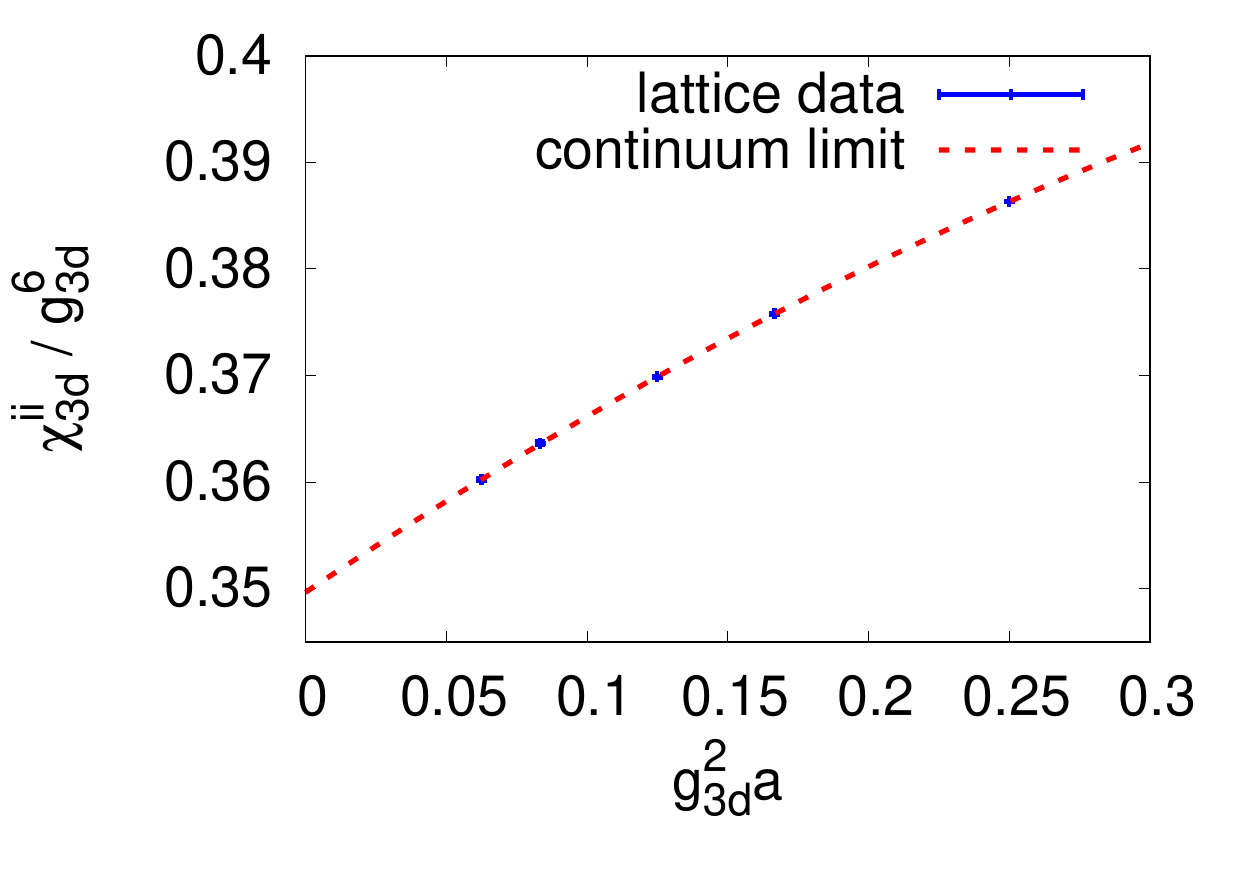} & 
		\includegraphics[scale=0.55]{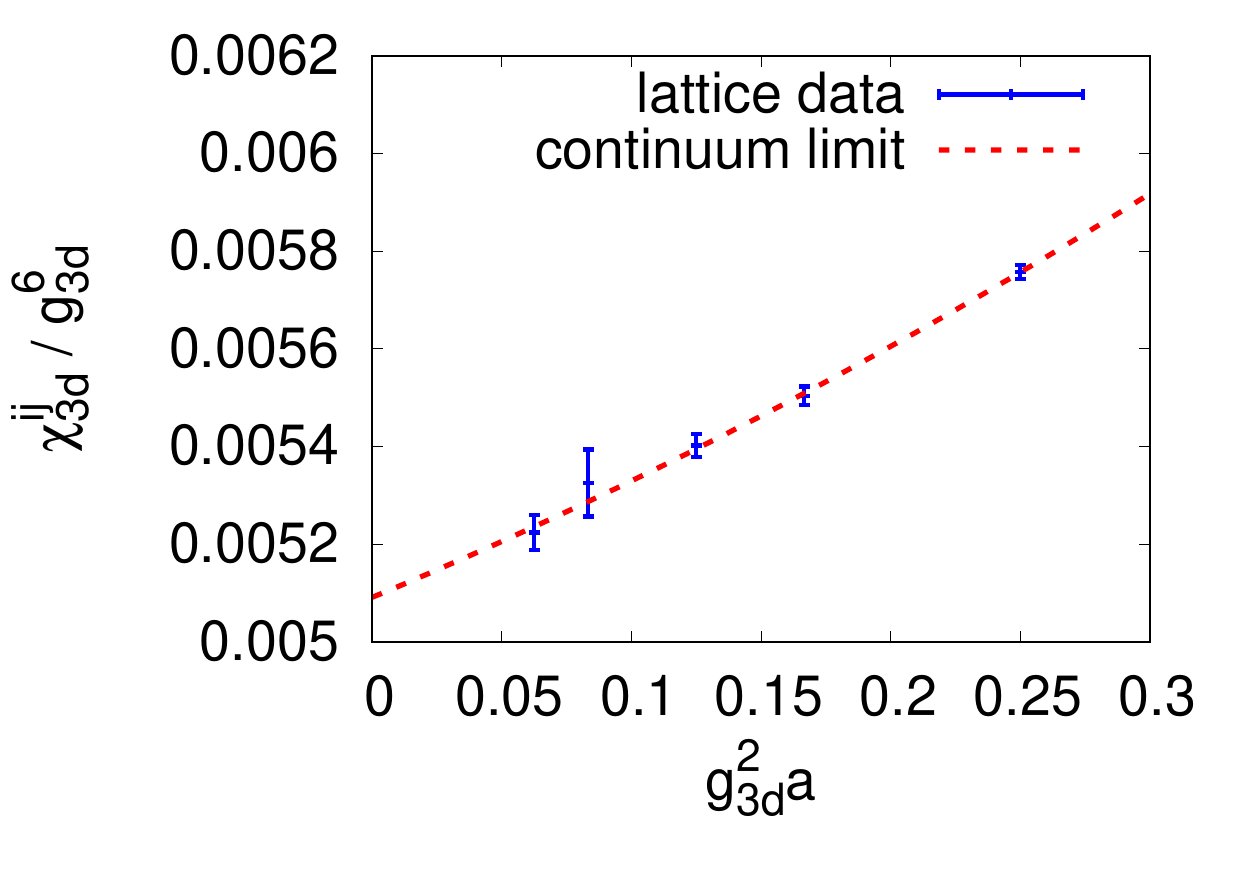} \\
		(a) $\chitd^{ii} / \gsix$ & (b) $\chitd^{ij} / \gsix$ \\
	\end{tabular}
	\caption{Continuum extrapolation of the diagonal and non-diagonal second order susceptibilities. Example at $T = 100~\mathrm{GeV}$, $\nf = 5$, and $z = 0.1$.}
	\label{cont_extr}
\end{figure}

One can argue on dimensional grounds that only $K_{1,1}$ and $K_{2,3}$ can receive divergent lattice corrections. In fact, they have been calculated \cite{Kajantie:1997tt,Laine:1997dy} and read
\begin{align}
K_{1,1}^\mathrm{cont} &= K_{1,1}^\mathrm{latt} - \frac{\tilde{c}_1}{\gsqa} - \tilde{c}_2 (\tilde{c}_2' - \ln \gsqa) + \OO(\gsqa) \\
K_{2,3}^\mathrm{cont} &= K_{2,3}^\mathrm{latt} - \bar{c}_2 (\bar{c}_2' - \ln \gsqa) + \OO(\gsqa) \, ,
\end{align} 
with
\begin{align}
\tilde{c}_1 &\approx 6 \times 0.1684873399 \\
\tilde{c}_2 &= \frac{3}{2 \pi^3}  \\
\tilde{c}_2' &\approx 0.6796(1) + \ln 6 \\
\bar{c}_2 &= \frac{5}{16 \pi^2} \\
\bar{c}_2' &\approx 0.08848010 + \ln 6 \, .
\end{align}
All other condensates follow
\begin{equation}
K^\mathrm{cont} = K^\mathrm{latt} + \OO(\gsqa) \, .
\end{equation}
Given the very mild dependence of the condensates, especially the higher cumulants, on the lattice cutoff $a$, full removal of all $\OO(a)$-lattice effects is not crucial for our work
  (see Fig.~\ref{cont_extr}) -- especially of the condensates for the higher cumulants -- on the lattice cutoff $a$, a
The line in the $(x,y)$-plane on which EQCD describes real-world QCD lies in the supercooled phase of EQCD \cite{Kajantie:1997tt}. However, the metastability of EQCD diminishes as one approaches the crossover temperature in full QCD. Since our smallest temperature in Table~\ref{tab:matching} is too close to the crossover temperature, we applied a bit of mass reweighting to the $z=0.0, 0.025, 0.05$ points at $T=277~\mathrm{MeV}$ to ensure that the simulation remains in the symmetric phase. This procedure additionally inflates the errors of these points.

Our lattice EQCD simulation program is modified from the one used in \cite{Moore:2019lua} and is based on the openQCD-1.6-package \cite{openQCD} by Martin Lüscher. A combined update of 2 heatbath sweeps followed by 9 over-relaxation sweeps in a checkerboard ordering through the entire volume is applied. The scalar kinetic term in \eqref{eq:EQCD_latt_action} is incorporated into the gauge sector update via a Metropolis step. The quartic scalar interaction also requires an accept/reject step in the scalar heatbath update. Both heatbath and over-relaxation are concluded by a final Metropolis step for the cubic term. 

In order to extrapolate our results to the continuum, we simulate each of the six temperatures in full QCD from Table~\ref{tab:matching} times at six different values for $z$ and at five different lattice spacings each:
\begin{equation}
\gsqa = \frac{1}{4} ,\, \frac{1}{6} ,\, \frac{1}{8} ,\, \frac{1}{12} ,\, \frac{1}{16} .
\end{equation}
In total this makes $180$ distinct simulations. The simulation parameters and the respective reached statistical power are tabulated in Appendix~\ref{app:sim_params}. Due to the full $\OO(a)$-improvement in EQCD not being known in our setup, we combine the condensates to the different susceptibilities at finite $a$ and extrapolate the susceptibilities to the continuum instead of the single condensates. As long as there are more condensates involved into the computation of a susceptibility than different variants of the susceptibility exist, this procedure is preferred since it minimizes the number of interpolations. We found it to be sufficient to fit a quadratic curve for each susceptibility, so that each fit has five data points and three fit parameters. An example for the continuum extrapolation is displayed in Fig.~\ref{cont_extr}. We see from here that the overall dependence on the lattice cutoff is mild and the data can be well described with a quadratic curve in $\gsqa$. 
Since EQCD possesses a mass gap, finite volume effects are exponentially suppressed. Therefore, it is sufficient to choose a volume that satisfies the empirical bound \cite{Hietanen:2008xb}
\begin{equation}
\frac{V}{a^3} \geq \left( \frac{6}{\gsqa} \right)^3 \, .
\end{equation}
To this end, we keep the physical volume fixed at 
\begin{equation}
\gsix \times V = 8^3 \, .
\end{equation}

\section{Results}
\label{sec:results}
We provide our continuum-extrapolated, tabulated results in three-dimensional units in Appendix~\ref{app:tab_results}. Comparing to four-dimensional predictions from the lattice and perturbation theory requires an application of the matching procedure outlined in Sec.~\ref{sec:thermo_from_EQCD}. 

\subsection{Second derivatives}
\begin{figure}[htbp!] 
\centering 
	\begin{tabular}{ccc}
		\includegraphics[scale=0.55]{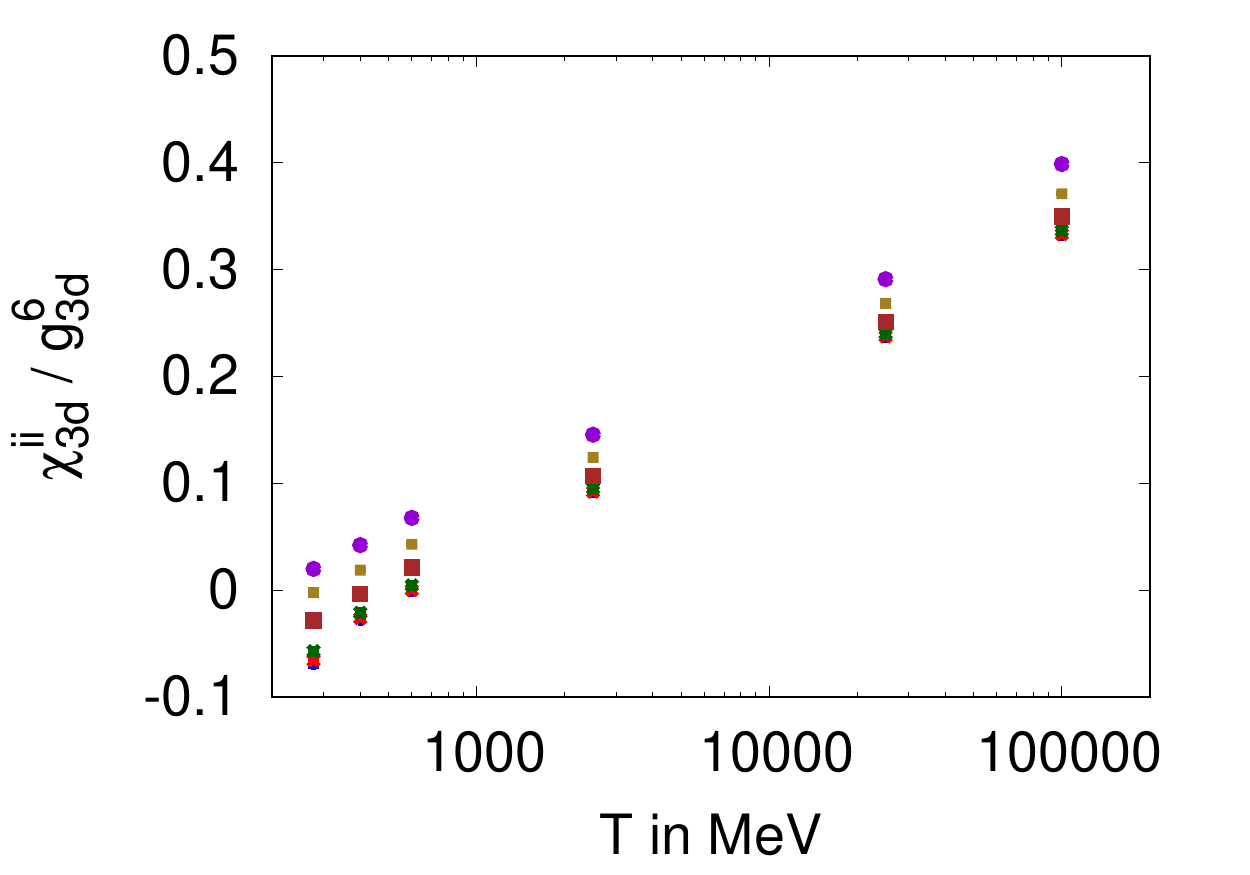} & 
		\includegraphics[scale=0.55]{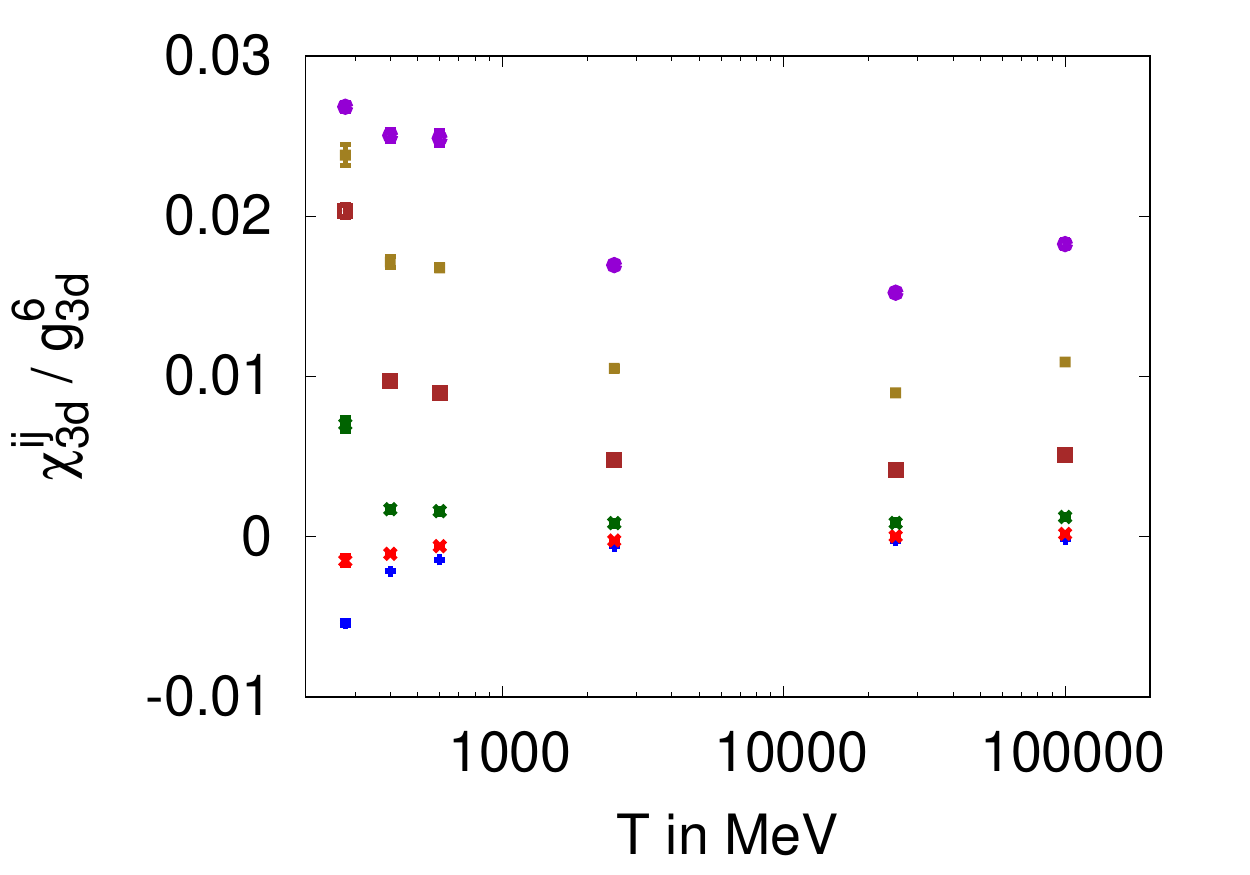} \\
		(a) & (b) \\
	\end{tabular}
	\caption{Second order susceptibilities in three-dimensional units as functions of the temperature. Blue points correspond to $z=0.0$, red to $z=0.025$, green to $z=0.05$, brown to $z=0.1$, olive to $z=0.15$, and violet to $z=0.2$.}
	\label{T_two}
\end{figure}

We start reviewing our simulation results with the qualitatively already known data for the quadratic quark number susceptibility, evaluated at different $\nf$ than in \cite{Hietanen:2008tv}. 
The quadratic susceptibilities are plotted as a function of the temperature in Fig.~\ref{T_two}. For the diagonal susceptibility, we essentially observe an exponential dependence on the temperature which is generated by the exponential temperature dependence of the screening mass term $y$ through the scale setting.
At lower temperatures close to $\Tc$, the diagonal susceptibility becomes slightly negative, a feature that is already known from \cite{Hietanen:2008tv}. Increasing the chemical potential $\bmu$ results in a mild increase of $\chitd^{ii}$, the curves are strictly ordered by their respective $z$.
As expected, the off-diagonal quark number susceptibility $\chitd^{ij}$ is numerically suppressed by about one order of magnitude compared to its diagonal counterpart $\chitd^{ii}$. The off-diagonal susceptibility vanishes at $\bmu=0$ for temperatures far away from the pseudocritical temperature of QCD. Coming closer to the crossover, the quality of the perturbative dimensional reduction decreases, so a deviation from the expected value of $0$ can be explained as an artifact of a poor dimensional reduction.
Deviations from $0$ set in as one turns on the chemical potential, also preserving the hierarchy in $z$ observed in the diagonal susceptibility.

\begin{figure}[htbp!] 
\centering 
	\begin{tabular}{cc}
		\includegraphics[scale=0.55]{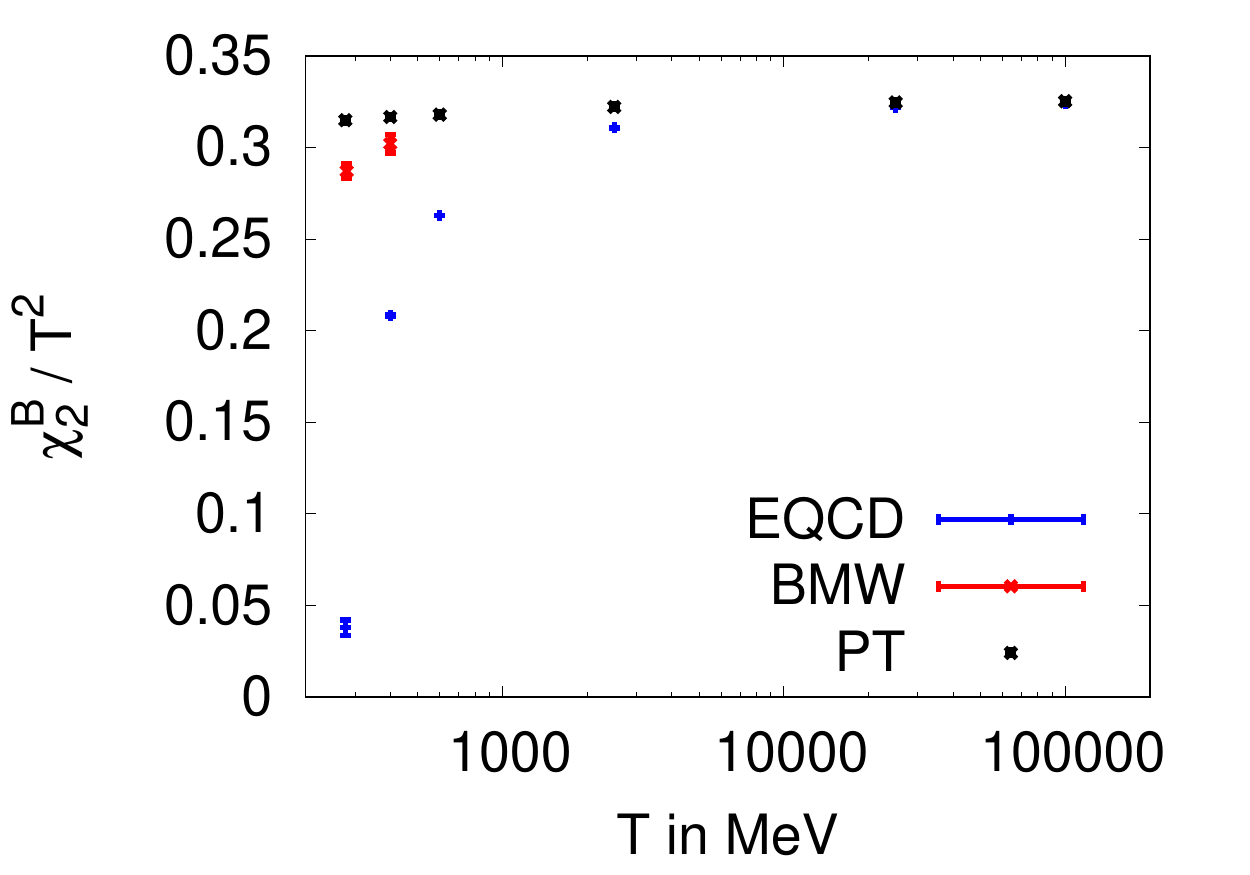} & 
		\includegraphics[scale=0.55]{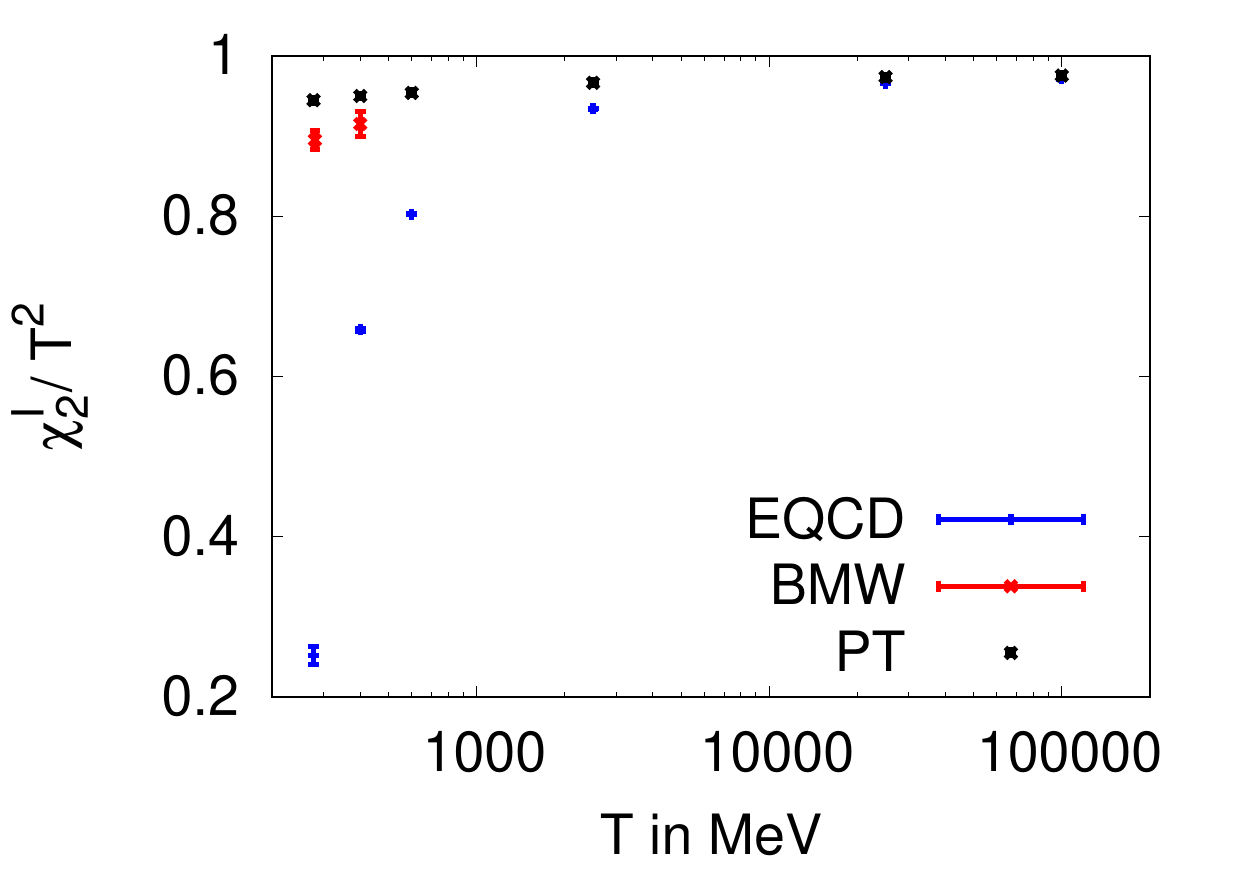} \\
		(a) & (b) \\
		\includegraphics[scale=0.55]{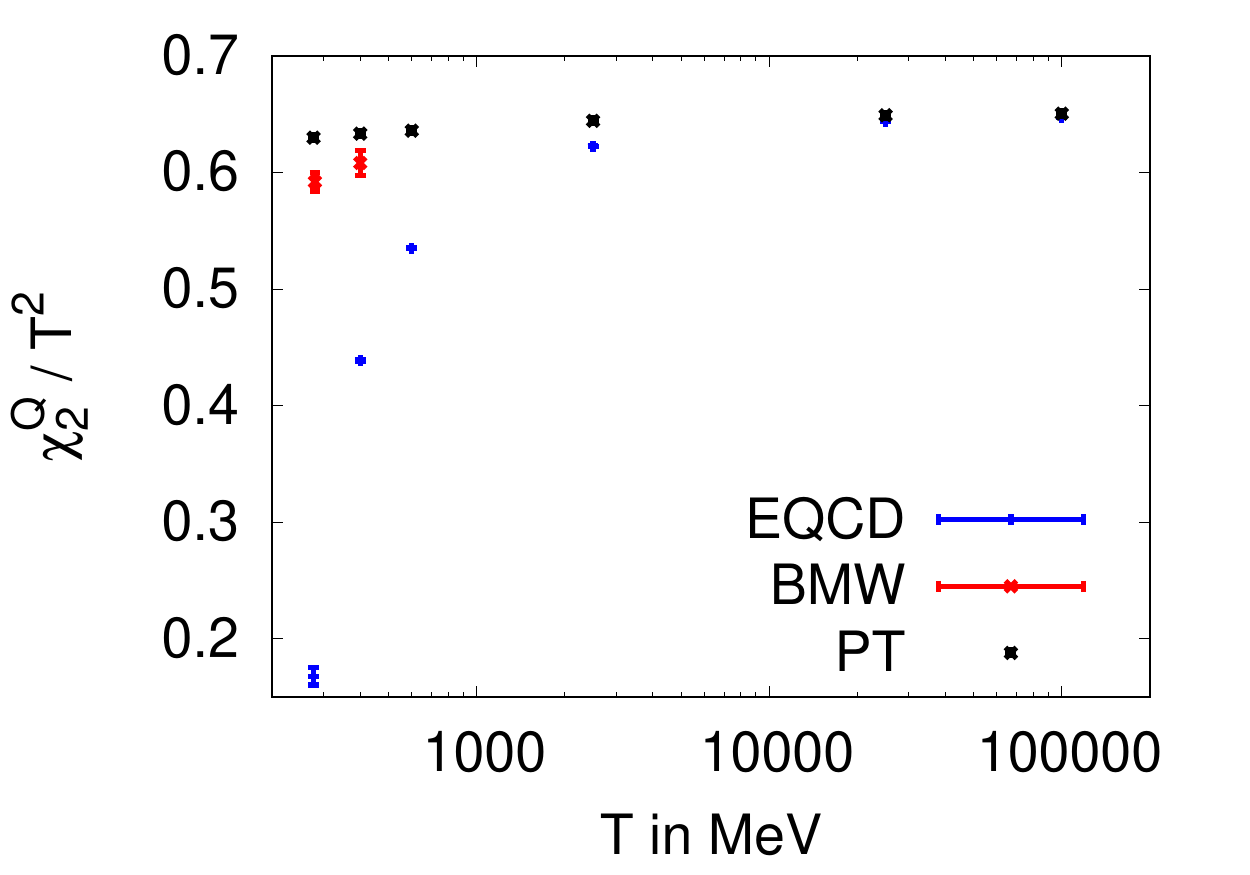} &
		\includegraphics[scale=0.55]{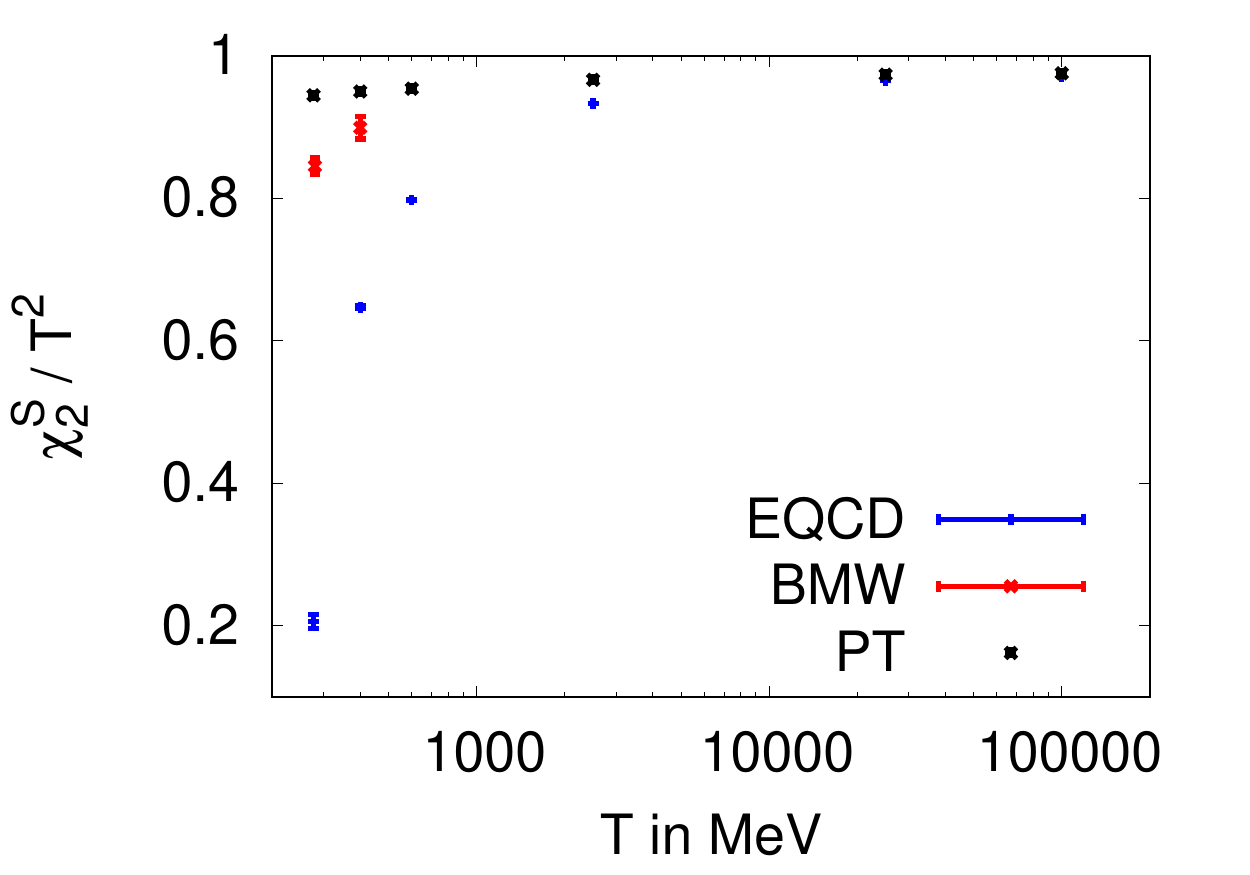} \\
		(c) & (d) \\
		\includegraphics[scale=0.55]{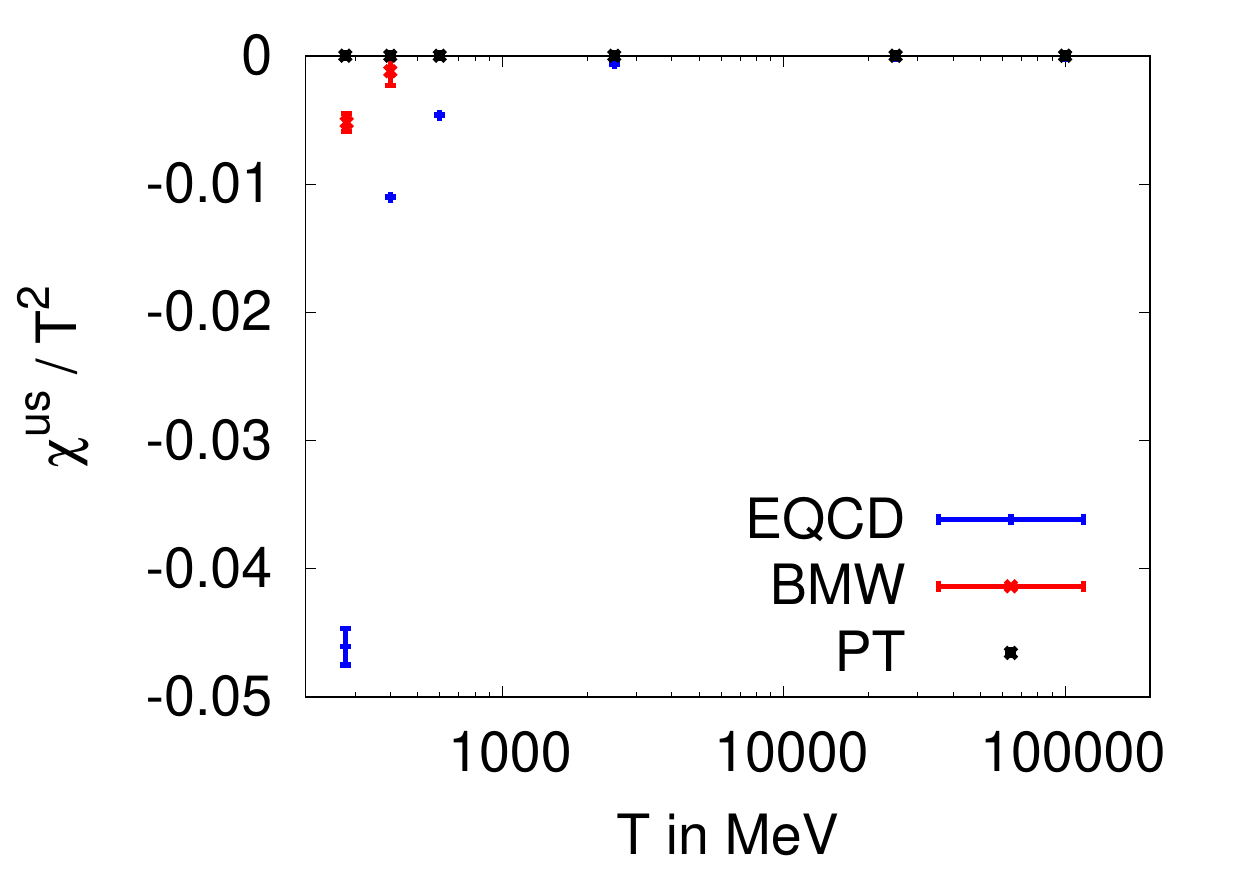} & \\ 
		(e)  &  \\
	\end{tabular}
	\caption{Comparison of lattice results \cite{Borsanyi:2011sw}, perturbative results \cite{Ipp:2006ij}, and our results for susceptibilities $\chi_2^B$ (a), $\chi_2^I$ (b),
 	$\chi_2^Q$ (c), $\chi_2^S$ (d), and $\chi^\text{us}$ (e) at $\bmu = 0$. }
	\label{comp_latt_PT}
\end{figure}

An important benchmark for our data is the comparison to full QCD lattice data \cite{Borsanyi:2011sw} on the small-T side and full QCD perturbation theory on the large-T side in Fig.~\ref{comp_latt_PT}. To this end, we have to apply the matching procedure outlined on Sec.~\ref{sec:thermo_from_EQCD}. Since we use a different number of massless quark flavors $\nf$ for different temperatures, it does not seem sensible to draw the perturbative prediction as a line. Therefore, we computed the values at the same temperatures as our lattice EQCD data. Comprehensive lattice data for all conserved charges ($B$, $I$, $Q$, and $S$) is only available for the quadratic susceptibilities, and we analyze their behavior before moving on to the higher cumulants.
The conserved-charge-susceptibilities are related to our diagonal and off-diagonal quark number susceptibilities via
\begin{align}
\chi_2^\mathrm{B} &= \frac{1}{3} \left( \chi_{4\mathrm{d}}^{ii} + 2 \chi_{4\mathrm{d}}^{ij} \right) \\
\chi_2^\mathrm{Q} &= \frac{2}{3} \left( \chi_{4\mathrm{d}}^{ii} - \chi_{4\mathrm{d}}^{ij} \right) \\
\chi_2^\mathrm{I} &= \frac{1}{2} \left( \chi_{4\mathrm{d}}^{ii} - \chi_{4\mathrm{d}}^{ij} \right) \\
\chi_2^\mathrm{S} &= \chi^{ii} \\
\chi^{\mathrm{us}} &= \chi^{ij} \, .
\end{align}
We see that agreement with lattice data is not even reached at $T=400~\mathrm{MeV}$, although the gap between our values and the full-QCD lattice prediction shrinks by a good margin as temperature increases. This seems to contrast the earlier experience from \cite{Hietanen:2008xb}. However, there are two important differences between the present work and \cite{Hietanen:2008tv}. On the one hand, we use a $\overline{\mathrm{MS}}$-scale of $\LMSb=341~\mathrm{MeV}$ from $2+1$-flavor lattice QCD in contrast to the much lower $\LMSb=245~\mathrm{MeV}$ from $2$-flavor lattice QCD in \cite{Hietanen:2008tv}. Therefore, our smallest temperature features a much lower $T / \LMSb$-ratio than the corresponding temperature in \cite{Hietanen:2008tv} and consequently lies much closer to the QCD crossover, where dimensional reduction ceases to work.
Moreover, our computation includes three massless flavors of quarks instead of two. At small temperatures, neglecting the strange quark mass might actually not be a valid approximation and causes a power series of extra corrections in $m_\mathrm{s}/T$.
As expected, our results approach the perturbative result with increasing temperature and good agreement with the perturbative result is reached at $T=25~\mathrm{GeV}$, which does not contradict earlier results \cite{Hietanen:2008tv}. As the dimensional reduction technique itself is expected to work beyond $T \gtrsim 2 \Tc$ \cite{Laine:2003ay}, the conclusion must be that nonperturbative physics plays an important role in EQCD for temperatures below $T=25~\mathrm{GeV}$.
Another peculiar feature of fig.~\ref{comp_latt_PT} is that the purely perturbative predictions are closer to the fourdimensional lattice results than our lattice EQCD results matched to full QCD. We suspect that this is due to the using a 1 loop running coupling for the matching to EQCD in \eqref{eq:match_gsq}, whereas \cite{Ipp:2006ij} features a 2 loop running coupling. While there should not be any difference formally since the matching at $\OO(g^4)$ is incomplete, anyways, there is a big difference in the numerical values of $g^2$ at low temperatures; the 2 loop $g^2$ is about $20\%$ smaller than its 1 loop equivalent at $T=277~\mathrm{MeV}$.

\subsection{Third derivatives}
\begin{figure}[htbp!] 
\centering 
	\begin{tabular}{cc}
		\includegraphics[scale=0.55]{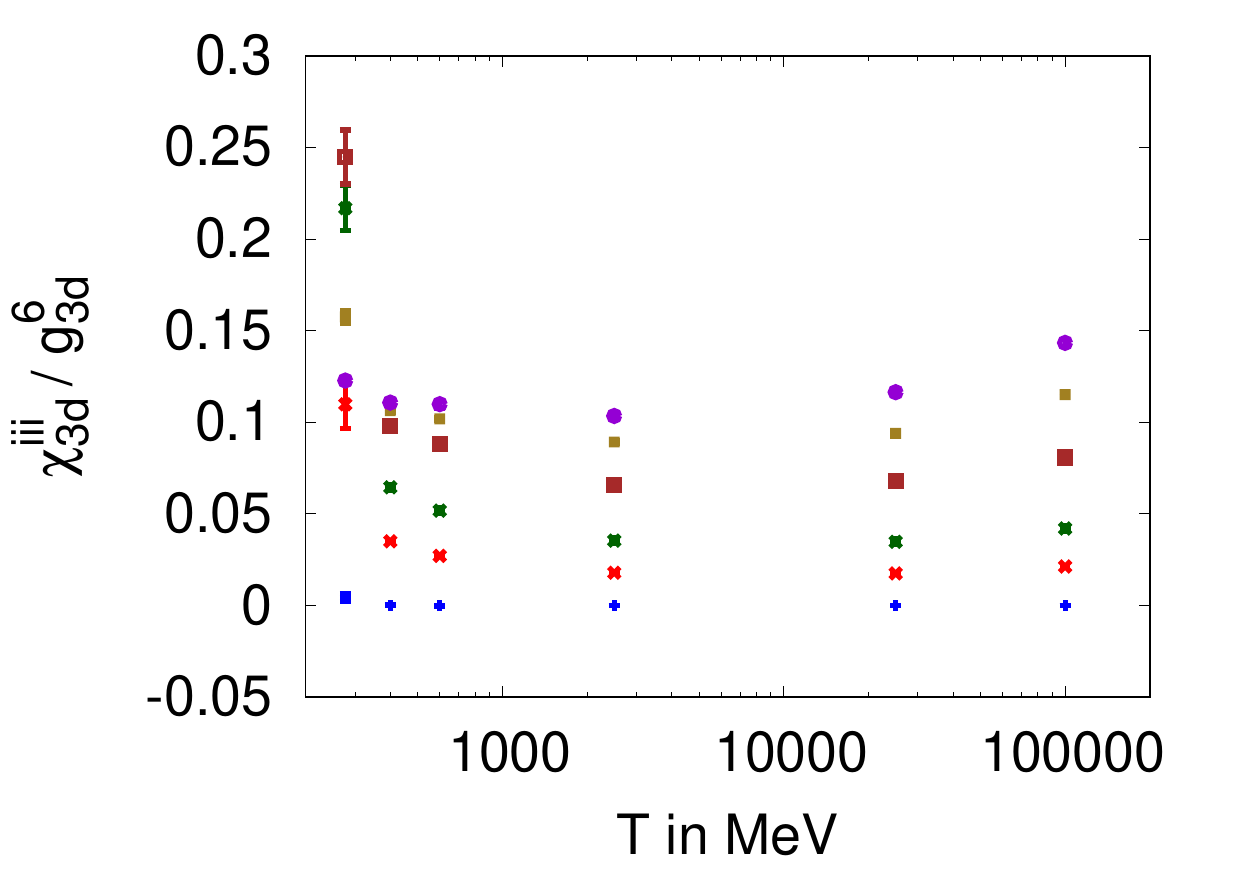} & 
		\includegraphics[scale=0.55]{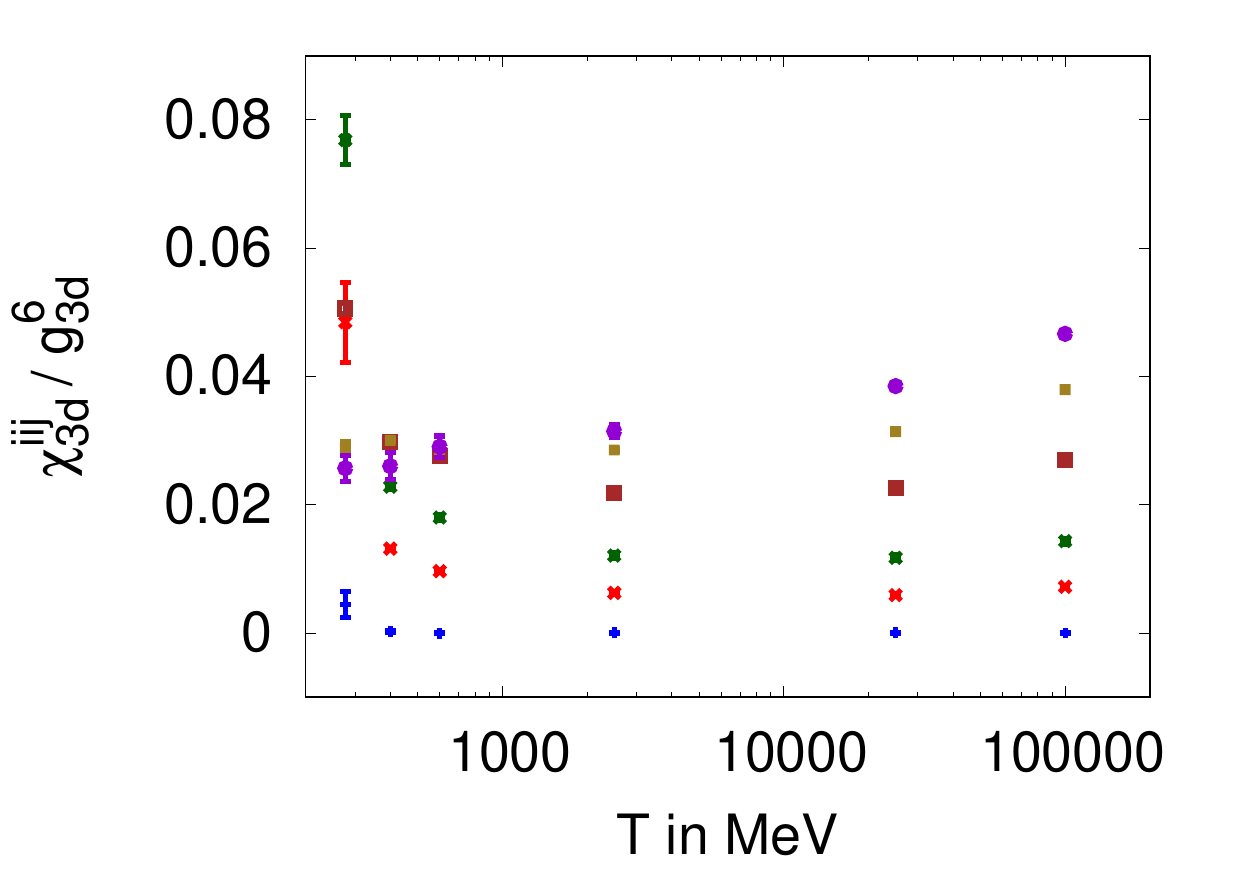} \\
		(a) & (b) \\
		\includegraphics[scale=0.55]{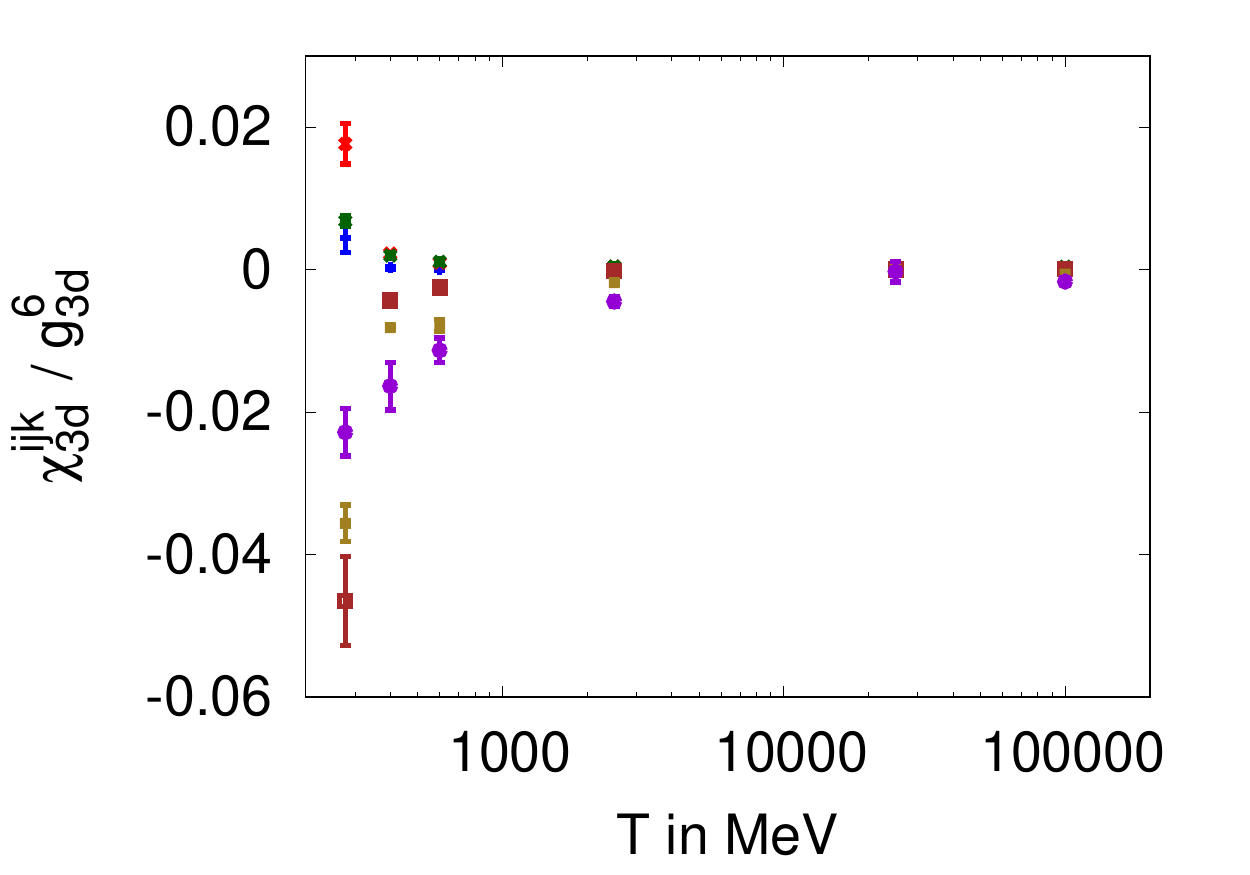} \\
		(c) \\
	\end{tabular}
	\caption{Third derivatives $\chi^{iii}$ (a), $\chi^{iij}$ (b) and $\chi^{ijk}$ (c) in three-dimensional units as functions of the temperature. Blue points correspond to $z=0.0$, red to $z=0.025$, green to $z=0.05$, brown to $z=0.1$, olive to $z=0.15$, and violet to $z=0.2$.}
	\label{T_three}
\end{figure}

Third derivatives are odd in $\bmu$, therefore, they should vanish at $\bmu=0$. Moreover, \eqref{eq:ac_cubic_sus} suggests that at $\bmu=0$, all three variants of the derivative should be numerically equal. Both features can be found in Fig.~\ref{T_three}, consistency with zero for the $z=0$ data is indeed realized at a quite exact level. 
Moreover, the derivatives in three-dimensional units contain an overall factor of $\pi$ from $\bmu = \frac{\mu}{\pi T}$, while the more natural scaling would be $\frac{\mu}{T}$. Therefore, we expect the overall order of magnitude of the third derivative to be higher than the second one, due to this extra factor of $\pi$, also translating to the ratios of the other higher cumulants in three-dimensional units.
Fig.~\ref{T_three} furthermore shows clearly that the error of data points closer to $\Tc$ drastically increases, a feature for which the small mass reweighting of some of these data points cannot be entirely blamed.

While it becomes evident from the data that there is a certain lower bound in temperature below which the dimensional reduction is not valid any more, the situation for large $z$ is somewhat different. Even though a value of $z=0.1$ at $\nf=3$ corresponds to $\mu \approx T$, which is definitely outside of the range of validity of our dimensional reduction procedure, there is no inherent evidence from the data for the breakdown of the $\mu \ll T$-approximation that went into the scale setting in \eqref{eq:match_gsq}. Where temperatures close to the pseudocritcal temperature of full QCD cause the error bars to blow up, a large value of $z$ just strongly favors one of the two possible signs of the $\Tr \, \Phi^3$ operator in the EQCD action \eqref{eq:EQCD_cont_action} and therefore reduces the fluctuations in the system, leading to smaller error bars at comparable numerical effort.

\subsection{Fourth derivatives}
\begin{figure}[htbp!] 
\centering 
	\begin{tabular}{cc}
		\includegraphics[scale=0.5]{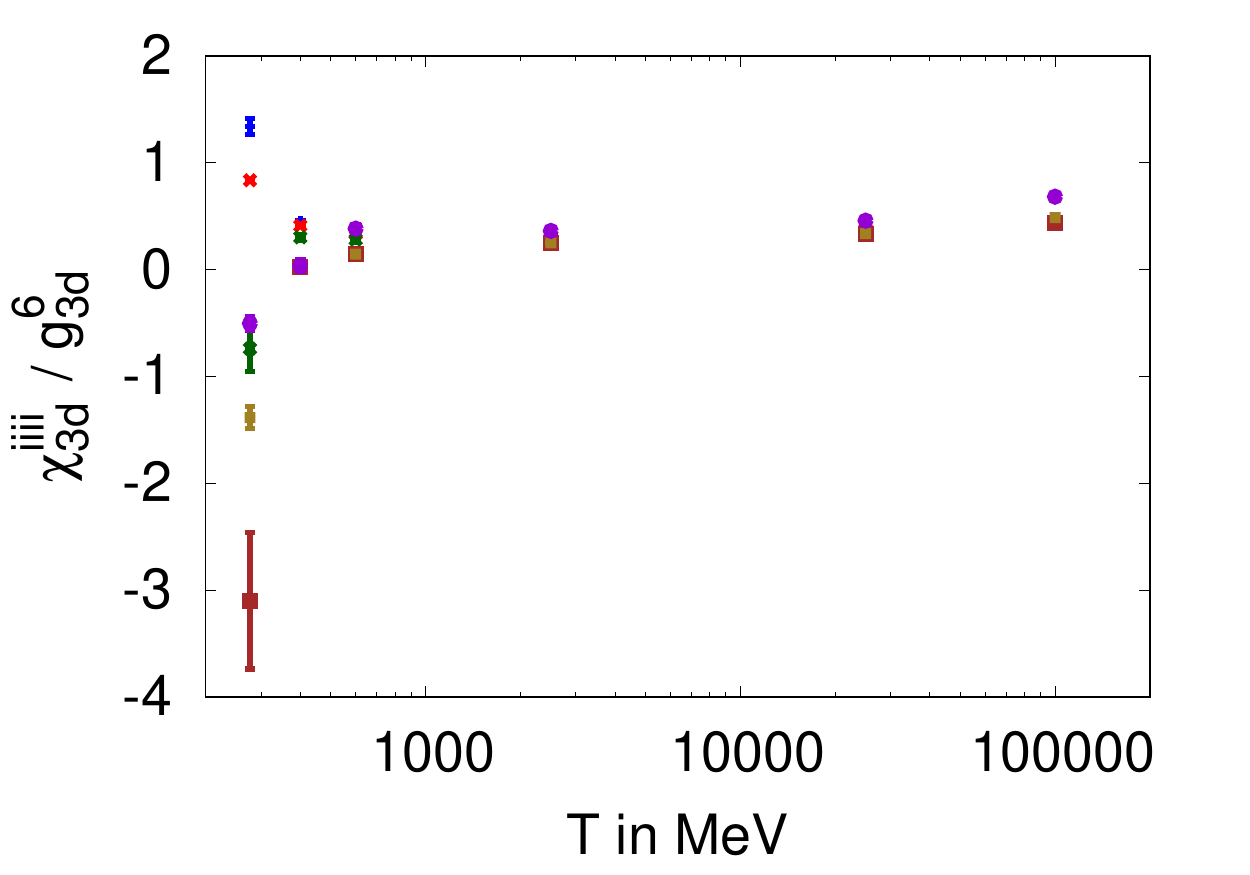} & 
		\includegraphics[scale=0.5]{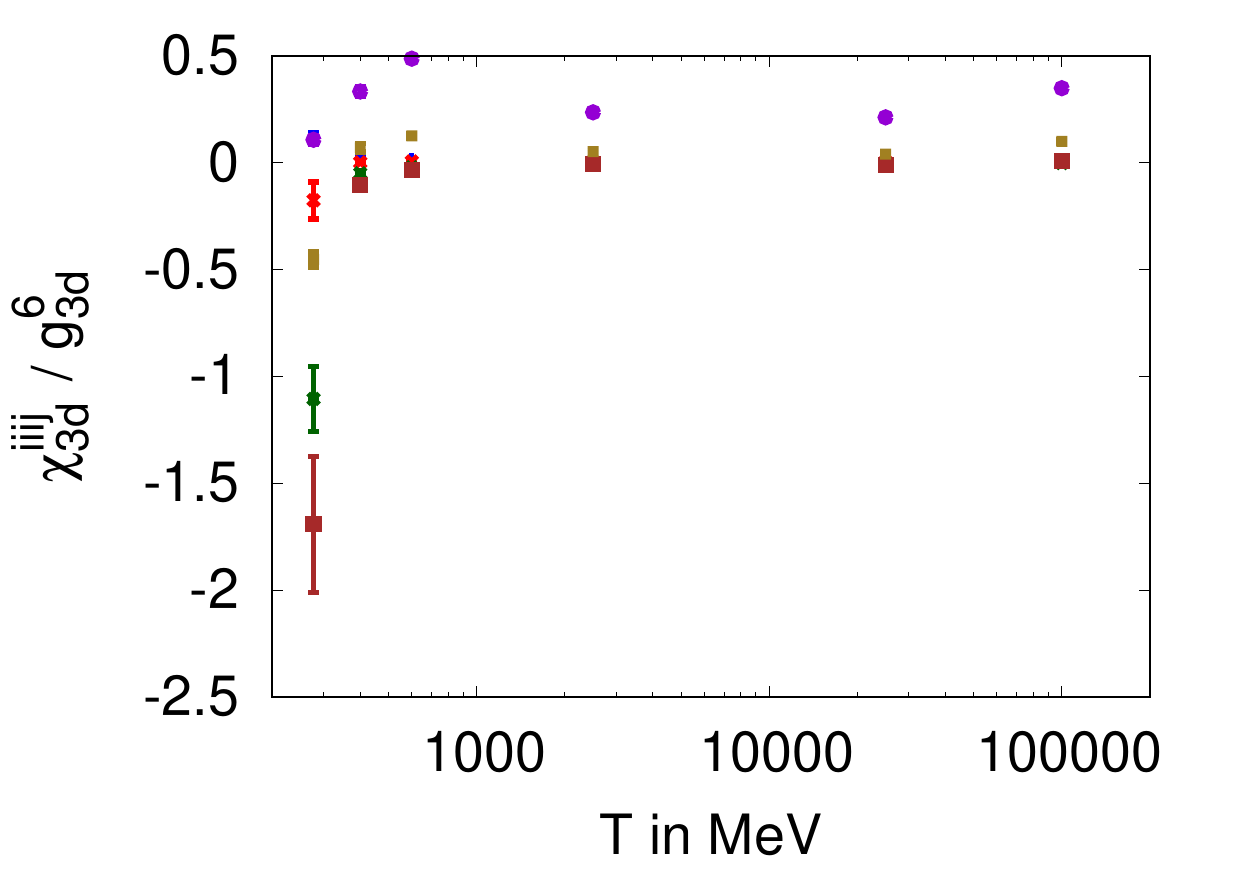} \\
		(a) & (b) \\
		\includegraphics[scale=0.5]{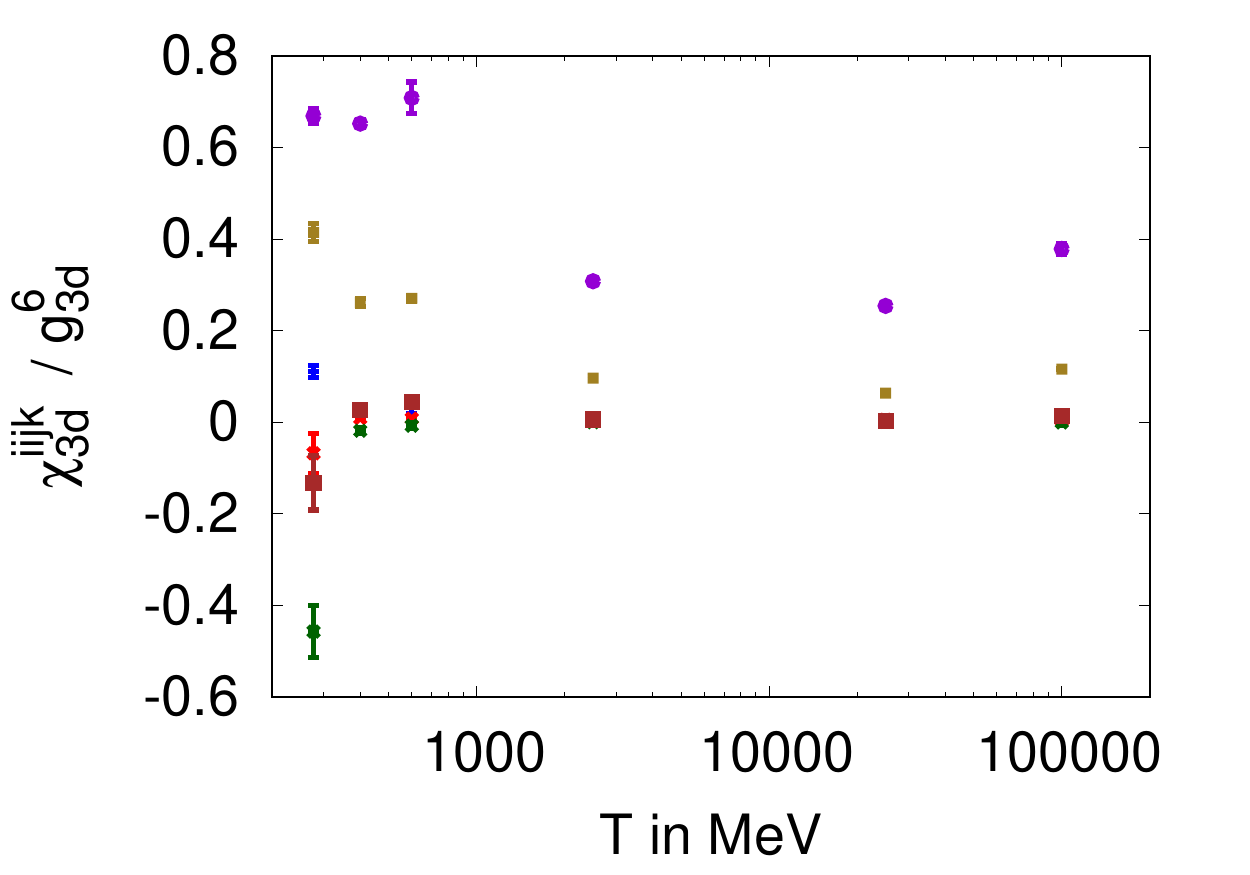} &
		\includegraphics[scale=0.5]{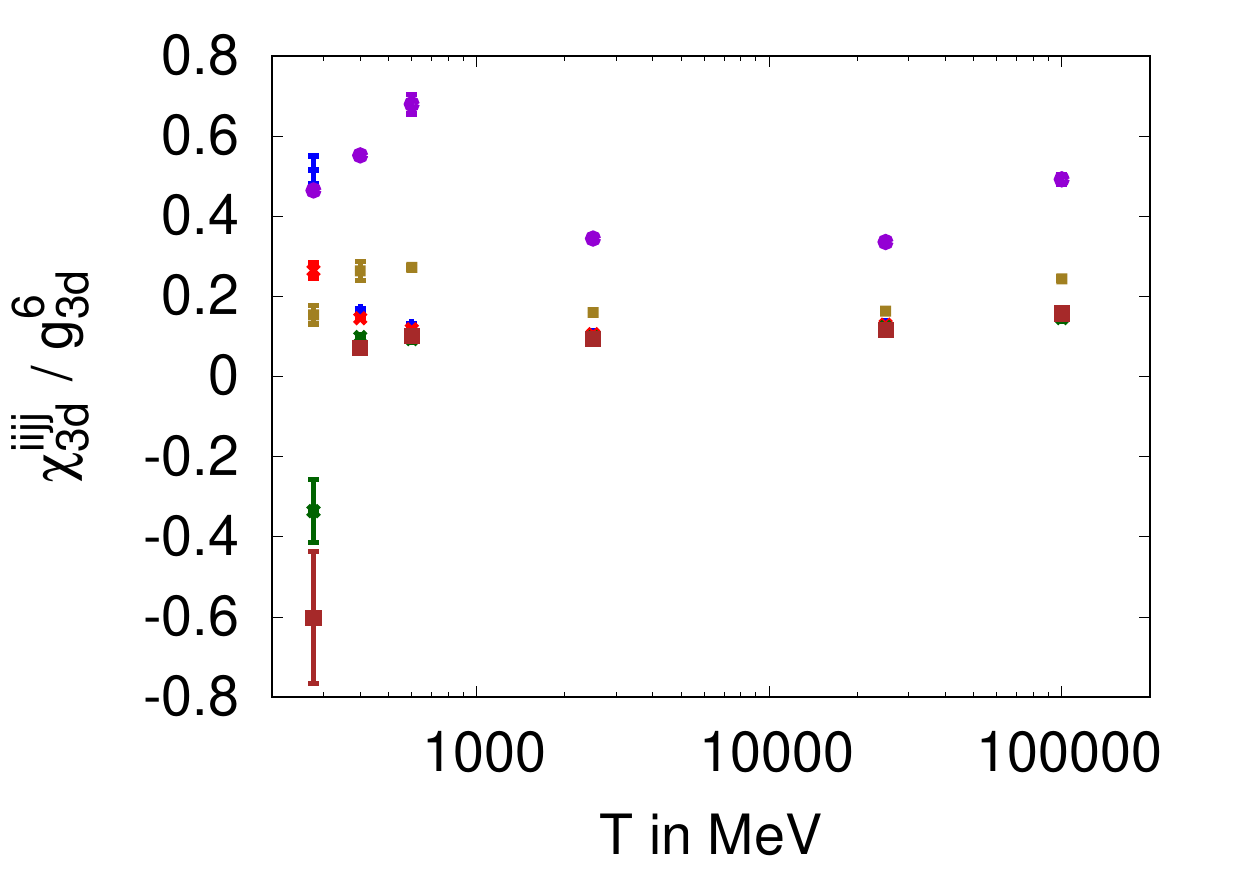} \\
		(c) & (d) \\
		\includegraphics[scale=0.5]{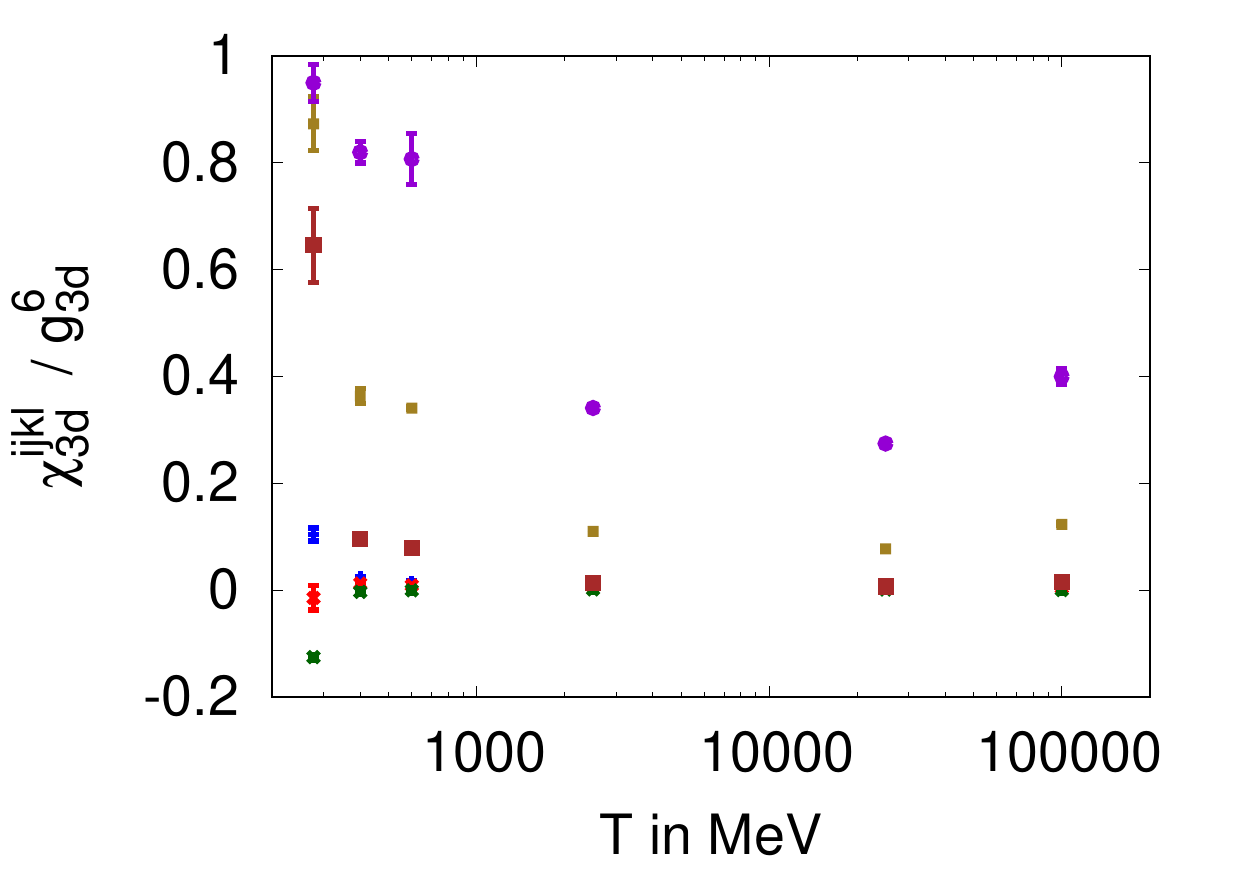} & \\ 
		(e)  &  \\
	\end{tabular}
	\caption{Fourth order susceptibilities $\chi^{iiii}$ (a), $\chi^{iiij}$ (b), $\chi^{iijk}$ (c), $\chi^{iijj}$ (d), and $\chi^{ijkl}$ (e) in three-dimensional units as functions of the temperature. Blue points correspond to $z=0.0$, red to $z=0.025$, green to $z=0.05$, brown to $z=0.1$, olive to $z=0.15$, and violet to $z=0.2$.}
	\label{T_four}
\end{figure}

The fourth derivative is again even in $\bmu$, therefore, the fully diagonal susceptibility $\chitd^{iiii}$ and its bi-diagonal counterpart $\chitd^{iijj}$ do have a finite expectation value even at $\bmu = 0$, which also reflects in our results in Fig.~\ref{T_four}. Despite the lack of a clear physical interpretation, we still display $\chitd^{ijkl}$ at temperatures with $\nf > 3$, since it gives an idea about the part of \eqref{eq:ac_quartic_sus} that solely contains condensates that generically appear in fourth derivatives.
A rough scaling by an additional factor of $\pi$ compared to the third derivatives is maintained here, too. 
Following the trend of the third derivatives, the data at $T = 277~\mathrm{MeV}$ starts to become ever more noisy, whereas the error bars for all other temperatures are barely visible.

\subsection{Fifth derivatives}
\begin{figure}[htbp!] 
\centering 
	\begin{tabular}{ccc}
		\includegraphics[scale=0.42]{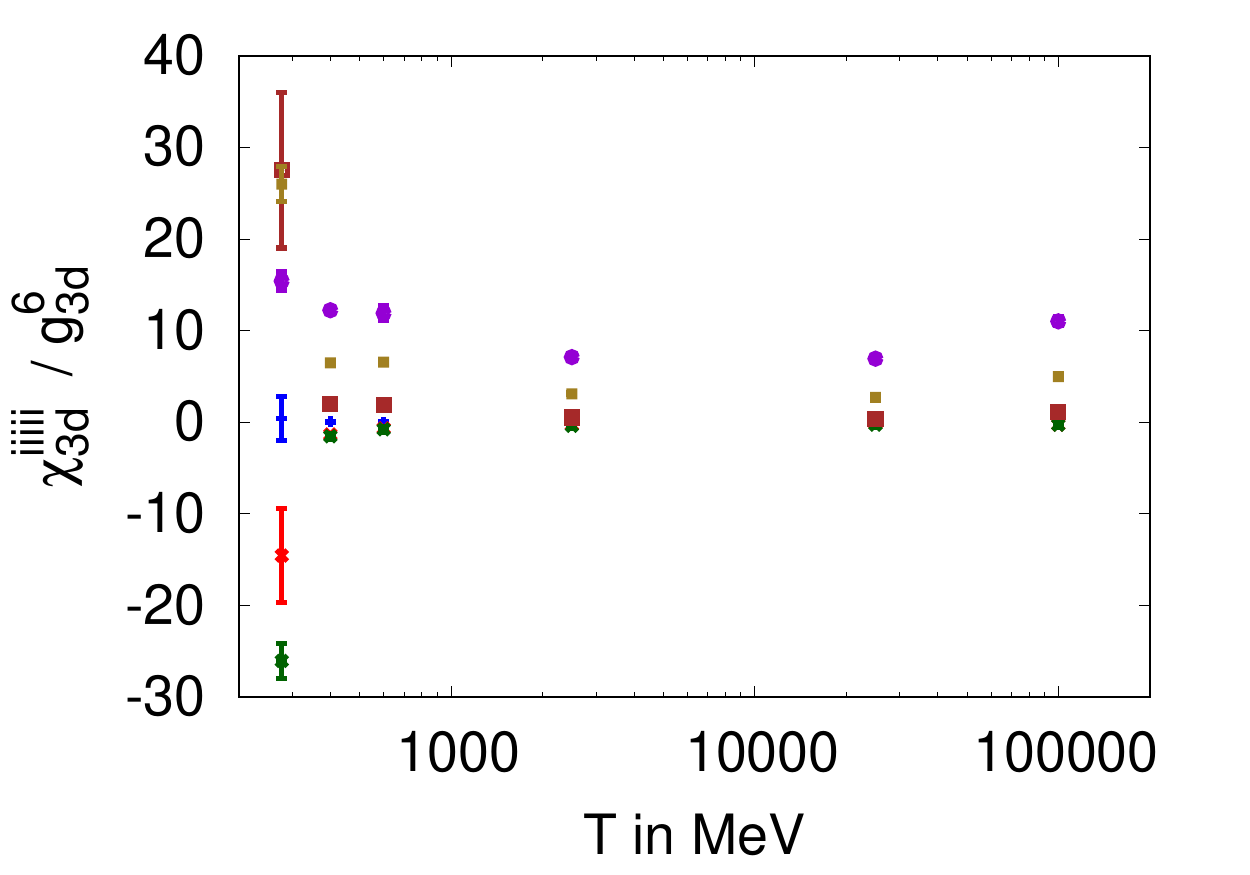} & 
		\hspace{-6mm}\includegraphics[scale=0.42]{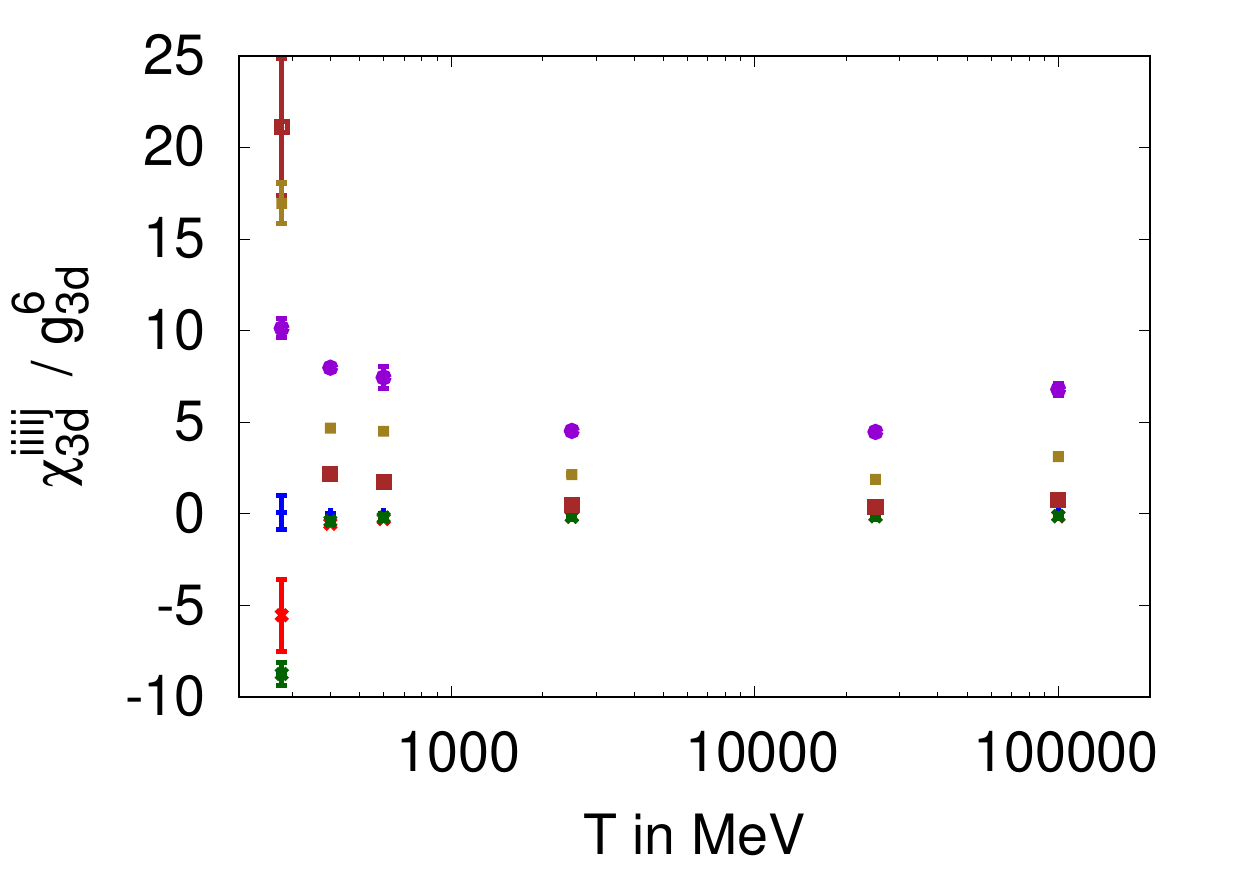} &
		\hspace{-6mm}\includegraphics[scale=0.42]{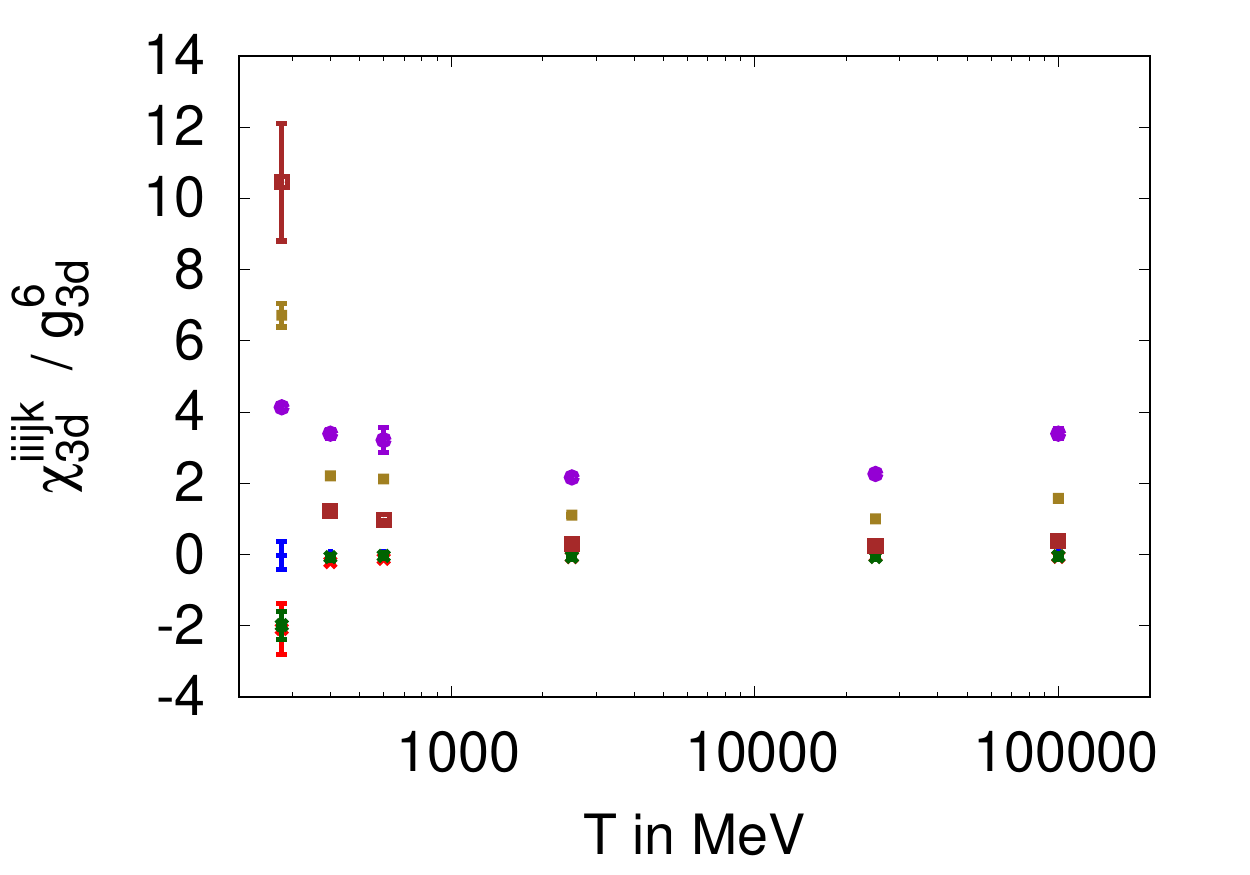} \\
		(a) & (b) & (c) \\
		\includegraphics[scale=0.42]{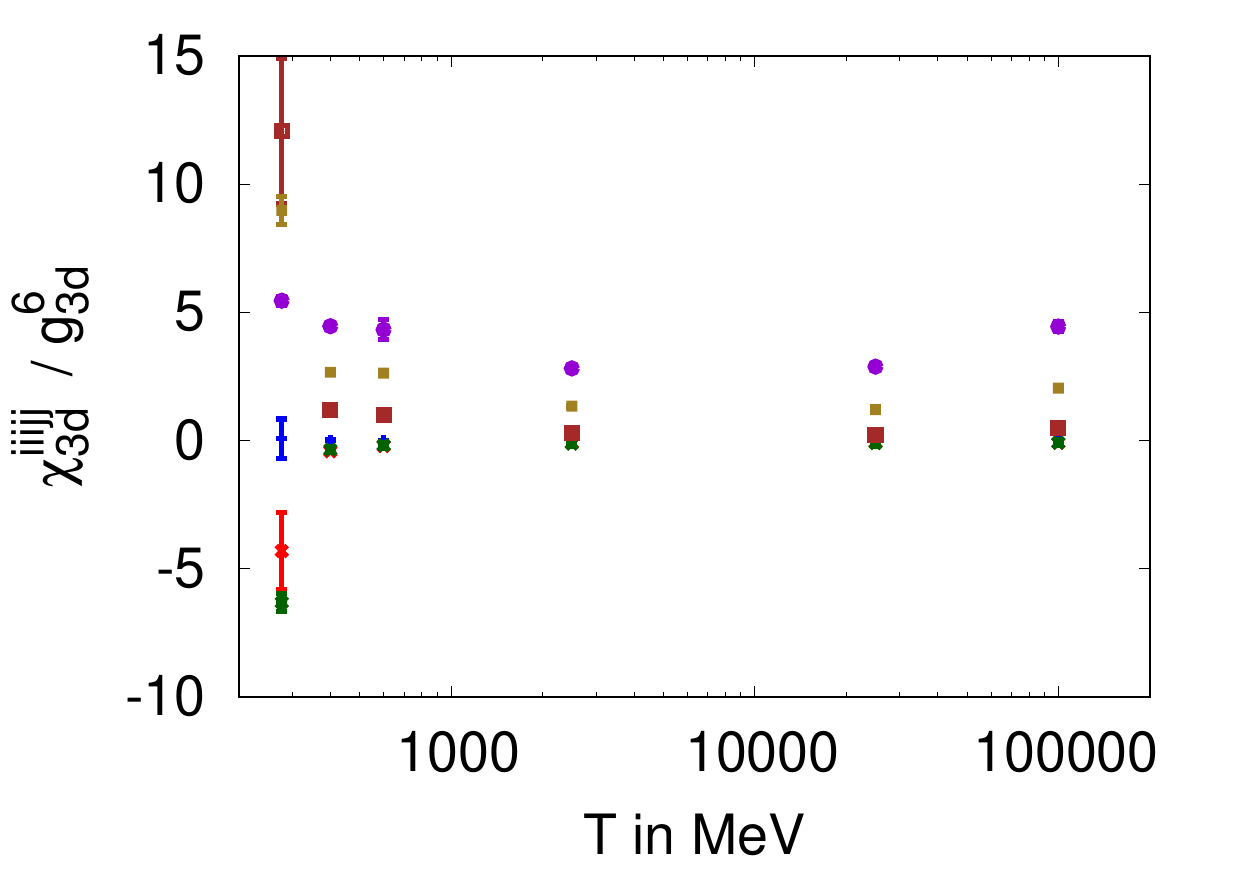} &
		\hspace{-6mm}\includegraphics[scale=0.42]{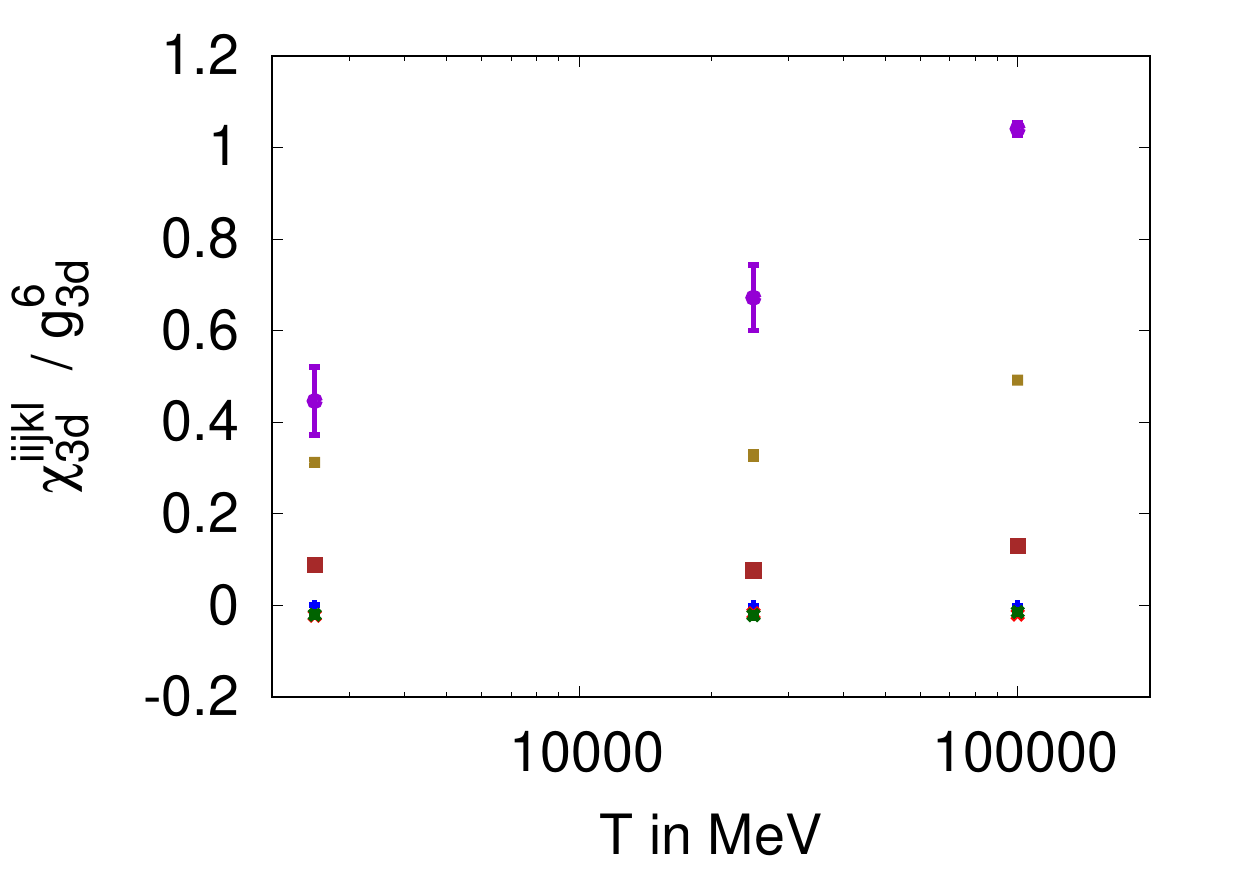} &
		\hspace{-6mm}\includegraphics[scale=0.42]{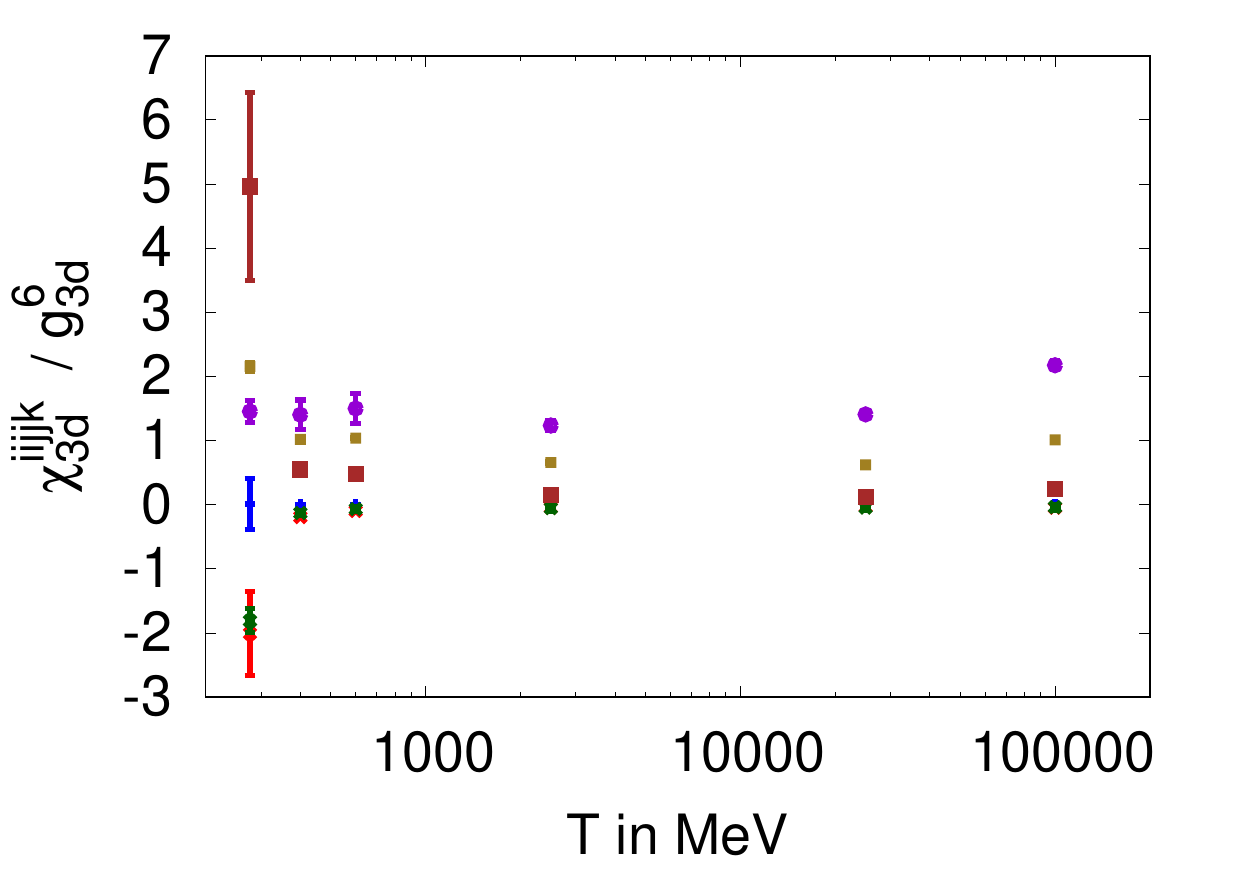} \\ 
		(d) & (e) & (f) \\
		\includegraphics[scale=0.42]{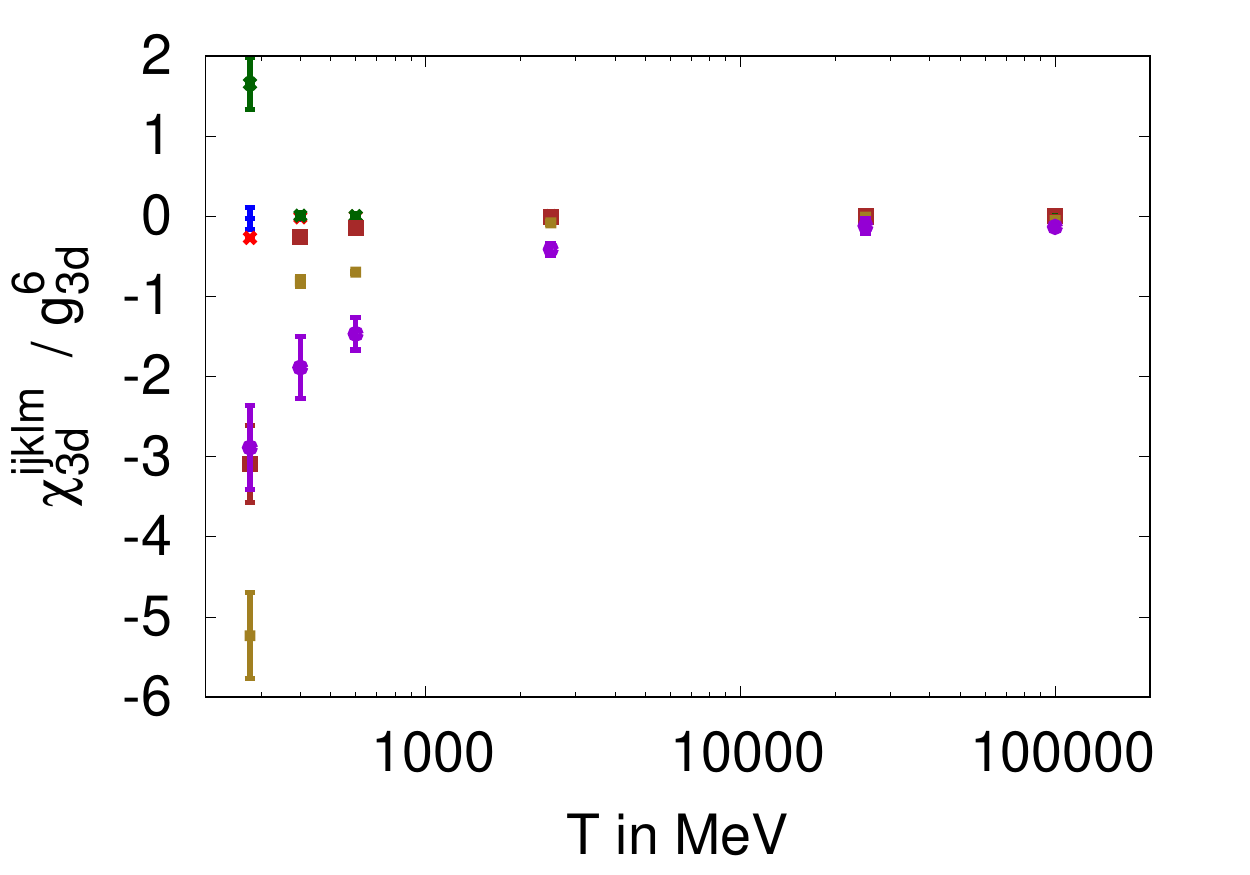} &               
		& \\ 
		(j) &  &  \\
	\end{tabular}
	\caption{All distinct fifth order susceptibilities (a)--(j) in three-dimensional units as functions of the temperature. Blue points correspond to $z=0.0$, red to $z=0.025$, green to $z=0.05$, brown to $z=0.1$, olive to $z=0.15$, and violet to $z=0.2$.}
	\label{T_five}
\end{figure}

Fifth derivatives are again odd in $\bmu$ and therefore expected to feature a vanishing expectation value at $\bmu=0$, as one can see from our results in Fig.~\ref{T_five}.
Observing another relative increase of $\pi$ in the order of magnitude, the consistency with $0$ of the $z=0$-data is still very precise.
Aside from the generically fifth-derivative-contribution $\chitd^{ijklm}$, Fig.~\ref{T_five} only shows data from scenarios in which the respective derivative is well-defined, i.e.\ $\chitd^{iijkl}$ is only displayed at temperatures that correspond to at least $\nf=4$. Even though without physical meaning, we still display the numerical values of the derivatives omitted in the plot in App.~\ref{app:tab_results}.

\subsection{Sixth derivatives}
Just as for the previous derivatives, we observe a relative factor of $\pi$ in the order of magnitude of the sixth derivative. The relative errors of $T=277~\mathrm{MeV}$ increased once more, only physically well-defined derivatives are plotted other than the fully off-diagonal part $\chitd^{ijklmn}$.

\begin{figure}[htbp!] 
\centering 
	\begin{tabular}{ccc}
		\includegraphics[scale=0.42]{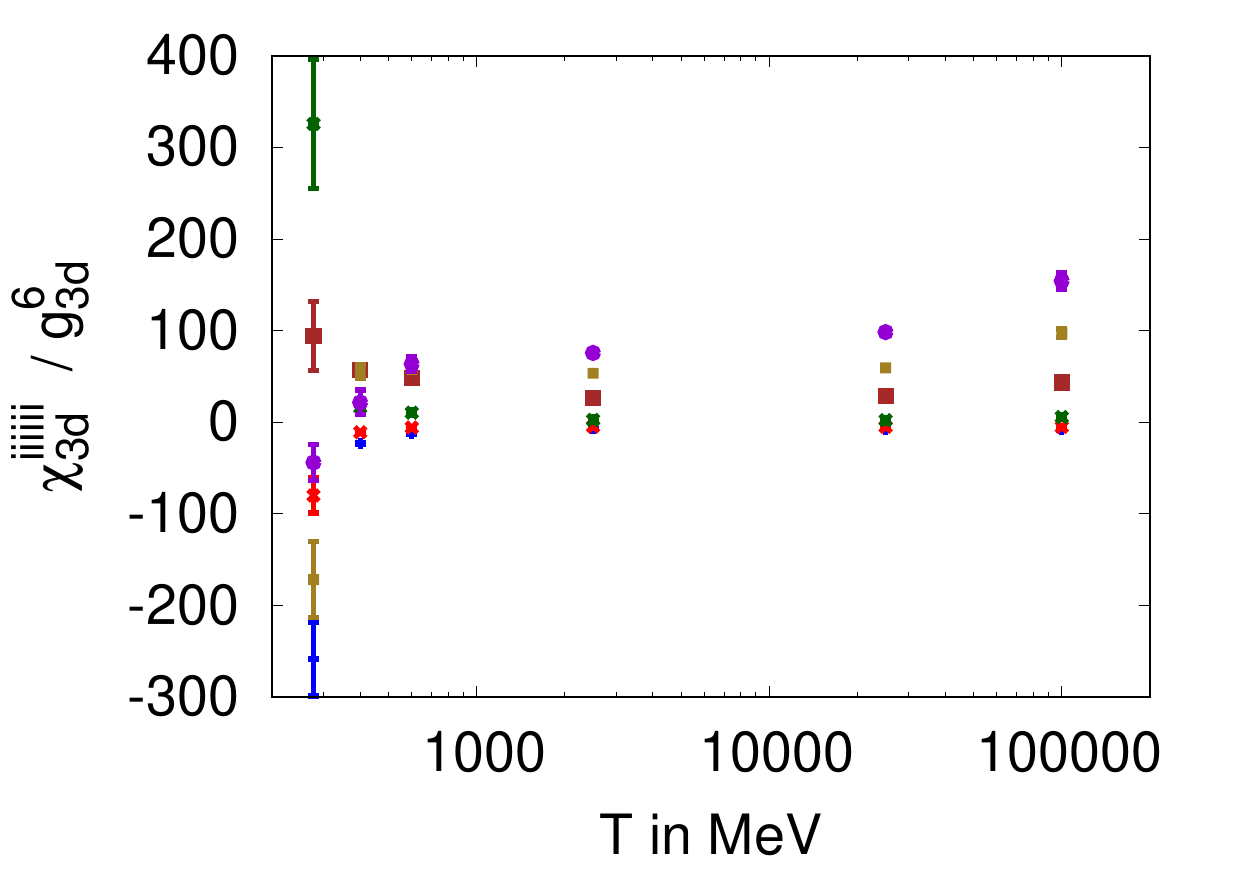} & 
		\hspace{-6mm}\includegraphics[scale=0.42]{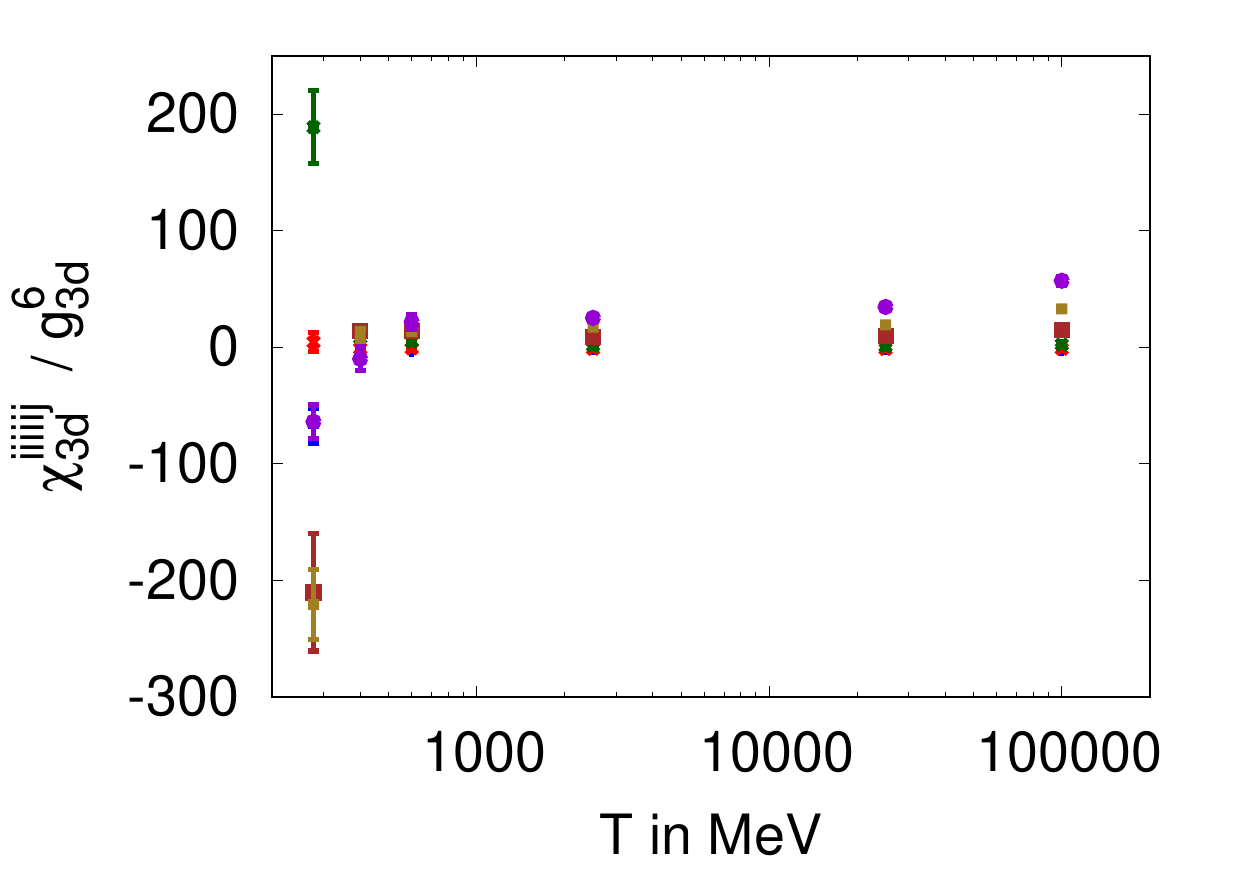} &
		\hspace{-6mm}\includegraphics[scale=0.42]{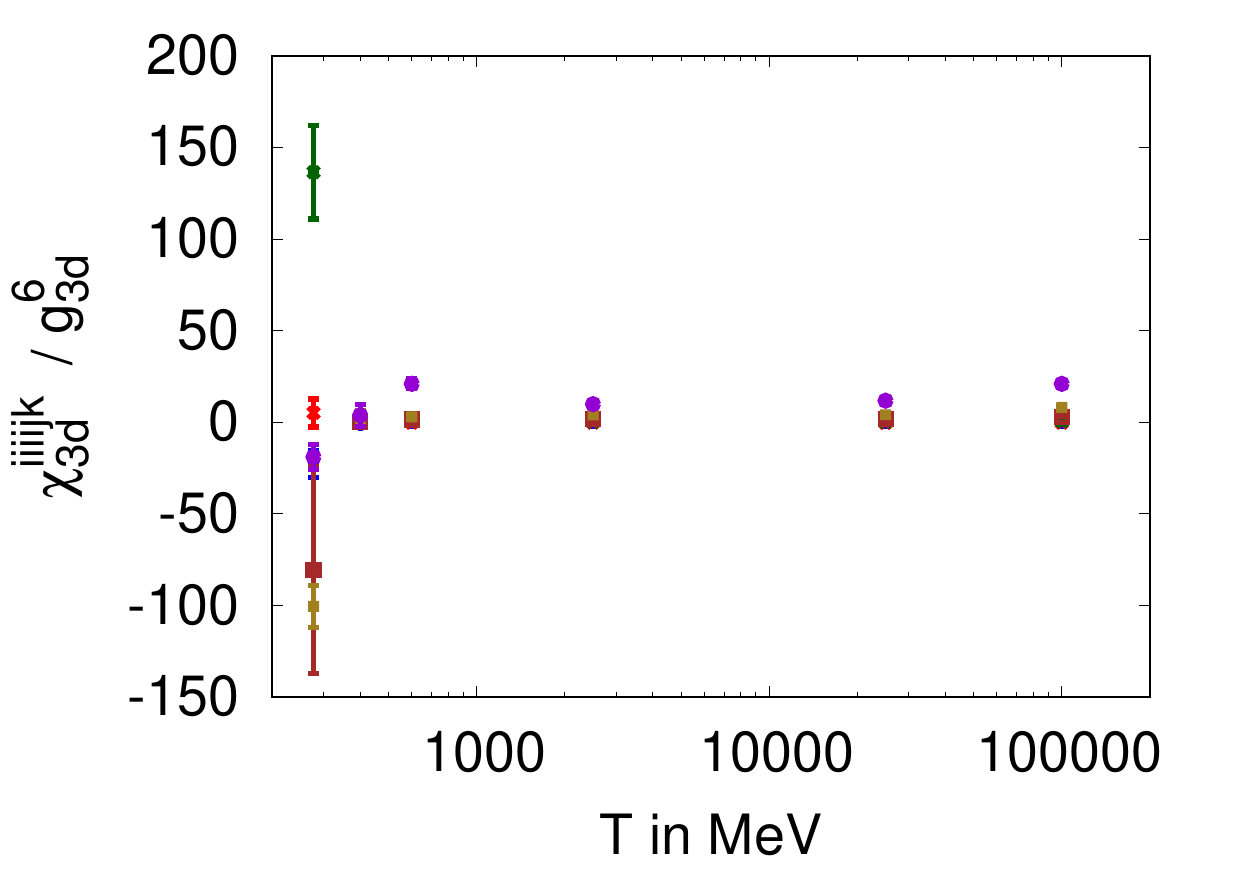} \\
		(a) & (b) & (c) \\
		\includegraphics[scale=0.42]{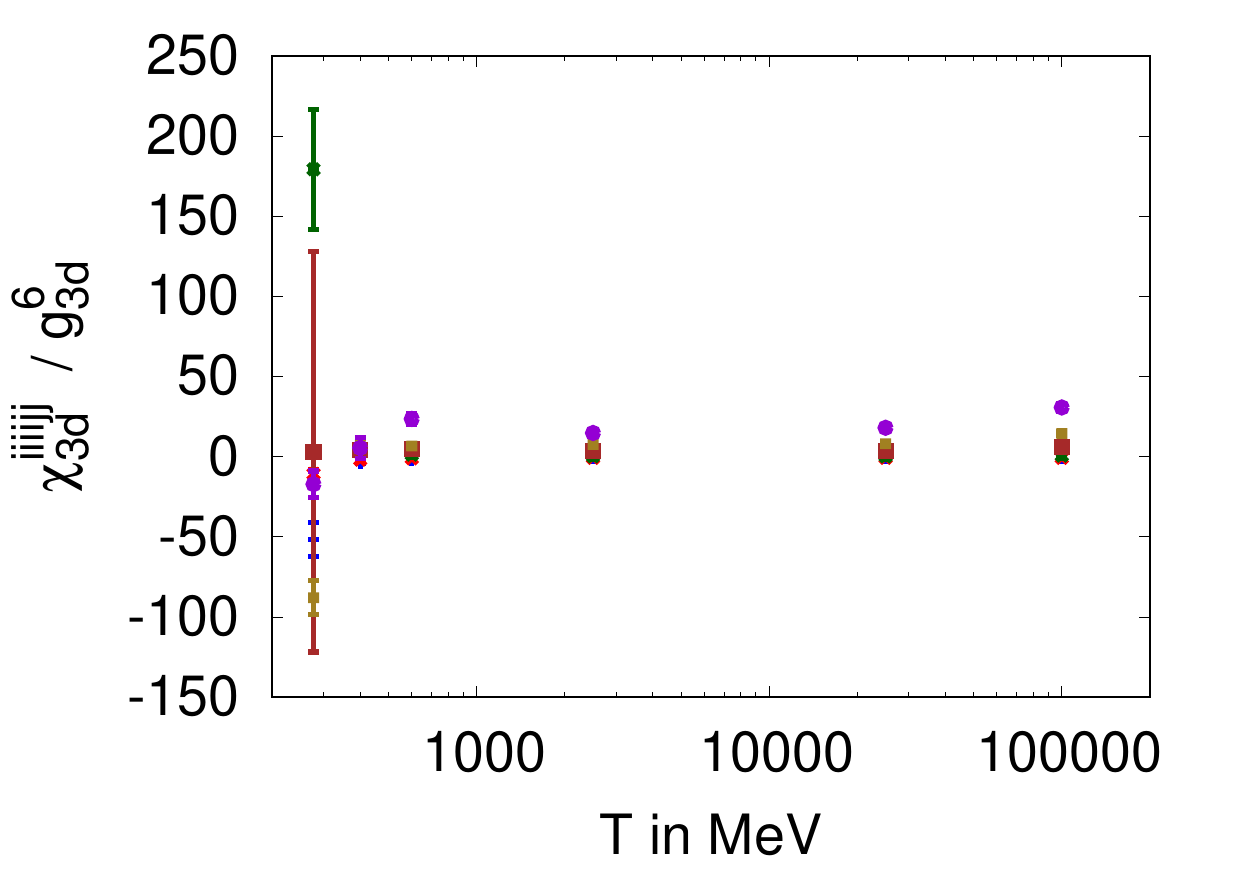} &
		\hspace{-6mm}\includegraphics[scale=0.42]{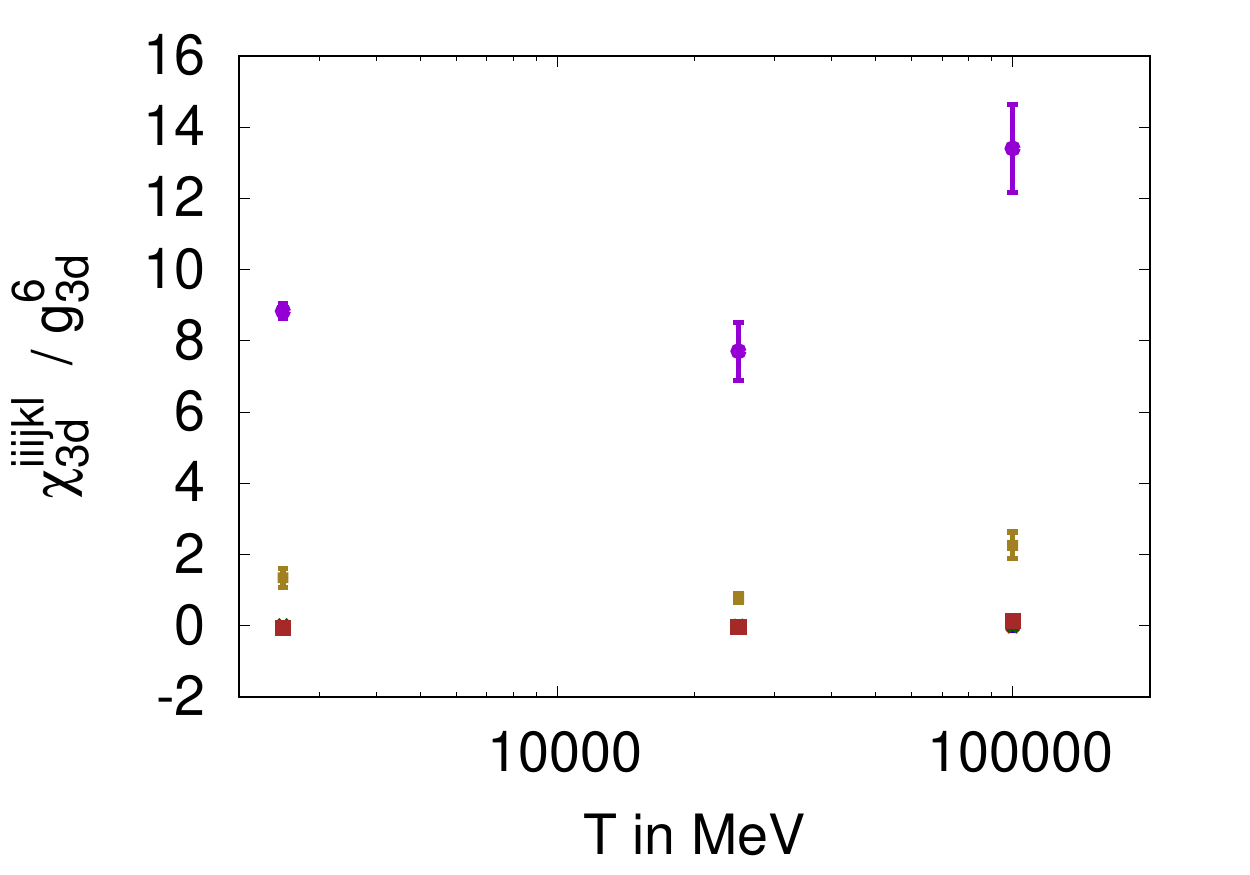} &
		\hspace{-6mm}\includegraphics[scale=0.42]{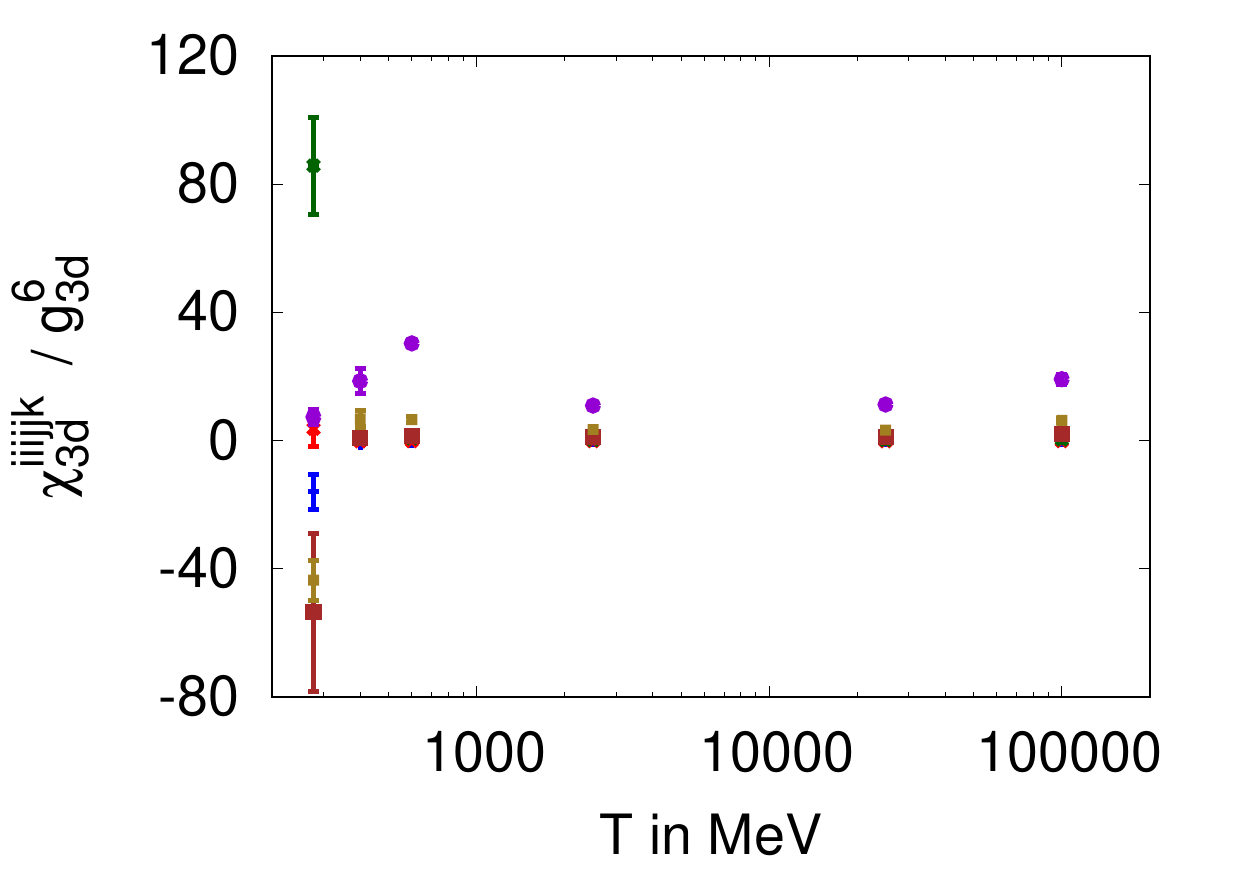} \\ 
		(d) & (e) & (f) \\
		\includegraphics[scale=0.42]{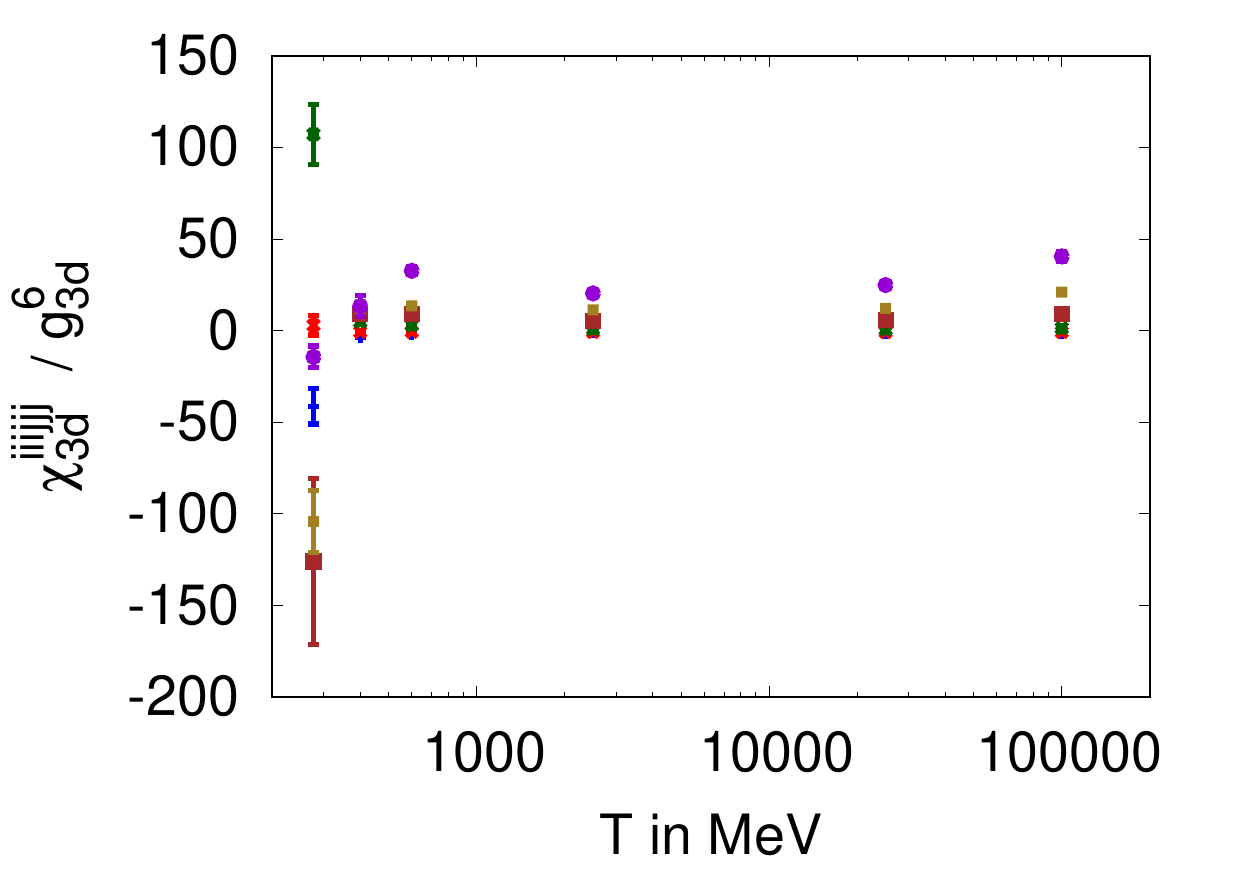} & 
		\hspace{-6mm}\includegraphics[scale=0.42]{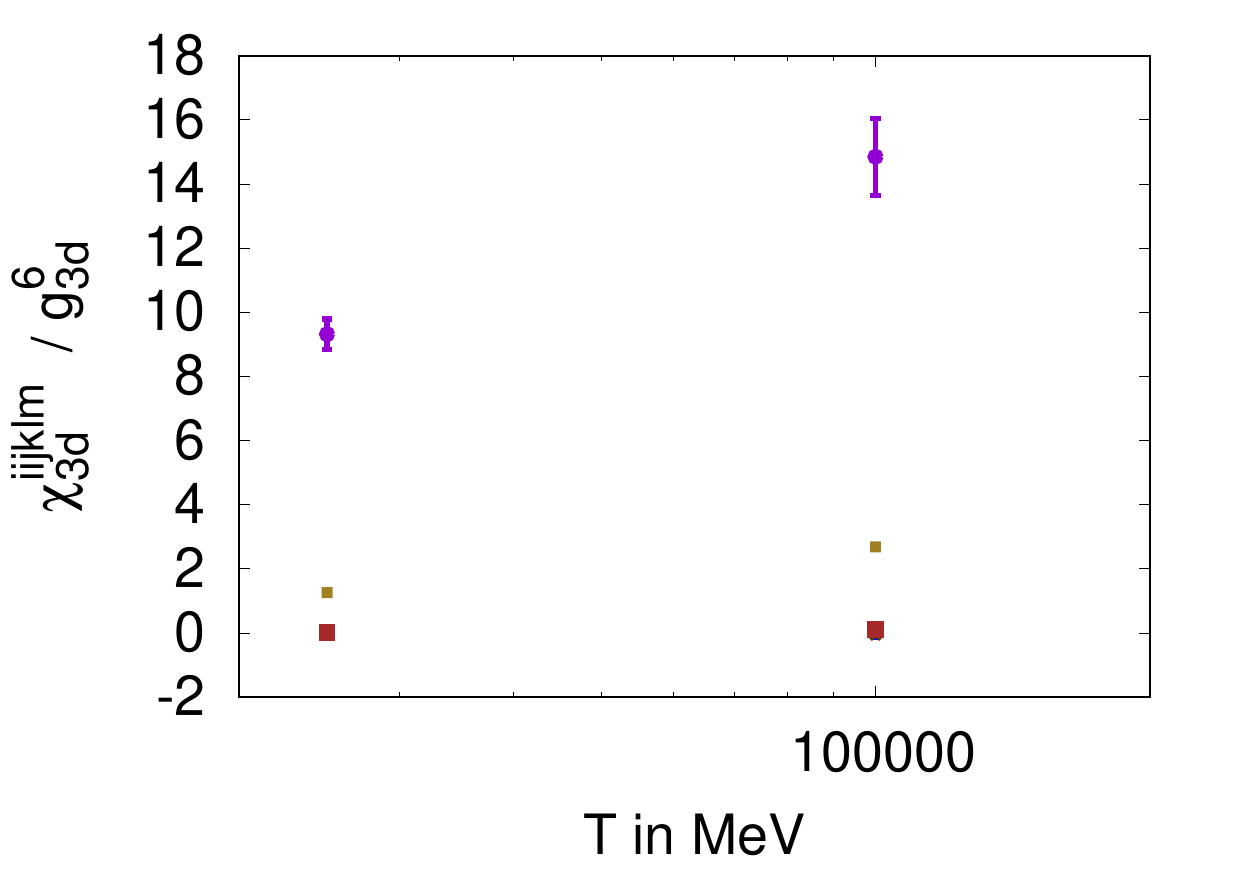} &
		\hspace{-6mm}\includegraphics[scale=0.42]{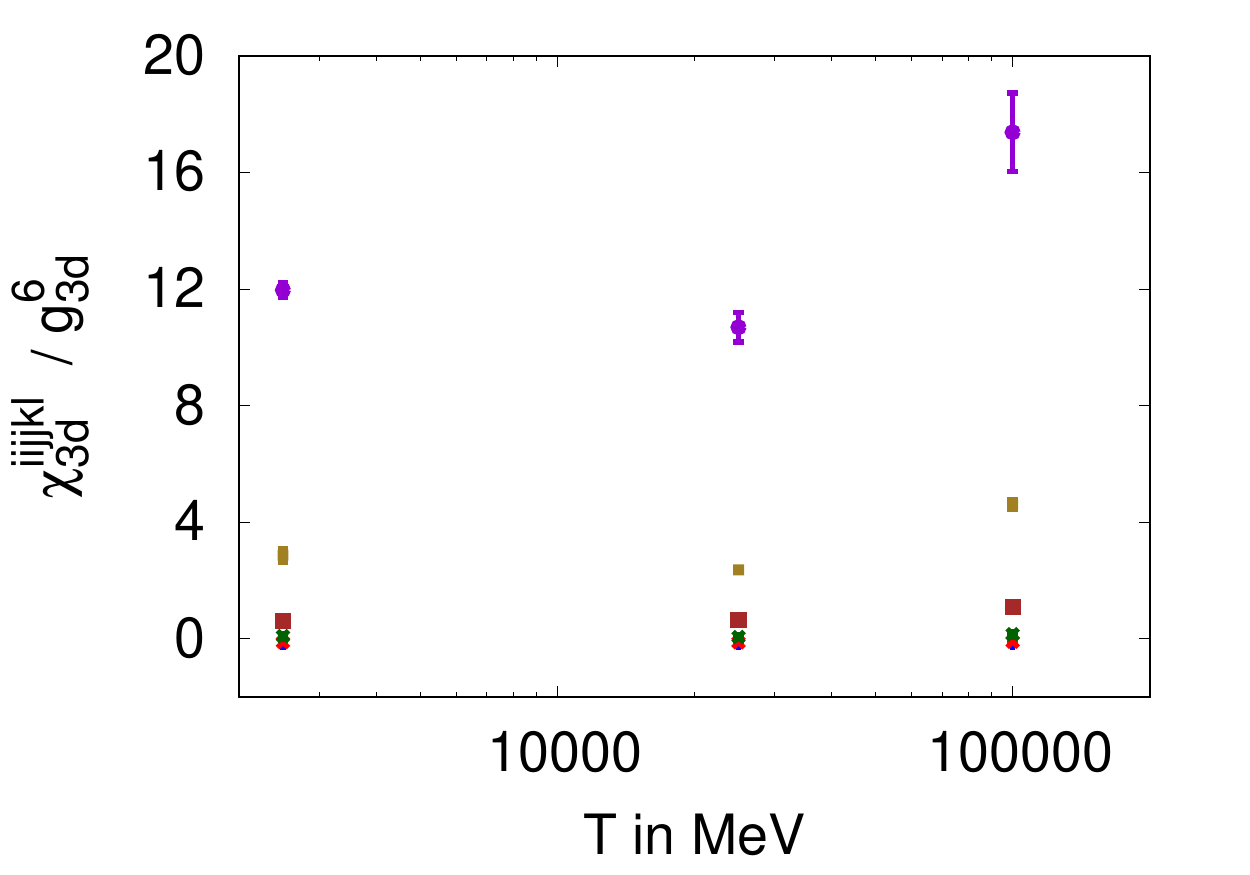} \\
		(g) & (h) & (i) \\
		\includegraphics[scale=0.42]{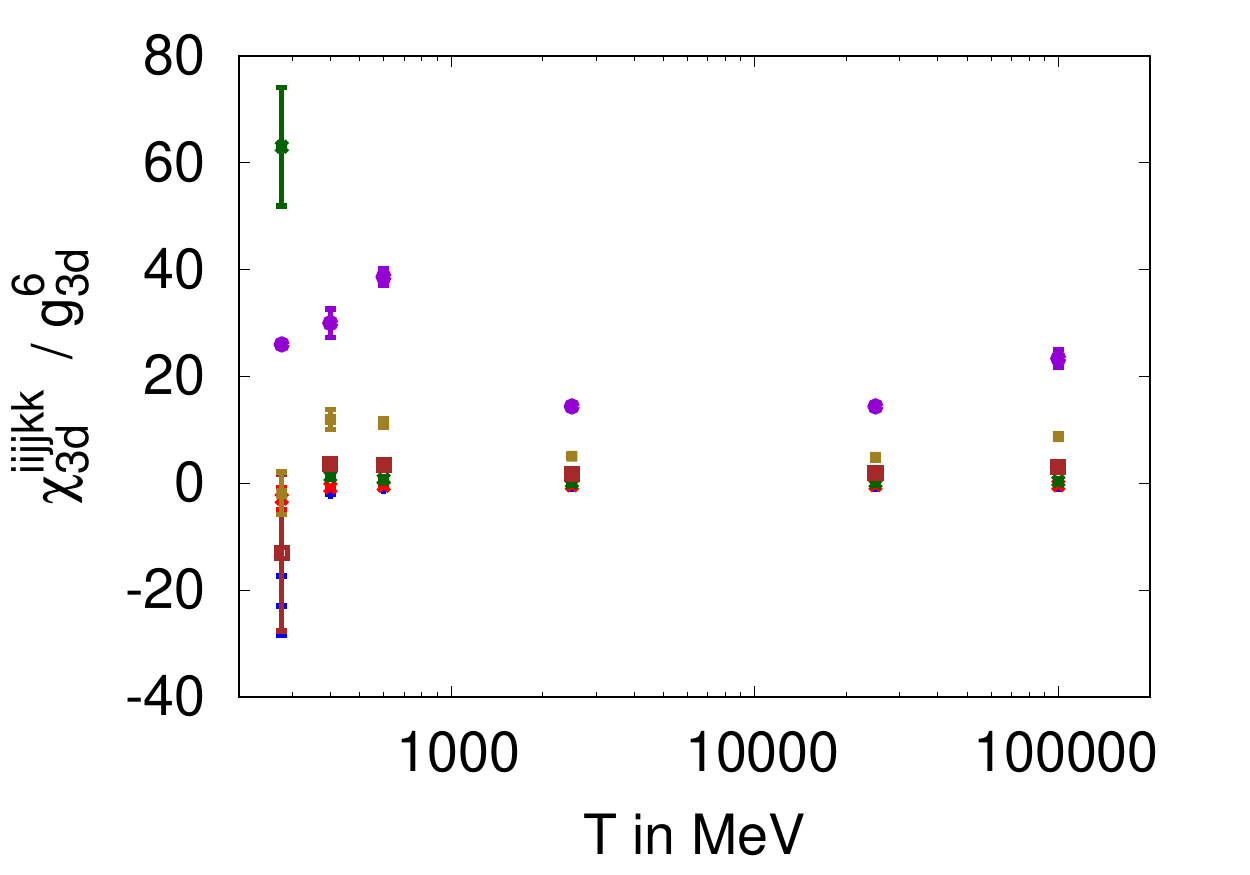} &
		\hspace{-6mm}\includegraphics[scale=0.42]{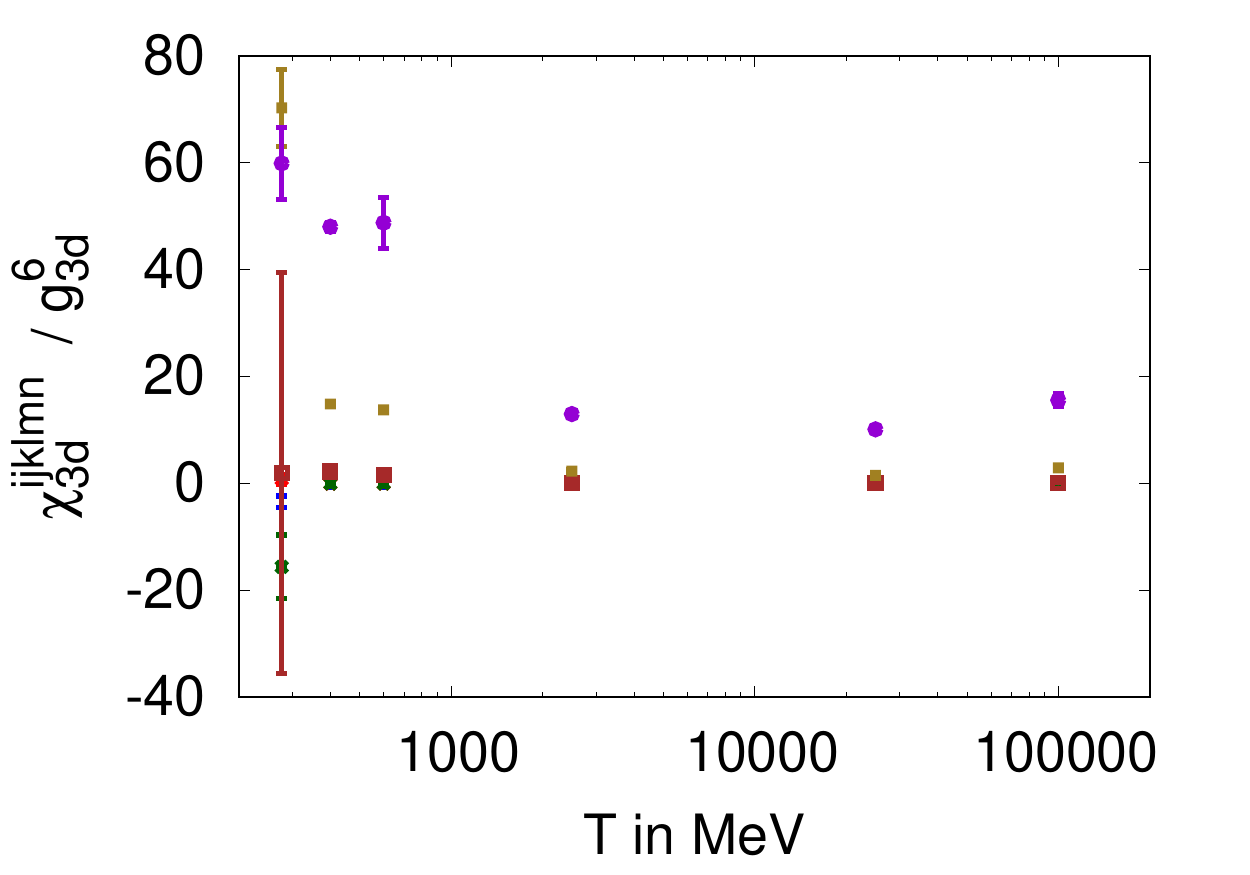} &                \\ 
		(j) & (k) &  \\
	\end{tabular}
	\caption{Distinct sixth order susceptibilities (a)--(k) in three-dimensional units as functions of the temperature. Blue points correspond to $z=0.0$, red to $z=0.025$, green to $z=0.05$, brown to $z=0.1$, olive to $z=0.15$, and violet to $z=0.2$.}
	\label{T_six}
\end{figure}

\subsection{Comparison of higher derivatives to perturbation theory}
\begin{figure}[htbp!] 
\centering 
	\begin{tabular}{cc}
		\includegraphics[scale=0.5]{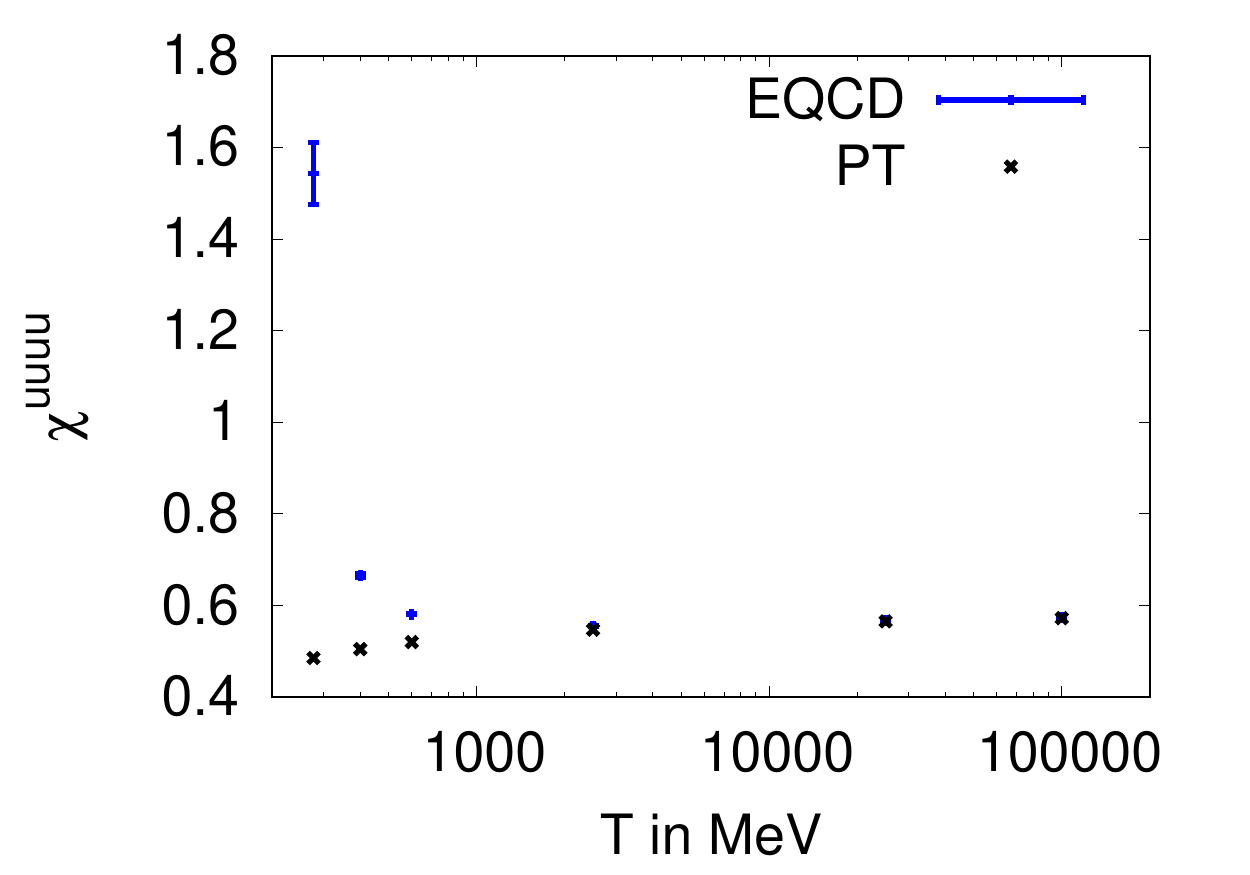} & 
		\includegraphics[scale=0.5]{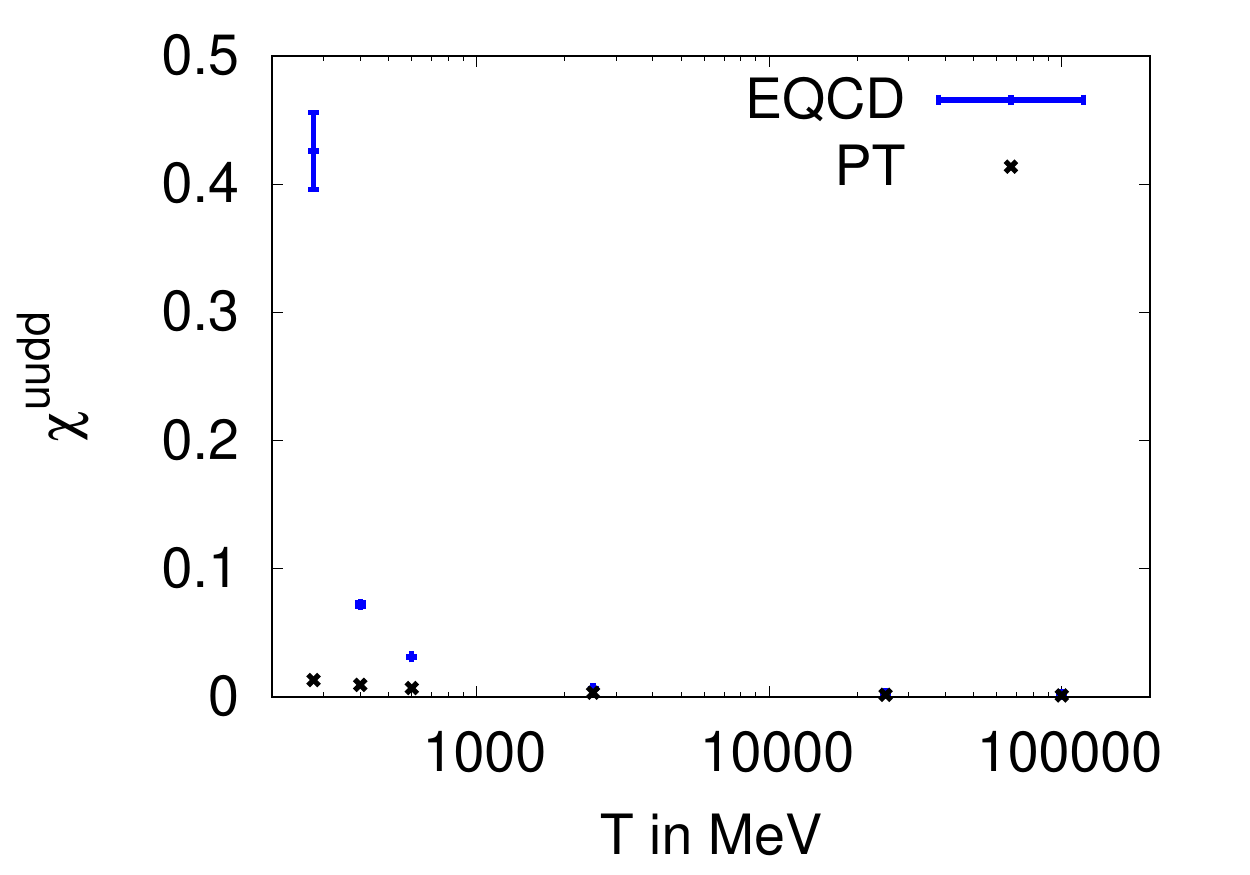} \\
		(a) & (b) \\
		\includegraphics[scale=0.5]{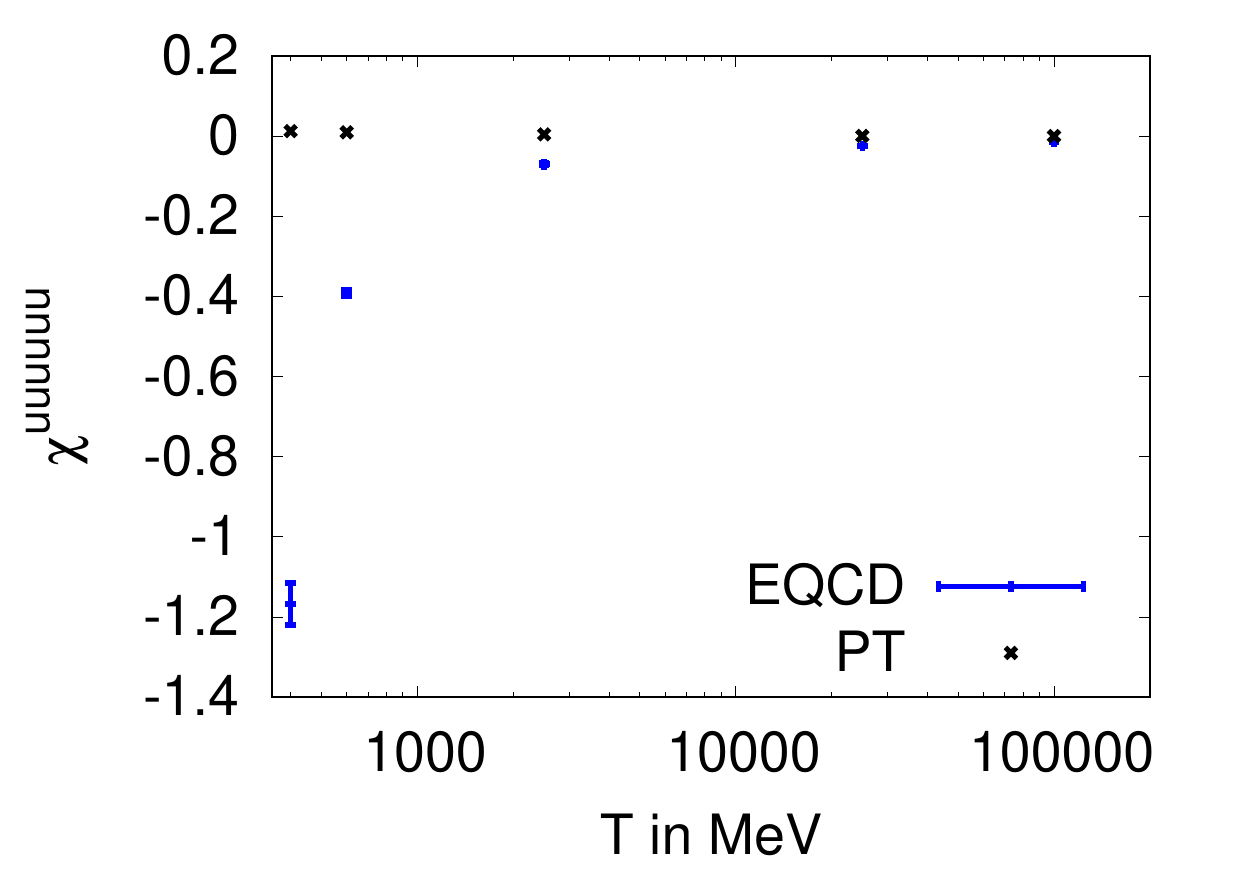} &
		\includegraphics[scale=0.5]{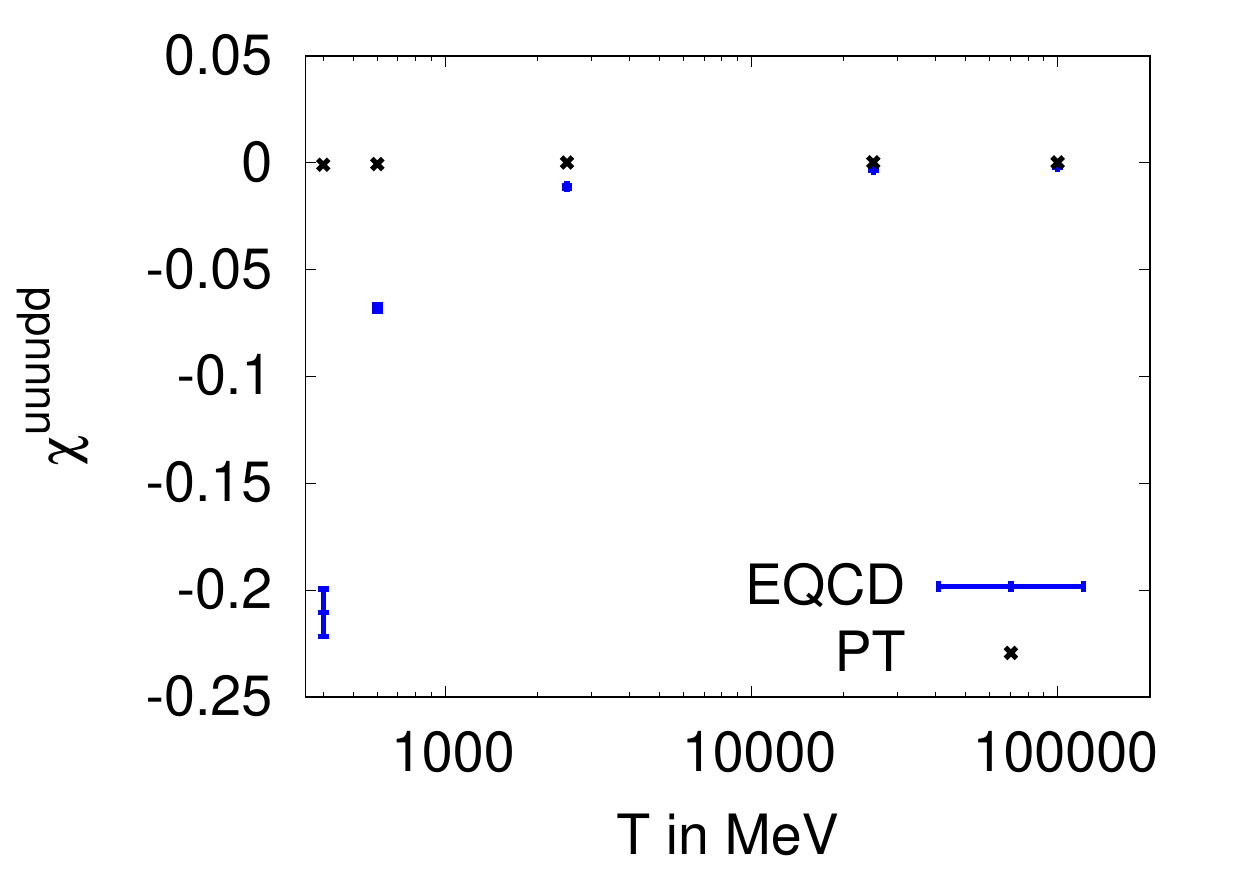} \\
		(c) & (d) \\
		\includegraphics[scale=0.5]{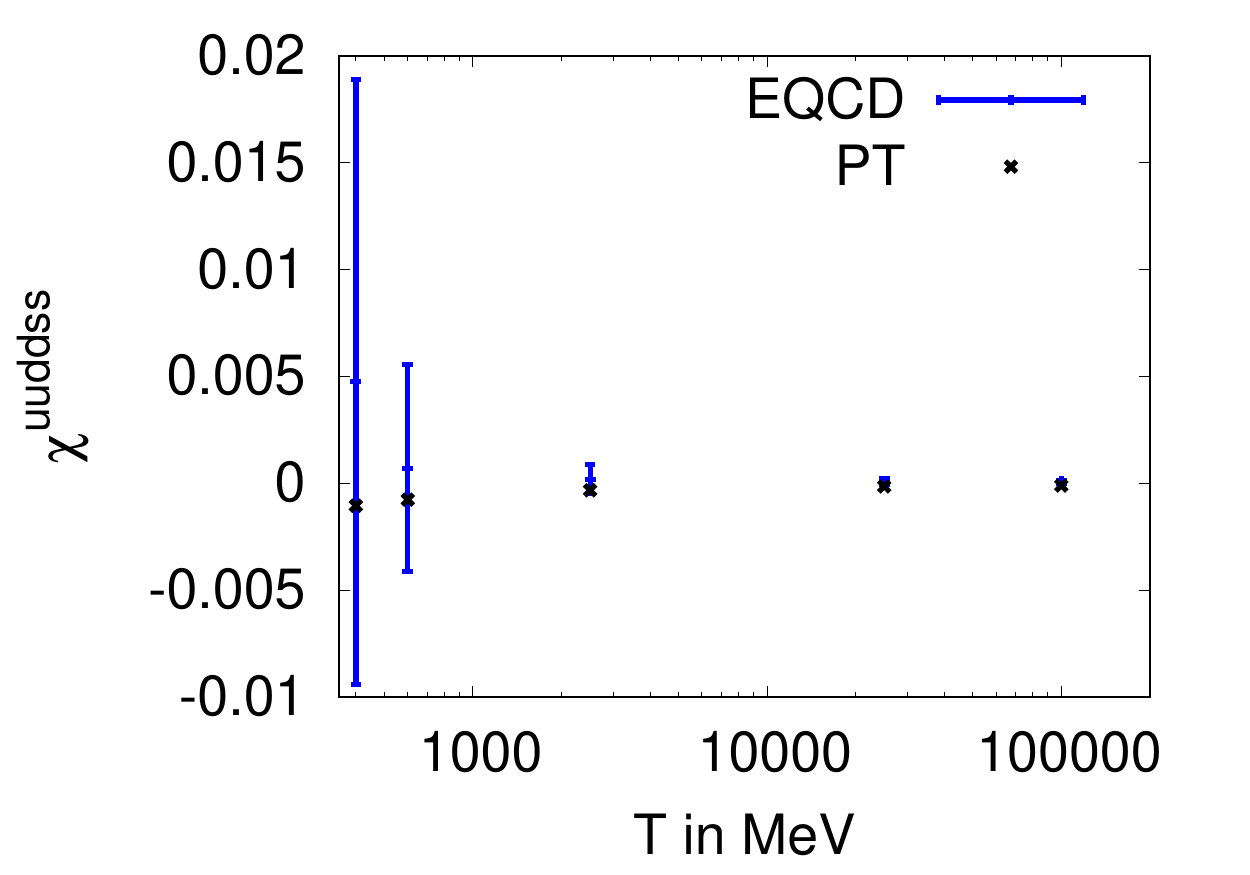} & 
		\includegraphics[scale=0.5]{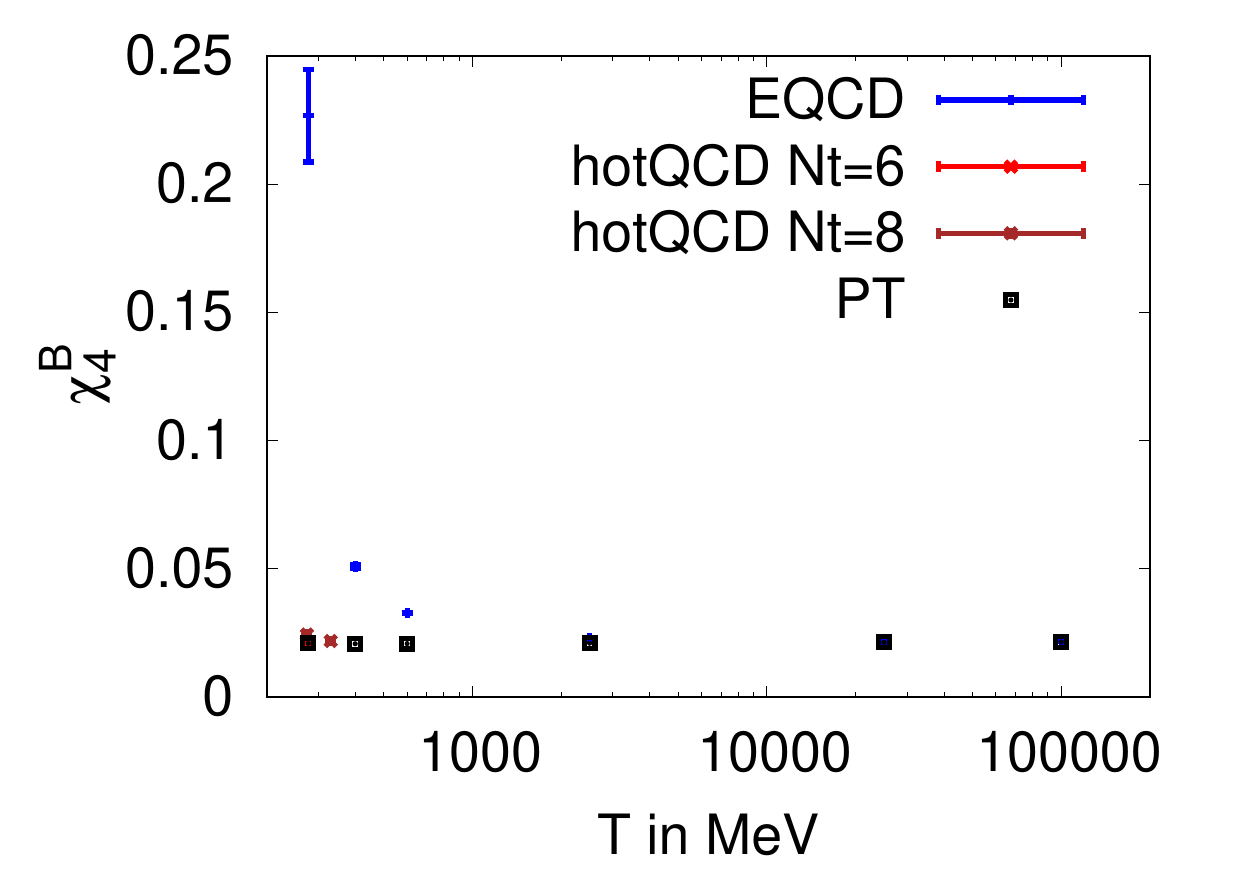}\\ 
		(e)  &  (f)\\
	\end{tabular}
	\caption{Comparison of perturbative results \cite{Ipp:2006ij}, hotQCD lattice results \cite{Bazavov:2017dus}, and our results for quartic and sextic susceptibilities at $\bmu=0$.}
	\label{comp_higher_PT}
\end{figure}

Beyond second derivatives, comparing our data for the higher derivatives to their respective perturbative predictions provides an important benchmark.
Restricting ourselves to $\bmu=0$, we plot all susceptibilities which are not expected to vanish for symmetry reasons in Fig.~\ref{comp_higher_PT}. Just like for the second derivatives, the discrepancy near the QCD crossover temperature is substantial. As one increases the temperature, the two curves approach each other. 
The lowest temperature was not even included in the plots of the sixth order derivatives since it essentially consists of a huge error bar and would only prevent the reader from noticing details in the plot at higher temperatures.  
The higher derivatives feature a quicker convergence towards the perturbative solution in general.
We translate our quartic susceptibilities to derivatives with respect to the baryon chemical potential $\mu_\mathrm{B}$ via
\begin{equation}
\chi^{B}_4 = \frac{1}{27} \left( \chi^{iiii} + 8 \chi^{iiij} + 12 \chi^{iijk} + 6 \chi^{iijj} \right) \, .
\end{equation}
Technically, the above expression is only exact for $\nf=3$. However, if applying it consistently to perturbative results and EQCD results, both relying on the same number of massless flavors, our data and perturbation theory are still comparable, even at higher $\nf$.

Fig.~\ref{comp_higher_PT}(f) compares our data to perturbation theory and lattice data \cite{Bazavov:2017dus}. Since the lattice data is only available at two different $N_\mathrm{t}$, an extrapolation to the continuum is not possible. However, we face the same significant deviations of our low-temperature results from lattice computations. The margin is so big here that it is excluded that this can be tracked down to remaining lattice-cutoff effects of the full QCD data. On the contrary, the reasons for this deviation seem to be the same as for the big deviations from the lattice data for the second derivative: neglecting the strange quark mass, poor scale fixing, and insufficient accuracy in the matching of the effective theory close to the pseudocritical temperature.

\section{Conclusion}
\label{sec:concl}
EQCD is an effective description of high-temperature QCD which can be robustly derived using perturbation theory. It fully includes the non-perturbative soft thermal physics, and offers an economical way to study its effects in Monte Carlo simulations which are numerically less costly than full 4-dimensional lattice simulations. It also has a dramatically milder sign problem than full QCD.
In the present work, we have derived formulae for higher derivatives of the pressure with respect to the quark chemical potential up to sixth order. We have studied these operators on the lattice and provided continuum-extrapolated results in six cases corresponding to physical scenarios at different temperatures $T$ and number of massless quark flavors $\nf$. 
Our data was compared to perturbation theory at high temperatures and full-QCD lattice simulations at low temperatures close to the crossover temperature $\Tc$ of full QCD at vanishing quark chemical potential.
For temperatures reasonably well above $\Tc$, we were able to extract a clear signal on the lattice for all derivatives up to sixth order; a clean extrapolation to the continuum is possible. Agreement with perturbative results sets in around $ T \approx 25 ~ \mathrm{GeV}$.

Our results shed light into the limitations of the dimensional reduction procedure in general, and the perturbative treatment of EQCD.
The data at $\nf=3$ show substantial deviations from $\nf=2+1$ full QCD lattice data. We conclude that the dimensional reduction procedure at $\OO(g^3)$-accuracy does not work for temperatures below $400~\mathrm{MeV}$, probably even higher temperatures. In particular, the reasons for this are threefold:
First, the quality of the dimensional reduction is poor around the pseudocritical temperature of full QCD, which is where full QCD-lattice data is available. Especially the matching of the chemical potential-term of full QCD to the three-dimensional $z$  constrains the accuracy of the matching to $\OO(g^3)$. The substantially better agreement of purely perturbative results (containing EQCD treatment for the soft sector) with fourdimensional lattice predictions hints that already a -- formally incomplete -- usage of a higher loop running coupling could accelerate the convergence of our fully matched lattice EQCD results towards fourdimensional lattice predictions drastically.
Second, we neglected the mass of the strange quark, which turns out not to be a valid approximation up to $T \approx 400~\mathrm{MeV}$. A finite strange quark mass could be incorporated into the dimensional reduction procedure from full QCD to EQCD \cite{Laine:2019uua}, even though that would partly break the symmetry of the derivatives in flavor space.
Third, and somewhat related to the previous point, we set the scale by a value for $\LMSb$ for $2+1$ flavor full QCD, which assumes a finite strange quark mass and therefore deviates from the $\nf = 3$-flavor results that would be more rigorous to use in our case. 

Under the caveat of these improvements, lattice EQCD simulations of higher order cumulants could deliver a valuable tool for the determination of the equation of state of nuclear matter at temperatures of a few times the pseudocritical temperature and moderate quark chemical potential.

\section*{Acknowledgments}
\label{sec:Aknwldg}
We thank Aleksi Vuorinen and Mikko Laine for fruitful discussions on the present work, in particular the perturbative aspects. We highly appreciated Olaf Kaczmarek and Frithjof Karsch providing us with raw data for the higher cumulants from the lattice. Moreover, we thank Ari Hietanen for his initial work on the analytical expressions of the third and fourth derivatives from EQCD.
The support of the Academy of Finland grants 308791 and 320123 is acknowledged.

\appendix

\section{Simulation parameters}
\label{app:sim_params}
Table \ref{sim_params} shows the simulation parameters at which we conducted out three-dimensional Monte-Carlo simulations. A total computation budget of $430,000~\mathrm{CPU}\times\mathrm{hrs}$ was utilized. Due to bad core allocation on the local cluster, some of the statistics of different temperature or $z$, but equal $\gsqa$, differ a bit. As already pointed out, we applied a small amount of mass reweighting to the $z=0.0, 0.025, 0.05$ data points at $T = 277~\mathrm{MeV}$ at all considered lattice spacings to stabilize the system in the supercooled phase. The reweighting allowed a computation at $y_\mathrm{comp}=0.375$.

\begin{table}[htbp!] 
\centering
{\tiny
\begin{tabular}{|c|c|c|c|c|c|}	
\hline
$\gsqa$ & $x$ & $y_0$ & $z$ & $V$ & statistics \\
\hline
$1/4, 1/6, 1/8, 1/12, 1/16$ & $0.1143767$ & $0.3733574$ & $0.0$ & $32^3, 48^3, 64^3, 96^3, 128^3$ & $1227640, 632100, 523420, 252380, 230120$ \\
$1/4, 1/6, 1/8, 1/12, 1/16$ & $0.1143767$ & $0.3733574$ & $0.025$ & $32^3, 48^3, 64^3, 96^3, 128^3$ & $1077660, 621940, 472540, 293720, 246080$ \\
$1/4, 1/6, 1/8, 1/12, 1/16$ & $0.1143767$ & $0.3733574$ & $0.05$ & $32^3, 48^3, 64^3, 96^3, 128^3$ & $1259060, 605380, 481740, 284080, 251800$ \\
$1/4, 1/6, 1/8, 1/12, 1/16$ & $0.1143767$ & $0.3733574$ & $0.1$ & $32^3, 48^3, 64^3, 96^3, 128^3$ & $1026300, 627440, 496320, 314060, 261820$ \\
$1/4, 1/6, 1/8, 1/12, 1/16$ & $0.1143767$ & $0.3733574$ & $0.15$ & $32^3, 48^3, 64^3, 96^3, 128^3$ & $1098280, 648460, 536000, 289840, 164600$ \\
$1/4, 1/6, 1/8, 1/12, 1/16$ & $0.1143767$ & $0.3733574$ & $0.2$ & $32^3, 48^3, 64^3, 96^3, 128^3$ & $1098920, 650240, 541740, 306080, 273100$ \\
\hline
$1/4, 1/6, 1/8, 1/12, 1/16$ & $0.09618573$ & $0.4361834$ & $0.0$ & $32^3, 48^3, 64^3, 96^3, 128^3$ & $1177760, 633280, 538700, 319880, 248320$ \\
$1/4, 1/6, 1/8, 1/12, 1/16$ & $0.09618573$ & $0.4361834$ & $0.025$ & $32^3, 48^3, 64^3, 96^3, 128^3$ & $1096740, 649880, 538840, 295400, 231500$ \\
$1/4, 1/6, 1/8, 1/12, 1/16$ & $0.09618573$ & $0.4361834$ & $0.05$ & $32^3, 48^3, 64^3, 96^3, 128^3$ & $1098280, 651780, 538760, 282420, 162940$ \\
$1/4, 1/6, 1/8, 1/12, 1/16$ & $0.09618573$ & $0.4361834$ & $0.1$ & $32^3, 48^3, 64^3, 96^3, 128^3$ & $1107220, 648700, 536120, 279660, 245120$ \\
$1/4, 1/6, 1/8, 1/12, 1/16$ & $0.09618573$ & $0.4361834$ & $0.15$ & $32^3, 48^3, 64^3, 96^3, 128^3$ & $1101740, 650160, 53240, 292720, 243980$ \\
$1/4, 1/6, 1/8, 1/12, 1/16$ & $0.09618573$ & $0.4361834$ & $0.2$ & $32^3, 48^3, 64^3, 96^3, 128^3$ & $1101740, 650160, 532400, 292720, 243980$ \\
\hline
$1/4, 1/6, 1/8, 1/12, 1/16$ & $0.08182539$ & $0.5055096$ & $0.0$ & $32^3, 48^3, 64^3, 96^3, 128^3$ & $1272640, 979620, 539800, 287060, 210940$ \\
$1/4, 1/6, 1/8, 1/12, 1/16$ & $0.08182539$ & $0.5055096$ & $0.025$ & $32^3, 48^3, 64^3, 96^3, 128^3$ & $1161060, 664900, 492600, 193580, 253700$ \\
$1/4, 1/6, 1/8, 1/12, 1/16$ & $0.08182539$ & $0.5055096$ & $0.05$ & $32^3, 48^3, 64^3, 96^3, 128^3$ & $1099440, 651020, 535260, 308380, 265500$ \\
$1/4, 1/6, 1/8, 1/12, 1/16$ & $0.08182539$ & $0.5055096$ & $0.1$ & $32^3, 48^3, 64^3, 96^3, 128^3$ & $1096340, 649840, 536620, 304440, 254560$ \\
$1/4, 1/6, 1/8, 1/12, 1/16$ & $0.08182539$ & $0.5055096$ & $0.15$ & $32^3, 48^3, 64^3, 96^3, 128^3$ & $1101140, 415120, 522760, 304920, 253740$ \\
$1/4, 1/6, 1/8, 1/12, 1/16$ & $0.08182539$ & $0.5055096$ & $0.2$ & $32^3, 48^3, 64^3, 96^3, 128^3$ & $1091620, 636820, 468260, 309900, 245920$ \\
\hline
$1/4, 1/6, 1/8, 1/12, 1/16$ & $0.04466692$ & $0.7981834$ & $0.0$ & $32^3, 48^3, 64^3, 96^3, 128^3$ & $888240, 446580, 232320, 312360, 260140$ \\
$1/4, 1/6, 1/8, 1/12, 1/16$ & $0.04466692$ & $0.7981834$ & $0.025$ & $32^3, 48^3, 64^3, 96^3, 128^3$ & $1097200, 649900, 536400, 221940, 85160$ \\
$1/4, 1/6, 1/8, 1/12, 1/16$ & $0.04466692$ & $0.7981834$ & $0.05$ & $32^3, 48^3, 64^3, 96^3, 128^3$ & $1095460, 649960, 534580, 294620, 246120$ \\
$1/4, 1/6, 1/8, 1/12, 1/16$ & $0.04466692$ & $0.7981834$ & $0.1$ & $32^3, 48^3, 64^3, 96^3, 128^3$ & $1101780, 650800, 814480, 239280, 239720$ \\
$1/4, 1/6, 1/8, 1/12, 1/16$ & $0.04466692$ & $0.7981834$ & $0.15$ & $32^3, 48^3, 64^3, 96^3, 128^3$ & $1094760, 650720, 536720, 241880, 179340$ \\
$1/4, 1/6, 1/8, 1/12, 1/16$ & $0.04466692$ & $0.7981834$ & $0.2$ & $32^3, 48^3, 64^3, 96^3, 128^3$ & $910680, 600100, 546580, 272380, 241120$ \\
\hline
$1/4, 1/6, 1/8, 1/12, 1/16$ & $0.02372764$ & $1.250949$ & $0.0$ & $32^3, 48^3, 64^3, 96^3, 128^3$ & $920380, 684900, 544300, 281420, 224680$ \\
$1/4, 1/6, 1/8, 1/12, 1/16$ & $0.02372764$ & $1.250949$ & $0.025$ & $32^3, 48^3, 64^3, 96^3, 128^3$ & $1102120, 649580, 536620, 124440, 188120$ \\
$1/4, 1/6, 1/8, 1/12, 1/16$ & $0.02372764$ & $1.250949$ & $0.05$ & $32^3, 48^3, 64^3, 96^3, 128^3$ & $1099420, 648700, 536280, 293940, 246760$ \\
$1/4, 1/6, 1/8, 1/12, 1/16$ & $0.02372764$ & $1.250949$ & $0.1$ & $32^3, 48^3, 64^3, 96^3, 128^3$ & $1097060, 650400, 474980, 169780, 260860$ \\
$1/4, 1/6, 1/8, 1/12, 1/16$ & $0.02372764$ & $1.250949$ & $0.15$ & $32^3, 48^3, 64^3, 96^3, 128^3$ & $1045720, 641860, 536300, 189620, 90720$ \\
$1/4, 1/6, 1/8, 1/12, 1/16$ & $0.02372764$ & $1.250949$ & $0.2$ & $32^3, 48^3, 64^3, 96^3, 128^3$ & $1018700, 238860, 148320, 256360, 191560$ \\
\hline
$1/4, 1/6, 1/8, 1/12, 1/16$ & $0.01995365$ & $1.497731$ & $0.0$ & $32^3, 48^3, 64^3, 96^3, 128^3$ & $1098300, 628160, 510340, 275100, 250740$ \\
$1/4, 1/6, 1/8, 1/12, 1/16$ & $0.01995365$ & $1.497731$ & $0.025$ & $32^3, 48^3, 64^3, 96^3, 128^3$ & $840600, 665080, 554880, 262820, 176880$ \\
$1/4, 1/6, 1/8, 1/12, 1/16$ & $0.01995365$ & $1.497731$ & $0.05$ & $32^3, 48^3, 64^3, 96^3, 128^3$ & $1246540, 658280, 537820, 86120, 35700$ \\
$1/4, 1/6, 1/8, 1/12, 1/16$ & $0.01995365$ & $1.497731$ & $0.1$ & $32^3, 48^3, 64^3, 96^3, 128^3$ & $1097020, 649760, 535880, 71700, 235200$ \\
$1/4, 1/6, 1/8, 1/12, 1/16$ & $0.01995365$ & $1.497731$ & $0.15$ & $32^3, 48^3, 64^3, 96^3, 128^3$ & $1252160, 650720, 536080, 247040, 235580$ \\
$1/4, 1/6, 1/8, 1/12, 1/16$ & $0.01995365$ & $1.497731$ & $0.2$ & $32^3, 48^3, 64^3, 96^3, 128^3$ & $1096340, 650360, 536300, 134540, 257580$ \\
\hline
\end{tabular}
}
\caption{Simulation parameters of our $180$ runs.}
\label{sim_params}
\end{table}

\FloatBarrier

\section{Expressions for higher derivatives}
\label{app:express_higher_der}
In this appendix, we present the full analytically continued expressions for the higher derivatives. Each new derivative requires new condensates and new way of analytically continuing the lower condensates.
We denote the condensates with $K$. The first index is the sum of the two powers of $\Phi^2$ and $\Phi^3$. The second index runs from $1$ to $\mathrm{first index} + 1$ and refers to the power of $\Phi^3$ involved.
Symmetrization in indices are denoted by three dots.

\subsection{Fourth derivatives}
Continuing the condensates
\begin{align*}
K_{2,1} &= V \gsix \Big\langle 
	\left( \Phi^2 - \langle \Phi^2 \rangle \right)^2 \Big\rangle
	\approx a_{21,0} + a_{21,2} z^2 \\
K_{3,1} &\equiv V^2 \gE^{12} \Big\langle 
	\left( \Phi^2 - \langle \Phi^2 \rangle \right)^3 \Big\rangle 
\approx a_{31,0} + a_{31,2} z^2 \\
K_{3,2} &\equiv V^2 \gE^{12} \Big\langle 
	\left( \Phi^2 - \langle \Phi^2 \rangle \right)^2 
	\left( \Phi^3 - \langle \Phi^3 \rangle \right) \Big\rangle
= \frac{\partial K_{2,1}}{\partial (iz)}
\approx - 2 i z a_{21,2} \\
K_{3,3} &\equiv V^2 \gE^{12} \Big\langle 
	\left( \Phi^2 - \langle \Phi^2 \rangle \right) 
	\left( \Phi^3 - \langle \Phi^3 \rangle \right)^2 \Big\rangle \\
&= \frac{\partial K_{2,2}}{\partial (iz)} 
\approx - a_{22,1} + 3 a_{22,3} z^2 \\
K_{4,1} & \equiv V^3 \gE^{18} \Big\langle
	\left( \Phi^2 - \langle \Phi^2 \rangle \right)^4 \Big\rangle
\approx a_{41,0} \\
K_{4,2} & \equiv V^3 \gE^{18} \Big\langle
	\left( \Phi^2 - \langle \Phi^2 \rangle \right)^3
	\left( \Phi^3 - \langle \Phi^3 \rangle \right) \Big\rangle
= \frac{\partial K_{3,1}}{\partial (iz)} 
\approx -2 i z a_{31,2} \\
K_{4,3} & \equiv V^3 \gE^{18} \Big\langle
	\left( \Phi^2 - \langle \Phi^2 \rangle \right)^2
	\left( \Phi^3 - \langle \Phi^3 \rangle \right)^2 \Big\rangle \\
&= \frac{\partial K_{3,2}}{\partial (iz)}
= \frac{\partial^2 K_{2,1}}{\partial^2 (iz)} 
\approx - 2 a_{21,2} \\
K_{4,4} & \equiv V^3 \gE^{18} \Big\langle
	\left( \Phi^2 - \langle \Phi^2 \rangle \right)
	\left( \Phi^3 - \langle \Phi^3 \rangle \right)^3 \Big\rangle \\
&= \frac{\partial K_{3,3}}{\partial (iz)} = \frac{\partial^2 K_{2,2}}{\partial^2(iz)} 
\approx -6 i z a_{22,3} \\
K_{4,5} & \equiv V^3 \gE^{18} \Big\langle
	\left( \Phi^3 - \langle \Phi^3 \rangle \right)^4 \Big\rangle \notag \\
&= \frac{\partial K_{3,4}}{\partial (iz)} = \frac{\partial^2 K_{2,3}}{\partial^2 (iz)} \approx -2 a_{23,2}  \, ,
\end{align*}
analytically yields the analytically continued expression for the fourth derivative
\begin{align}	\label{eq:ac_quartic_sus}
\left( \chitd^{ijkl} \right)^\mathrm{ac} =& \;
	16 y_2^4 \bmi \bmj \bmk \bml K_{4,1}(z_I) \notag \\
	&+ \frac{8 y_2^3}{3 \pi} \frac{z_R}{z_I} 
		\left( \bmi \bmj \bmk + \bmj \bmk \bml 
		+ \bmk \bml \bmi + \bml \bmi \bmj \right)
		K_{4,2}(z_I) \notag \\
	&- \frac{4 y_2^2}{9 \pi^2} 
		\left( \bmi \bmj + \bmi \bmk + \bmi \bml
		+ \bmj \bmk + \bmj \bml + \bmk \bml \right)
		K_{4,3}(z_I) \notag \\
	&- \frac{2 y_2}{(3 \pi)^3}
		\left( \bmi + \bmj + \bmk + \bml \right) 
		\frac{z_R}{z_I} K_{4,4}(z_I)
	+ \frac{1}{(3 \pi)^4} K_{4,5}(z_I) \notag \\
	&- 8 y_2^3 
		\left( \delta_{ij} \bmk \bml + \delta_{ik} \bmj \bml
		+ \delta_{il} \bmj \bmk 
		\right. \notag \\
& \hspace{30pt} \left.
		+ \delta_{jk} \bmi \bml + \delta_{jl} \bmi \bmj 
		+ \delta_{kl} \bmi \bmj \right)
		\left( K_{3,1}(z_I) 
			+ \frac{z_R^2}{z_I} K_{4,2}(z_I) \right)	\notag \\
	&- \frac{4 y_2^2}{3 \pi}
		\left[ 
			\delta_{ij} (\bmk + \bml) 
			+ \delta_{ik} (\bmj + \bml)
			+ \delta_{il} (\bmj + \bmk)
			+ \delta_{jk} (\bmi + \bml) \right. \notag \\ 
	& \hspace{50pt} \left.
			+ \delta_{jl} (\bmi + \bmk)
			+ \delta_{kl} (\bmi + \bmj) 
		\right] \frac{z_R}{z_I} K_{3,2}(z_I) \notag \\
	&+ \frac{2 y_2}{(3 \pi)^2} 
		\left( \delta_{ij} + \delta_{ik} + \delta_{il}
		+ \delta_{jk} + \delta_{jl} + \delta_{kl} \right) 
		\left( 
			K_{3,3}(z_I) + \frac{z_R^2}{z_I} K_{4,4}(z_I) 
		\right) \notag \\
	&+ 4 y_2^2
		\left( \delta_{ij} \delta_{kl} 
			+ \delta_{ik} \delta_{jl}
			+ \delta_{il} \delta_{jk} \right) 
		\left( K_{2,1}(z_I) + \frac{z_R^2}{z_I} K_{3,2}(z_I) \right) \, .
\end{align}
We see that $K_{3,2}$ is the first derivative of $K_{2,1}$ with respect to $iz$, $K_{4,2}$ the first derivative of $K_{3,1}$, and $K_{4,4}$ the first derivative of $K_{3,3}$. Therefore, the three named condensates of lower order behave non-trivially under the analytic continuation.

\subsection{Fifth derivatives}
Analytically continuing the condensates as
\begin{align*}
K_{3,1} &= V^2 \gE^{12} \left\langle 
			\left( \Phi^2 - \langle \Phi^2 \rangle \right)^3  					\right\rangle
		\approx a_{31,0} + a_{31,2} z^2 \\
K_{3,2} &= V^2 \gE^{12} \left\langle 
			\left( \Phi^2 - \langle \Phi^2 \rangle \right)^2 
			\left( \Phi^3 - \langle \Phi^3 \rangle \right) 					\right\rangle 
		\approx -2 i z a_{21,1}\\
K_{4,1} &= V^3 \gE^{18} \left\langle 
			\left( \Phi^2 - \langle \Phi^2 \rangle \right)^4
		\right\rangle 
		\approx a_{41,0} + a_{41,2} z^2 \\
K_{4,2} &= V^3 \gE^{18} \left\langle 
			\left( \Phi^2 - \langle \Phi^2 \rangle \right)^3 
			\left( \Phi^3 - \langle \Phi^3 \rangle \right) 					\right\rangle 
		\approx -2 i z a_{31,2} \\
K_{4,3} &= V^3 \gE^{18} \left\langle 
			\left( \Phi^2 - \langle \Phi^2 \rangle \right)^2 
			\left( \Phi^3 - \langle \Phi^3 \rangle \right)^2				\right\rangle 
		\approx a_{43,0} + a_{43,2} z^2 \\
K_{4,4} &= V^3 \gE^{18} \left\langle 
			\left( \Phi^2 - \langle \Phi^2 \rangle \right) 
			\left( \Phi^3 - \langle \Phi^3 \rangle \right)^3				\right\rangle 
		\approx  -2 i z a_{33,2} \\
K_{5,1} &= V^4 \gE^{24} \left\langle 
			\left( \Phi^2 - \langle \Phi^2 \rangle \right)^5
		\right\rangle
		\approx a_{51,0} \\
K_{5,2} &= V^4 \gE^{24} \left\langle 
			\left( \Phi^2 - \langle \Phi^2 \rangle \right)^4 
			\left( \Phi^3 - \langle \Phi^3 \rangle \right) 					\right\rangle
		\approx -2 i z a_{41,2} \\
K_{5,3} &= V^4 \gE^{24} \left\langle 
			\left( \Phi^2 - \langle \Phi^2 \rangle \right)^3 
			\left( \Phi^3 - \langle \Phi^3 \rangle \right)^2				\right\rangle 
		\approx -2 a_{31,2} \\
K_{5,4} &= V^4 \gE^{24} \left\langle 
			\left( \Phi^2 - \langle \Phi^2 \rangle \right)^2 
			\left( \Phi^3 - \langle \Phi^3 \rangle \right)^3				\right\rangle 
		\approx -2 i z a_{43,2} \\
K_{5,5} &= V^4 \gE^{24} \left\langle 
			\left( \Phi^2 - \langle \Phi^2 \rangle \right) 
			\left( \Phi^3 - \langle \Phi^3 \rangle \right)^4				\right\rangle 
		\approx -2 a_{33,2} \\
K_{5,6} &= V^4 \gE^{24} \left\langle 
			\left( \Phi^3 - \langle \Phi^3 \rangle \right)^5				\right\rangle
		\approx -2 i z a_{45,2} \, ,
\end{align*}
we arrive at the fifth order derivative expression
\begin{align}	\label{eq:ac_quintic_sus}
\left( \chitd^{ijklm} \right)^\mathrm{ac} &= 
		- (2 y_2)^5 \bmi \bmj \bmk \bml \bmm K_{5,1}
	- \frac{(2 y_2)^4}{3 \pi} (\bmi \bmj \bmk \bml + \dots)
		\frac{\zR}{\zI} K_{5,2}
\notag \\ & 
	+ \frac{(2 y_2)^3}{(3 \pi)^2} (\bmi \bmj \bmk + \dots)
		K_{5,3}
	+ \frac{(2 y_2)^2}{(3 \pi)^3} (\bmi \bmj + \dots)
		\frac{\zR}{\zI} K_{5,4}
\notag \\ & 
	- \frac{2 y_2}{(3 \pi)^4} (\bmi + \dots)
		K_{5,5}
	- \frac{1}{(3 \pi)^5} 
		\frac{\zR}{\zI} K_{5,6}
\notag \\ &
	+ (2 y_2)^4 (\bmi \bmj \bmk \delta_{lm} + \dots)
		\left( K_{4,1} + \frac{\zR^2}{\zI} 
			K_{5,2} \right)
	+ \frac{(2 y_2)^3}{3 \pi} (\bmi \bmj \delta_{lm} + \dots)
		\frac{\zR}{\zI} K_{4,2}
\notag \\ &
	- \frac{(2 y_2)^2}{(3 \pi)^2} (\bmi \delta_{lm} + \dots)
		\left( K_{4,3} + \frac{\zR^2}{\zI} 
			K_{5,4} \right)
	- \frac{2 y_2}{(3 \pi)^3} (\delta_{lm} + \dots)
		\frac{\zR}{\zI} K_{4,4}
\notag \\ &
	- (2 y_2)^3 (\bmi \delta_{jk} \delta_{lm} + \dots)
		\left( K_{3,1} + \frac{\zR^2}{\zI} 
			K_{4,2} \right)
\notag \\ &
	- \frac{(2 y_2)^2}{3 \pi} 
		(\delta_{jk} \delta_{lm} + \dots)
		\frac{\zR}{\zI} 
		\left( K_{3,2} - \frac{\zR^2}{3} 
			K_{5,4} \right) \, .
\end{align}
We see that the condensates $K_{5,2}$, $K_{5,4}$, $K_{4,2}$, and $K_{5,4}$ can be phrased as derivatives of $K_{4,1}$, $K_{4,3}$, $K_{3,1}$, and $K_{3,2}$, respectively, and therefore the latter four are nontrivially continued under $iz \to z$.

\subsection{Sixth derivatives}
The sixth order derivative contains the condensates
\begin{align*}
K_{3,1} &= V^2 \gE^{12} \left\langle  
			\left( \Phi^2 - \langle \Phi^2 \rangle \right)^3
		\right\rangle \\
K_{4,1} &= V^3 \gE^{18} \left\langle  
			\left( \Phi^2 - \langle \Phi^2 \rangle \right)^4
		\right\rangle \\
K_{4,2} &= V^3 \gE^{18} \left\langle  
			\left( \Phi^2 - \langle \Phi^2 \rangle \right)^3
			\left( \Phi^3 - \langle \Phi^3 \rangle \right)
		\right\rangle \\
K_{4,3} &= V^3 \gE^{18} \left\langle  
			\left( \Phi^2 - \langle \Phi^2 \rangle \right)^2
			\left( \Phi^3 - \langle \Phi^3 \rangle \right)^2
		\right\rangle \\
K_{5,1} &= V^4 \gE^{24} \left\langle  
			\left( \Phi^2 - \langle \Phi^2 \rangle \right)^5
		\right\rangle \\
K_{5,2} &= V^4 \gE^{24} \left\langle  
			\left( \Phi^2 - \langle \Phi^2 \rangle \right)^4
			\left( \Phi^3 - \langle \Phi^3 \rangle \right)
		\right\rangle \\
K_{5,3} &= V^4 \gE^{24} \left\langle  
			\left( \Phi^2 - \langle \Phi^2 \rangle \right)^3
			\left( \Phi^3 - \langle \Phi^3 \rangle \right)^2
		\right\rangle \\
K_{5,4} &= V^4 \gE^{24} \left\langle  
			\left( \Phi^2 - \langle \Phi^2 \rangle \right)^2
			\left( \Phi^3 - \langle \Phi^3 \rangle \right)^3
		\right\rangle \\
K_{5,5} &= V^4 \gE^{24} \left\langle  
			\left( \Phi^2 - \langle \Phi^2 \rangle \right)
			\left( \Phi^3 - \langle \Phi^3 \rangle \right)^4
		\right\rangle \\
K_{6,1} &= V^5 \gE^{30} \left\langle  
			\left( \Phi^2 - \langle \Phi^2 \rangle \right)^6
		\right\rangle \\
K_{6,2} &= V^5 \gE^{30} \left\langle  
			\left( \Phi^2 - \langle \Phi^2 \rangle \right)^5
			\left( \Phi^3 - \langle \Phi^3 \rangle \right)
		\right\rangle \\
K_{6,3} &= V^5 \gE^{30} \left\langle  
			\left( \Phi^2 - \langle \Phi^2 \rangle \right)^4
			\left( \Phi^3 - \langle \Phi^3 \rangle \right)^2
		\right\rangle \\
K_{6,4} &= V^5 \gE^{30} \left\langle  
			\left( \Phi^2 - \langle \Phi^2 \rangle \right)^3
			\left( \Phi^3 - \langle \Phi^3 \rangle \right)^3
		\right\rangle \\
K_{6,5} &= V^5 \gE^{30} \left\langle  
			\left( \Phi^2 - \langle \Phi^2 \rangle \right)^2
			\left( \Phi^3 - \langle \Phi^3 \rangle \right)^4
		\right\rangle \\
K_{6,6} &= V^5 \gE^{30} \left\langle  
			\left( \Phi^2 - \langle \Phi^2 \rangle \right)
			\left( \Phi^3 - \langle \Phi^3 \rangle \right)^5
		\right\rangle \\
K_{6,7} &= V^5 \gE^{30} \left\langle  
			\left( \Phi^3 - \langle \Phi^3 \rangle \right)^6
		\right\rangle \, ,
\end{align*}
which can be combined to the analytically continued expression for the sixth order derivative
\begin{align}	\label{eq:ac_sextic_sus}
\left( \chitd^{ijklmn} \right)^\mathrm{ac} =&
	+ (2 y_2)^6 \bmi \bmj \bmk \bml \bmm \bmn
		K_{6,1}
	+ \frac{(2 y_2)^5}{3 \pi} 
		(\bmi \bmj \bmk \bml \bmm + \dots)
		\frac{\zR}{\zI} K_{6,2}
\notag \\ & 
	- \frac{(2 y_2)^4}{(3 \pi)^2} 
		(\bmi \bmj \bmk \bml + \dots)
		K_{6,3}
	- \frac{(2 y_2)^3}{(3 \pi)^3} 
		(\bmi \bmj \bmk + \dots)
		\frac{\zR}{\zI} K_{6,4}
\notag \\ &
	+ \frac{(2 y_2)^2}{(3 \pi)^4} 
		(\bmi \bmj + \dots) 
		K_{6,5}
	+ \frac{2 y_2}{(3 \pi)^5} 
		(\bmi + \dots) 
		\frac{\zR}{\zI} K_{6,6}
	- \frac{1}{(3 \pi)^6} 
		K_{6,7}
\notag \\ &
	- (2 y_2)^5 (\bmi \bmj \bmk \bml \delta_{mn} + \dots)
		\left( K_{5,1} + \frac{\zR^2}{\zI}
			 K_{6,2} \right)
\notag \\ &
	- \frac{(2 y_2)^4}{3 \pi} 
		(\bmi \bmj \bmk \delta_{mn} + \dots)
		\frac{\zR}{\zI} K_{5,2}
\notag \\ &
	+ \frac{(2 y_2)^3}{(3 \pi)^2} 
		(\bmi \bmj \delta_{mn} + \dots)
		\left( K_{5,3}  + \frac{\zR^2}{\zI}
			 K_{6,4} \right)
	+ \frac{(2 y_2)^2}{(3 \pi)^3} 
		(\bmi \delta_{mn} + \dots)
		\frac{\zR}{\zI} K_{5,4}
\notag \\ &
	- \frac{2 y_2}{(3 \pi)^4} 
		(\delta_{mn} + \dots)
		\left( K_{5,5}  + \frac{\zR^2}{\zI}
			 K_{6,6} \right)
\notag \\ &
	+ (2 y_2)^4 (\bmi \bmj \delta_{kl} \delta_{mn} + \dots)
		\left( K_{4,1}  + \frac{\zR^2}{\zI}
			 K_{5,2} \right)
\notag \\ &
	+ \frac{(2 y_2)^3}{3 \pi} 
		(\bmi \delta_{kl} \delta_{mn} + \dots)
		\frac{\zR}{\zI} 
		\left( K_{4,2} - \frac{\zR^2}{3} 
		K_{6,4} \right)
\notag \\ &
	- \frac{(2 y_2)^2}{(3 \pi)^2} 
		(\delta_{kl} \delta_{mn} + \dots)
		\left( K_{4,3}  + \frac{\zR^2}{\zI}
			 K_{5,4} \right)
\notag \\ &
	- (2 y_2)^3 (\delta_{ij} \delta_{kl} \delta_{mn} + \dots) 
		\left( K_{3,1} 
		+ \frac{\zR^2}{\zI} K_{4,2} 
		- \frac{\zR^4}{3 \zI} K_{6,4} \right) 
\, .
\end{align}
Beyond the usual identification of condensates that correspond to first or second derivatives of other condensates with respect to $iz$, we see the connection of three condensates for the first time via 
\begin{equation}
K_{6,4} = \frac{\partial^2 K_{4,2}}{\partial^2 (iz)} = \frac{\partial^3 K_{3,1}}{\partial^3 (iz)} \, .
\end{equation}

\FloatBarrier

\section{Tabulated results}
\label{app:tab_results}
In this appendix, we give tabulated numerical results for all (three-dimensional) cumulants that we computed, in units of the three-dimensional coupling $\gsq$. The number of different indices of some of the derivatives exceeds the number of massless quark flavor available. Therefore, these cumulants do not correspond to a physical scenario. However, $\chitd^{ijkl}$ for instance has no physical meaning in scenarios with $\nf=3$, but gives a good estimate how big the influence of the generically fourth-order condensates in \eqref{eq:ac_quartic_sus} is, which still contains valid, though not strictly physical, information.

\begin{table}[htbp!] 
\centering
{\tiny
\begin{tabular}{|c||c|c|c|c|c|c|}	
\hline
 & \multicolumn{6}{c|}{$T=277 ~ \mathrm{MeV} , \; \nf = 3$} \\
\hline
$z$ & $0.0$ & $0.025$ & $0.05$ & $0.1$ & $0.15$ & $0.2$ \\
\hline
$\chitd^{ii} / \gsix$ & $-0.0695(12)$ & $-0.0656(11)$ & $-0.05702(90)$ & $-0.02829(52)$ & $-0.00220(40)$ & $0.01994(21)$ \\
$\chitd^{ij} / \gsix$ & $-0.00537(17)$ & $-0.00149(31)$ & $0.00702(42)$ & $0.02032(41)$ & $0.02381(66)$ & $0.02682(27)$ \\
$\chitd^{iii} / \gsix$ & $0.0044(20)$ & $0.110(13)$ & $0.217(12)$ & $0.245(15)$ & $0.1575(36)$ & $0.1228(12)$ \\
$\chitd^{iij} / \gsix$ & $0.0044(20)$ & $0.0484(62)$ & $0.0769(38)$ & $0.05059(76)$ & $0.0288(11)$ & $0.0257(21)$  \\
$\chitd^{ijk} / \gsix$ & $0.0044(20)$ & $0.0176(28)$ & $0.00681(72)$ & $-0.0465(62)$ & $-0.0356(26)$ & $-0.0228(33)$ \\
$\chitd^{iiii} / \gsix$ & $1.338(79)$ & $0.835(11)$ & $-0.74(21)$ & $-3.10(64)$ & $-1.38(10)$ & $-0.504(67)$ \\
$\chitd^{iiij} / \gsix$ & $0.124(14)$ & $-0.177(86)$ & $-1.10(15)$ & $-1.69(32)$ & $-0.452(35)$ & $0.107(20)$ \\
$\chitd^{iijk} / \gsix$ & $0.111(13)$ & $-0.068(44)$ & $-0.457(57)$ & $-0.132(59)$ & $0.415(20)$ & $0.669(16)$ \\
$\chitd^{iijj} / \gsix$ & $0.515(35)$ & $0.265(21)$ & $-0.336(79)$ & $-0.60(17)$ & $0.154(23)$ & $0.46385(51)$ \\
$\chitd^{ijkl} / \gsix$ & $0.105(12)$ & $-0.014(23)$ & $-0.1250(42)$ & $0.645(69)$ & $0.872(50)$ & $0.949(34)$ \\
$\chitd^{iiiii} / \gsix$ & $0.4(24)$ & $-14.5(52)$ & $-26.0(19)$ & $27.5(85)$ & $26.0(19)$ & $15.41(98)$ \\
$\chitd^{iiiij} / \gsix$ & $0.08(93)$ & $-5.5(20)$ & $-8.73(61)$ & $21.1(37)$ & $17.0(11)$ & $10.13(53)$ \\
$\chitd^{iiijk} / \gsix$ & $-0.02(39)$ & $-2.08(71)$ & $-1.98(40)$ & $10.4(16)$ & $6.72(33)$ & $4.135(91)$  \\
$\chitd^{iiijj} / \gsix$ & $0.08(76)$ & $-4.3(15)$ & $-6.31(34)$ & $12.1(28)$ & $8.97(54)$ & $5.45(17)$ \\
$\chitd^{iijkl} / \gsix$ & $-0.03(22)$ & $-0.87(24)$ & $0.44(29)$ & $1.42(80)$ & $-1.25(25)$ & $-0.55(33)$ \\
$\chitd^{iijjk} / \gsix$ & $0.01(40)$ & $-2.01(66)$ & $-1.81(19)$ & $4.96(147)$ & $2.155(70)$ & $1.45(17)$ \\
$\chitd^{ijklm} / \gsix$ & $-0.03(14)$ & $-0.27(1)$ & $1.65(33)$ & $-3.09(48)$ & $-5.23(54)$ & $-2.89(52)$ \\
$\chitd^{iiiiii} / \gsix$ & $-258(40)$ & $-80(19)$ & $326(71)$ & $94(38)$ & $-172(412)$ & $-43(20)$ \\
$\chitd^{iiiiij} / \gsix$ & $-67(15)$ & $4.5(83)$ & $189(31)$ & $-210(50)$ & $-220(30)$ & $-64(14)$ \\
$\chitd^{iiiijk} / \gsix$ & $-22.9(72)$ & $5.1(76)$ & $136(26)$ & $-81(56)$ & $-100(11)$ & $-18.9(68)$ \\
$\chitd^{iiiijj} / \gsix$ & $-52(11)$ & $-10.8(27)$ & $179(37)$ & $3(125)$ & $-88(11)$ & $-17.1(82)$ \\
$\chitd^{iiijkl} / \gsix$ & $-8.0(41)$ & $3.5(49)$ & $68(13)$ & $-29(34)$ & $-14.2(26)$ & $19.35(32)$ \\
$\chitd^{iiijjk} / \gsix$ & $-16.0(54)$ & $3.7(54)$ & $86(15)$ & $-54(25)$ & $-43.5(62)$ & $7.4(23)$ \\
$\chitd^{iiijjj} / \gsix$ & $-41(10)$ & $3.0(54)$ & $107(16)$ & $-126(45)$ & $-104(17)$ & $-14.3(60)$ \\
$\chitd^{iijklm} / \gsix$ & $-4.2(27)$ & $1.6(23)$ & $12.1(21)$ & $-9(14)$ & $42(47)$ & $46.4(43)$ \\
$\chitd^{iijjkl} / \gsix$ & $-9.1(37)$ & $2.2(32)$ & $35.0(49)$ & $-26.2(69)$ & $13.4(36)$ & $33.6(21)$ \\
$\chitd^{iijjkk} / \gsix$ & $-22.9(56)$ & $-2.9(20)$ & $63(11)$ & $-13(15)$ & $-1.9(40)$ & $25.98(28)$ \\
$\chitd^{ijklmn} / \gsix$ & $-2.4(21)$ & $0.75(10)$ & $-15.6(60)$ & $2(37)$ & $70.3(72)$ & $59.9(68)$ \\
\hline
\hline
 & \multicolumn{6}{c|}{$T = 400~\mathrm{MeV} , \; \nf = 3$} \\
\hline
$z$ & $0.0$ & $0.025$ & $0.05$ & $0.1$ & $0.15$ & $0.2$  \\
\hline
$\chitd^{ii} / \gsix$  & $-0.02834(27)$ & $-0.02618(48)$ & $-0.02106(36)$ & $-0.003004(56)$ & $0.01892(11)$ & $0.04226(24)$ \\
$\chitd^{ij} / \gsix$ & $-0.0021364(69)$ & $-0.001061(28)$ & $0.001722(23)$ & $0.00969(15)$ & $0.01714(33)$ & $0.02504(35)$ \\
$\chitd^{iii} / \gsix$ & $0.00026(20)$ & $0.03510(36)$ & $0.06455(58)$ & $0.0982(14)$ & $0.10625(68)$ & $0.11074(73)$ \\
$\chitd^{iij} / \gsix$ & $0.00026(20)$ & $0.01318(15)$ & $0.022831(70)$ & $0.02985(68)$ & $0.03003(23)$ & $0.0260(21)$ \\
$\chitd^{ijk} / \gsix$ & $0.00026(20)$ & $0.002227(43)$ & $0.00197(23)$ & $-0.00434(35)$ & $-0.00813(28)$ & $-0.0164(34)$ \\
$\chitd^{iiii} / \gsix$ & $0.4493(71)$ & $0.4100(20)$ & $0.3013(61)$ & $0.021(14)$ & $0.012(22)$ & $0.033(56)$ \\
$\chitd^{iiij} / \gsix$ & $0.02723(71)$ & $0.0016(15)$ & $-0.0498(51)$ & $-0.1048(79)$ & $0.070(20)$ & $0.333(24)$ \\
$\chitd^{iijk} / \gsix$ & $0.02618(62)$ & $0.00899(76)$ & $-0.0183(11)$ & $0.0271(18)$ & $0.2612(90)$ & $0.6519(86)$ \\
$\chitd^{iijj} / \gsix$ & $0.1667(28)$ & $0.14508(64)$ & $0.0986(14)$ & $0.0705(38)$ & $0.263(24)$ & $0.5514(51)$ \\
$\chitd^{ijkl} / \gsix$ & $0.02565(57)$ & $0.01261(41)$ & $-0.00280(17)$ & $0.0964(17)$ & $0.364(14)$ & $0.819(20)$ \\
$\chitd^{iiiii} / \gsix$ & $0.043(43)$ & $-1.343(45)$ & $-1.528(75)$ & $1.981(92)$ & $6.49(18)$ & $12.23(21)$ \\
$\chitd^{iiiij} / \gsix$ & $0.008(16)$ & $-0.537(11)$ & $-0.398(12)$ & $2.179(28)$ & $4.680(78)$ & $7.975(78)$ \\
$\chitd^{iiijk} / \gsix$ & $-0.0010(72)$ & $-0.2087(25)$ & $-0.067(11)$ & $1.231(13)$ & $2.211(36)$ & $3.39(13)$ \\
$\chitd^{iiijj} / \gsix$ & $0.008(13)$ & $-0.4095(95)$ & $-0.350(11)$ & $1.182(11)$ & $2.664(51)$ & $4.459(93)$ \\
$\chitd^{iijkl} / \gsix$ & $-0.0015(37)$ & $-0.08068(31)$ & $-0.0186(53)$ & $0.234(23)$ & $0.196(46)$ & $-0.13(30)$ \\
$\chitd^{iijjk} / \gsix$ & $0.0015(67)$ & $-0.1902(33)$ & $-0.1291(35)$ & $0.550(13)$ & $1.018(38)$ & $1.40(23)$ \\
$\chitd^{ijklm} / \gsix$ & $-0.0017(20)$ & $-0.01653(94)$ & $0.0058(31)$ & $-0.264(31)$ & $-0.811(61)$ & $-1.89(38)$ \\
$\chitd^{iiiiii} / \gsix$ & $-22.67(99)$ & $-10.82(27)$ & $16.1(11)$ & $56.7(22)$ & $55.4(76)$ & $22(13)$ \\
$\chitd^{iiiiij} / \gsix$ & $-5.81(19)$ & $-1.18(16)$ & $8.07(41)$ & $14.0(13)$ & $10.1(60)$ & $-9.9(10)$ \\
$\chitd^{iiiijk} / \gsix$ & $-1.646(72)$ & $-0.247(40)$ & $2.138(93)$ & $0.2(11)$ & $2.8(38)$ & $3.8(60)$ \\
$\chitd^{iiiijj} / \gsix$ & $-4.04(21)$ & $-1.926(34)$ & $2.40(14)$ & $4.2(16)$ & $7.2(41)$ & $5.4(67)$ \\
$\chitd^{iiijkl} / \gsix$ & $-0.470(31)$ & $-0.0441(53)$ & $0.367(15)$ & $-1.05(54)$ & $6.5(20)$ & $23.6(28)$ \\
$\chitd^{iiijjk} / \gsix$ & $-1.186(52)$ & $-0.196(26)$ & $1.403(56)$ & $0.88(65)$ & $6.8(25)$ & $18.6(38)$ \\
$\chitd^{iiijjj} / \gsix$ & $-3.59(12)$ & $-0.748(96)$ & $4.82(24)$ & $9.28(64)$ & $12.0(37)$ & $13.3(58)$ \\
$\chitd^{iijklm} / \gsix$ & $-0.335(19)$ & $-0.0766(12)$ & $0.078(11)$ & $1.124(67)$ & $12.06(75)$ & $39.87(91)$ \\
$\chitd^{iijjkl} / \gsix$ & $-0.727(33)$ & $-0.145(13)$ & $0.668(19)$ & $1.55(19)$ & $10.8(13)$ & $33.3(18)$ \\
$\chitd^{iijjkk} / \gsix$ & $-1.891(90)$ & $-0.777(21)$ & $1.299(64)$ & $3.67(43)$ & $12.0(19)$ & $30.0(26)$ \\
$\chitd^{ijklmn} / \gsix$ & $-0.267(13)$ & $-0.09277(85)$ & $-0.067(16)$ & $2.21(29)$ & $14.85(20)$ & $48.01(84)$ \\
\hline
\end{tabular}
}
\caption{Results for the different derivatives of the EQCD pressure at $T=277,~400~\mathrm{MeV}$, number of massless quark flavors $\nf$, and chemical potentials $z =\frac{\nf \mu}{3 \pi^2 T}$; in units of the three-dimensional coupling $\gsq$.}
\label{res_tab_277_400}
\end{table}

\begin{table}[htbp!] 
\centering
{\tiny
\begin{tabular}{|c||c|c|c|c|c|c|}	
\hline
 & \multicolumn{6}{c|}{$T = 600 ~ \mathrm{MeV} , \; \nf = 3$} \\
\hline
$z$ & $0.0$ & $0.025$ & $0.05$ & $0.1$ & $0.15$ & $0.2$ \\
\hline
$\chitd^{ii} / \gsix$ & $-0.0695(12)$ & $-0.0656(11)$ & $-0.05702(90)$ & $-0.02829(52)$ & $-0.00220(40)$ & $0.01994(21)$ \\
$\chitd^{ij} / \gsix$ & $-0.0014415(69)$ & $-0.000577(19)$ & $0.001601(14)$ & $0.008940(10)$ & $0.01677(18)$ & $0.02486(45)$ \\
$\chitd^{iii} / \gsix$ & $-0.00001(12)$ & $0.02707(18)$ & $0.05181(76)$ & $0.08798(76)$ & $0.10183(62)$ & $0.1099(21)$ \\
$\chitd^{iij} / \gsix$ & $-0.00001(12)$ & $0.009646(84)$ & $0.01801(22)$ & $0.02760(36)$ & $0.02872(73)$ & $0.0291(17)$  \\
$\chitd^{ijk} / \gsix$ & $-0.00001(12)$ & $0.00095(12)$ & $0.001098(55)$ & $-0.00252(30)$ & $-0.00788(82)$ & $-0.01134(174)$ \\
$\chitd^{iiii} / \gsix$ & $0.3543(56)$ & $0.3342(21)$ & $0.2804(38)$ & $0.14193(89)$ & $0.144(24)$ & $0.383(36)$ \\
$\chitd^{iiij} / \gsix$ & $0.01800(33)$ & $0.00524(79)$ & $-0.02197(59)$ & $-0.0326(15)$ & $0.125(12)$ & $0.487(12)$ \\
$\chitd^{iijk} / \gsix$ & $0.01757(30)$ & $0.00800(45)$ & $-0.00742(15)$ & $0.04412(93)$ & $0.2707(62)$ & $0.708(35)$ \\
$\chitd^{iijj} / \gsix$ & $0.1296(20)$ & $0.11762(55)$ & $0.0933(12)$ & $0.10127(74)$ & $0.2720(53)$ & $0.679(25)$ \\
$\chitd^{ijkl} / \gsix$ & $0.01735(28)$ & $0.00919(29)$ & $-0.00006(28)$ & $0.0797(16)$ & $0.3408(27)$ & $0.807(47)$ \\
$\chitd^{iiiii} / \gsix$ & $0.035(28)$ & $-0.655(25)$ & $-0.765(15)$ & $1.878(11)$ & $6.569(16)$ & $11.90(84)$ \\
$\chitd^{iiiij} / \gsix$ & $0.0095(98)$ & $-0.2837(61)$ & $-0.1969(51)$ & $1.7377(49)$ & $4.5145(239)$ & $7.45(61)$ \\
$\chitd^{iiijk} / \gsix$ & $0.0017(36)$ & $-0.11629(48)$ & $-0.0334(21)$ & $0.9622(86)$ & $2.121(20)$ & $3.21(35)$ \\
$\chitd^{iiijj} / \gsix$ & $0.0082(78)$ & $-0.2089(52)$ & $-0.1756(41)$ & $0.9971(63)$ & $2.6345(99)$ & $4.32(39)$ \\
$\chitd^{iijkl} / \gsix$ & $0.0005(15)$ & $-0.04141(47)$ & $-0.01210(80)$ & $0.222(13)$ & $0.2415(91)$ & $0.09(19)$ \\
$\chitd^{iijjk} / \gsix$ & $0.0031(36)$ & $-0.0971(14)$ & $-0.0666(19)$ & $0.480(11)$ & $1.0390(88)$ & $1.50(24)$ \\
$\chitd^{ijklm} \gsix$ & $-0.00002(50)$ & $-0.00395(95)$ & $-0.00148(54)$ & $-0.149(16)$ & $-0.6979(98)$ & $-1.47(20)$ \\
$\chitd^{iiiiii} / \gsix$ & $-12.65(22)$ & $-5.74(23)$ & $10.51(34)$ & $48.13(18)$ & $57.7(30)$ & $63.6(85)$ \\
$\chitd^{iiiiij} / \gsix$ & $-3.399(61)$ & $-0.908(23)$ & $4.81(13)$ & $14.11(30)$ & $13.3(21)$ & $21.8(66)$ \\
$\chitd^{iiiijk} / \gsix$ & $-0.891(24)$ & $-0.219(12)$ & $1.067(28)$ & $1.59(33)$ & $3.0(13)$ & $21.0(26)$ \\
$\chitd^{iiiijj} / \gsix$ & $-2.138(42)$ & $-1.005(28)$ & $1.299(42)$ & $4.63(39)$ & $6.6(14)$ & $23.6(31)$ \\
$\chitd^{iiijkl} / \gsix$ & $-0.217(12)$ & $-0.054(11)$ & $0.0721(67)$ & $-0.28(19)$ & $5.75(82)$ & $32.0(13)$ \\
$\chitd^{iiijjk} / \gsix$ & $-0.643(18)$ & $-0.1673(94)$ & $0.707(19)$ & $1.60(21)$ & $6.57(98)$ & $30.29(78)$ \\
$\chitd^{iiijjj} / \gsix$ & $-2.098(39)$ & $-0.570(13)$ & $2.882(77)$ & $9.11(16)$ & $13.5(14)$ & $32.6(22)$ \\
$\chitd^{iijklm} / \gsix$ & $-0.1699(77)$ & $-0.0605(65)$ & $0.0146(29)$ & $0.982(59)$ & $11.09(58)$ & $43.1(36)$ \\
$\chitd^{iijjkl} / \gsix$ & $-0.395(12)$ & $-0.1155(67)$ & $0.3462(99)$ & $1.604(91)$ & $10.17(69)$ & $39.5(24)$ \\
$\chitd^{iijjkk} / \gsix$ & $-1.026(21)$ & $-0.4327(65)$ & $0.745(23)$ & $3.45(13)$ & $11.34(83)$ & $38.6(16)$ \\
$\chitd^{ijklmn} / \gsix$ & $-0.1466(56)$ & $-0.0637(46)$ & $-0.0142(10)$ & $1.614(70)$ & $13.76(50)$ & $48.7(48)$ \\
\hline
\hline
 & \multicolumn{6}{c|}{$T = 2.5~\mathrm{GeV} , \; \nf = 4$} \\
\hline
$z$ & $0.0$ & $0.025$ & $0.05$ & $0.1$ & $0.15$ & $0.2$ \\
\hline
$\chitd^{ii} / \gsix$ & $0.09130(10)$ & $0.09234(43)$ & $0.09535(17)$ & $0.106781(81)$ & $0.124285(64)$ & $0.145465(57)$  \\
$\chitd^{ij} / \gsix$ & $-0.0006043(77)$ & $-0.0002251(74)$ & $0.000860(14)$ & $0.004792(92)$ & $0.01051(11)$ & $0.01695(13)$ \\
$\chitd^{iii} / \gsix$ & $0.000067(60)$ & $0.01794(11)$ & $0.03541(25)$ & $0.06595(64)$ & $0.08928(79)$ & $0.1035(16)$ \\
$\chitd^{iij} / \gsix$ & $0.000067(60)$ & $0.006273(67)$ & $0.012091(92)$ & $0.02187(19)$ & $0.02853(23)$ & $0.03147(99)$  \\
$\chitd^{ijk} / \gsix$ & $0.000067(60)$ & $0.000435(63)$ & $0.000435(91)$ & $-0.00016(17)$ & $-0.00184(22)$ & $-0.00449(70)$ \\
$\chitd^{iiii} / \gsix$ & $0.3052(16)$ & $0.3022(11)$ & $0.2907(11)$ & $0.2502(71)$ & $0.2497(54)$ & $0.361(12)$ \\
$\chitd^{iiij} / \gsix$ & $0.01006(16)$ & $0.00608(32)$ & $-0.00259(23)$ & $-0.0082(31)$ & $0.0520(54)$ & $0.2354(38)$ \\
$\chitd^{iijk} / \gsix$ & $0.00990(18)$ & $0.00658(25)$ & $0.00018(13)$ & $0.0063(12)$ & $0.0963(33)$ & $0.3077(30)$ \\
$\chitd^{iijj} / \gsix$ & $0.679(25)$ & $0.10832(49)$ & $0.10528(46)$ & $0.09799(40)$ & $0.0924(27)$ & $0.1595(46)$ \\
$\chitd^{ijkl} / \gsix$ & $0.00982(19)$ & $0.00715(13)$ & $0.00152(10)$ & $0.01323(92)$ & $0.1100(49)$ & $0.3412(56)$ \\
$\chitd^{iiiii} / \gsix$ & $-0.016(11)$ & $-0.294(12)$ & $-0.390(12)$ & $0.526(65)$ & $3.100(98)$ & $7.12(16)$ \\
$\chitd^{iiiij} / \gsix$ & $-0.00248(54)$ & $-0.1438(20)$ & $-0.1587(65)$ & $0.497(28)$ & $2.146(53)$ & $4.53(11)$ \\
$\chitd^{iiijk} / \gsix$ & $0.0009(15)$ & $-0.0640(18)$ & $-0.0606(38)$ & $0.2897(123)$ & $1.109(21)$ & $2.167(85)$ \\
$\chitd^{iiijj} / \gsix$ & $-0.0025(15)$ & $-0.1016(17)$ & $-0.1183(44)$ & $0.296(19)$ & $1.349(33)$ & $2.813(96)$ \\
$\chitd^{iijkl} / \gsix$ & $0.0008(28)$ & $-0.02188(94)$ & $-0.0201(15)$ & $0.0891(42)$ & $0.3126(62)$ & $0.446(74)$ \\
$\chitd^{iijjk} / \gsix$ & $-0.00033(38)$ & $-0.04846(40)$ & $-0.0528(24)$ & $0.1583(91)$ & $0.658(13)$ & $1.236(81)$ \\
$\chitd^{ijklm} \gsix$ & $0.00069(37)$ & $-0.00089(72)$ & $0.00006(41)$ & $-0.0109(63)$ & $-0.086(13)$ & $-0.414(71)$ \\
$\chitd^{iiiiii} / \gsix$ & $-7.36(27)$ & $-4.75(11)$ & $2.64(11)$ & $26.42(72)$ & $53.6(15)$ & $75.7(19)$ \\
$\chitd^{iiiiij} / \gsix$ & $-2.1013(95)$ & $-1.185(76)$ & $1.411(66)$ & $8.99(25)$ & $17.31(45)$ & $25.3(12)$ \\
$\chitd^{iiiijk} / \gsix$ & $-0.4947(45)$ & $-0.265(20)$ & $0.308(16)$ & $1.66(15)$ & $4.04(30)$ & $9.91(46)$ \\
$\chitd^{iiiijj} / \gsix$ & $-1.150(54)$ & $-0.750(15)$ & $0.284(11)$ & $3.25(19)$ & $7.67(33)$ & $14.78(56)$  \\
$\chitd^{iiijkl} / \gsix$ & $-0.0809(12)$ & $-0.0357(53)$ & $0.0166(43)$ & $-0.06(10)$ & $1.35(26)$ & $8.84(21)$ \\
$\chitd^{iiijjk} / \gsix$ & $-0.3507(24)$ & $-0.189(14)$ & $0.203(11)$ & $1.141(98)$ & $3.46(26)$ & $10.94(28)$ \\
$\chitd^{iiijjj} / \gsix$ & $-1.2859(50)$ & $-0.726(46)$ & $0.844(40)$ & $5.43(16)$ & $11.30(29)$ & $20.37(60)$  \\
$\chitd^{iijklm} / \gsix$ & $-0.06852(87)$ & $-0.0368(27)$ & $0.0008(23)$ & $0.048(24)$ & $1.97(25)$ & $11.60(33)$ \\
$\chitd^{iijjkl} / \gsix$ & $-0.20666(28)$ & $-0.1132(79)$ & $0.0981(63)$ & $0.622(44)$ & $2.87(24)$ & $11.96(25)$ \\
$\chitd^{iijjkk} / \gsix$ & $-0.561(18)$ & $-0.3503(74)$ & $0.1850(89)$ & $1.744(72)$ & $5.08(21)$ & $14.41(27)$ \\
$\chitd^{ijklmn} / \gsix$ & $-0.0622(21)$ & $-0.0373(24)$ & $-0.0070(16)$ & $0.100(21)$ & $2.28(25)$ & $12.97(43)$ \\
\hline
\end{tabular}
}
\caption{Results for the different derivatives of the EQCD pressure at $T=600,~2500~\mathrm{MeV}$, number of massless quark flavors $\nf$, and chemical potentials $z =\frac{\nf \mu}{3 \pi^2 T}$; in units of the three-dimensional coupling $\gsq$.}
\label{res_tab_600_2500}
\end{table}

\begin{table}[htbp!] 
\centering
{\tiny
\begin{tabular}{|c||c|c|c|c|c|c|}	
\hline
 & \multicolumn{6}{c|}{$T = 25~\mathrm{GeV} , \; \nf = 5$} \\
\hline
$z$ & $0.0$ & $0.025$ & $0.05$ & $0.1$ & $0.15$ & $0.2$ \\
\hline
$\chitd^{ii} / \gsix$ & $0.23651(13)$ & $0.237460(60)$ & $0.240254(86)$ & $0.25109(12)$ & $0.26843(12)$ & $0.29107(30)$ \\
$\chitd^{ij} / \gsix$ & $-0.000241(15)$ & $0.000031(25)$ & $0.0008664(81)$ & $0.004170(72)$ & $0.008975(50)$ & $0.015226(46)$ \\
$\chitd^{iii} / \gsix$ & $0.000040(47)$ & $0.01758(22)$ & $0.03480(23)$ & $0.06778(89)$ & $0.09402(15)$ & $0.1164(18)$ \\
$\chitd^{iij} / \gsix$ & $0.000040(47)$ & $0.00591(15)$ & $0.01173(11)$ & $0.02260(29)$ & $0.03139(11)$ & $0.03849(52)$ \\
$\chitd^{ijk} / \gsix$ & $0.000040(47)$ & $0.00008(13)$ & $0.00020(10)$ & $0.00001(15)$ & $0.00009(21)$ & $-0.0003(15)$ \\
$\chitd^{iiii} / \gsix$ & $0.3834(27)$ & $0.3769(31)$ & $0.3674(10)$ & $0.3286(41)$ & $0.3356(84)$ & $0.457(19)$ \\
$\chitd^{iiij} / \gsix$ & $0.00730(16)$ & $0.00565(22)$ & $0.0010(11)$ & $-0.0108(17)$ & $0.0400(34)$ & $0.2119(51)$  \\
$\chitd^{iijk} / \gsix$ & $0.00721(17)$ & $0.00555(11)$ & $0.00126(47)$ & $0.0020(11)$ & $0.06387(59)$ & $0.25421(48)$ \\
$\chitd^{iijj} / \gsix$ & $0.1326(10)$ & $0.1293(11)$ & $0.12341(30)$ & $0.1150(23)$ & $0.1628(28)$ & $0.3353(33)$ \\
$\chitd^{ijkl} / \gsix$ & $0.00717(17)$ & $0.005355(60)$ & $0.00142(11)$ & $0.00724(94)$ & $0.07721(88)$ & $0.2744(24)$ \\
$\chitd^{iiiii} / \gsix$ & $-0.014(18)$ & $-0.190(12)$ & $-0.253(32)$ & $0.341(60)$ & $2.71(10)$ & $6.94(28)$ \\
$\chitd^{iiiij} / \gsix$  & $-0.0028(35)$ & $-0.1063(31)$ & $-0.1325(85)$ & $0.390(38)$ & $1.890(34)$ & $4.478(72)$ \\
$\chitd^{iiijk} / \gsix$ & $-0.00048(23)$ & $-0.0511(15)$ & $-0.0615(30)$ & $0.240(20)$ & $1.005(12)$ & $2.265(27)$ \\
$\chitd^{iiijj} / \gsix$ & $-0.0028(35)$ & $-0.0722(26)$ & $-0.0916(73)$ & $0.226(24)$ & $1.212(23)$ & $2.883(35)$ \\
$\chitd^{iijkl} / \gsix$ & $-0.000345(21)$ & $-0.01705(50)$ & $-0.0206(11)$ & $0.0764(56)$ & $0.3264(86)$ & $0.672(71)$ \\
$\chitd^{iijjk} / \gsix$ & $-0.0011(12)$ & $-0.0355(11)$ & $-0.0443(29)$ & $0.126(12)$ & $0.6217(78)$ & $1.409(40)$ \\
$\chitd^{ijklm} \gsix$ & $-0.000250(74)$ & $-0.000037(63)$ & $-0.00016(18)$ & $-0.0056(18)$ & $-0.014(11)$ & $-0.122(95)$ \\
$\chitd^{iiiiii} / \gsix$ & $-7.54(34)$ & $-4.85(21)$ & $2.35(15)$ & $28.3(13)$ & $59.36(67)$ & $98.4(12)$ \\
$\chitd^{iiiiij} / \gsix$ & $-2.2021(49)$ & $-1.360(16)$ & $1.014(40)$ & $9.71(38)$ & $19.29(28)$ & $34.62(21)$ \\
$\chitd^{iiiijk} / \gsix$ & $-0.48368(97)$ & $-0.3027(23)$ & $0.2121(43)$ & $1.876(28)$ & $4.11(13)$ & $11.80(91)$ \\
$\chitd^{iiiijj} / \gsix$ & $-1.124(68)$ & $-0.740(38)$ & $0.281(31)$ & $3.62(11)$ & $8.10(20)$ & $18.0(11)$  \\
$\chitd^{iiijkl} / \gsix$ & $-0.0479(17)$ & $-0.0332(18)$ & $0.0050(57)$ & $-0.034(30)$ & $0.78(13)$ & $7.71(81)$ \\
$\chitd^{iiijjk} / \gsix$ & $-0.33542(95)$ & $-0.2102(17)$ & $0.1400(33)$ & $1.265(24)$ & $3.238(84)$ & $11.24(70)$ \\
$\chitd^{iiijjj} / \gsix$ & $-1.3369(25)$ & $-0.8259(95)$ & $0.607(25)$ & $5.85(24)$ & $12.17(18)$ & $24.827(97)$ \\
$\chitd^{iijklm} / \gsix$ & $-0.0417(16)$ & $-0.0278(14)$ & $-0.0012(22)$ & $0.0189(59)$ & $1.256(34)$ & $9.32(47)$ \\
$\chitd^{iijjkl} / \gsix$ & $-0.1871(11)$ & $-0.1177(13)$ & $0.0679(25)$ & $0.6550(206)$ & $2.366(35)$ & $10.69(50)$ \\
$\chitd^{iijjkk} / \gsix$ & $-0.545(22)$ & $-0.353(14)$ & $0.1590(82)$ & $1.880(71)$ & $4.882(40)$ & $14.40(55)$ \\
$\chitd^{ijklmn} / \gsix$ & $-0.0385(16)$ & $-0.0250(12)$ & $-0.0039(17)$ & $0.045(19)$ & $1.493(16)$ & $10.13(31)$ \\
\hline
\hline
 & \multicolumn{6}{c|}{$T = 100~\mathrm{GeV} , \; \nf = 5$}  \\
\hline
$z$ & $0.0$ & $0.025$ & $0.05$ & $0.1$ & $0.15$ & $0.2$ \\
\hline
$\chitd^{ii} / \gsix$ & $0.331747(77)$ & $0.33303(13)$ & $0.33638(12)$ & $0.349623(71)$ & $0.370942(57)$ & $0.39872(15)$ \\
$\chitd^{ij} / \gsix$ & $-0.000137(13)$ & $0.0001848(86)$ & $0.001242(39)$ & $0.005091(44)$ & $0.010907(53)$ & $0.01825(27)$ \\
$\chitd^{iii} / \gsix$ & $0.000015(97)$ & $0.021282(24)$ & $0.04211(28)$ & $0.08081(74)$ & $0.1151(12)$ & $0.1433(18)$ \\
$\chitd^{iij} / \gsix$ & $0.000015(97)$ & $0.007202(86)$ & $0.01434(27)$ & $0.02703(22)$ & $0.03797(30)$ & $0.04665(31)$ \\
$\chitd^{ijk} / \gsix$ & $0.000015(97)$ & $0.00016(13)$ & $0.00041(28)$ & $0.000120(47)$ & $-0.00063(18)$ & $-0.00169(51)$ \\
$\chitd^{iiii} / \gsix$ & $0.4589(10)$ & $0.4513(22)$ & $0.4368(17)$ & $0.434(15)$ & $0.484(18)$ & $0.681(15)$  \\
$\chitd^{iiij} / \gsix$ & $0.00657(20)$ & $0.00403(29)$ & $-0.0018(14)$ & $0.0094(43)$ & $0.0995(75)$ & $0.348(10)$  \\
$\chitd^{iijk} / \gsix$ & $0.00652(22)$ & $0.00440(19)$ & $-0.00028(51)$ & $0.0136(14)$ & $0.11606(32)$ & $0.378(12)$  \\
$\chitd^{iijj} / \gsix$ & $0.15734(41)$ & $0.15338(84)$ & $0.14587(58)$ & $0.1569(62)$ & $0.2434(48)$ & $0.492(13)$ \\
$\chitd^{ijkl} / \gsix$ & $0.00649(22)$ & $0.00445(25)$ & $0.000476(62)$ & $0.01535(43)$ & $0.1235(29)$ & $0.400(16)$   \\
$\chitd^{iiiii} / \gsix$ & $0.038(17)$ & $-0.257(15)$ & $-0.278(70)$ & $1.08(12)$ & $4.99(28)$ & $11.01(50)$ \\
$\chitd^{iiiij} / \gsix$ & $0.0071(35)$ & $-0.1295(13)$ & $-0.114(12)$ & $0.742(20)$ & $3.13(13)$ & $6.80(31)$  \\
$\chitd^{iiijk} / \gsix$  & $-0.00046(47)$ & $-0.0587(17)$ & $-0.0438(33)$ & $0.3937(47)$ & $1.580(45)$ & $3.39(13)$ \\
$\chitd^{iiijj} / \gsix$ & $0.0074(34)$ & $-0.0906(20)$ & $-0.085(13)$ & $0.478(22)$ & $2.044(82)$ & $4.45(18)$  \\
$\chitd^{iijkl} / \gsix$ & $-0.000158(34)$ & $-0.01976(70)$ & $-0.0141(15)$ & $0.12950(48)$ & $0.4919(49)$ & $1.040(14)$ \\
$\chitd^{iijjk} / \gsix$ & $0.0024(11)$ & $-0.04342(37)$ & $-0.0375(40)$ & $0.2456(75)$ & $1.010(28)$ & $2.176(69)$ \\
$\chitd^{ijklm} \gsix$ & $0.00005(25)$ & $-0.00031(21)$ & $0.00074(62)$ & $-0.0025(19)$ & $-0.054(24)$ & $-0.136(45)$ \\
$\chitd^{iiiiii} / \gsix$ & $-7.93(56)$ & $-4.96(20)$ & $5.78(22)$ & $43.4(10)$ & $97.2(47)$ & $154.3(90)$ \\
$\chitd^{iiiiij} / \gsix$ & $-2.442(32)$ & $-1.254(68)$ & $2.35(18)$ & $14.76(27)$ & $33.0(20)$ & $57.1(37)$ \\
$\chitd^{iiiijk} / \gsix$ & $-0.5204(72)$ & $-0.264(17)$ & $0.486(56)$ & $3.08(11)$ & $8.06(98)$ & $21.0(18)$  \\
$\chitd^{iiiijj} / \gsix$ & $-1.14(10)$ & $-0.763(44)$ & $0.718(49)$ & $5.901(223)$ & $14.5(13)$ & $30.7(24)$  \\
$\chitd^{iiijkl} / \gsix$ & $-0.0366(34)$ & $-0.0179(39)$ & $0.011(16)$ & $0.131(38)$ & $2.26(37)$ & $13.4(12)$ \\
$\chitd^{iiijjk} / \gsix$ & $-0.3579(54)$ & $-0.182(12)$ & $0.323(37)$ & $2.083(69)$ & $6.345(566)$ & $19.2(15)$ \\
$\chitd^{iiijjj} / \gsix$ & $-1.478(20)$ & $-0.760(41)$ & $1.41(11)$ & $8.90(16)$ & $21.0(11)$ & $40.5(26)$  \\
$\chitd^{iijklm} / \gsix$ & $-0.0339(24)$ & $-0.0187(18)$ & $0.0018(47)$ & $0.1093(90)$ & $2.686(74)$ & $14.9(12)$ \\
$\chitd^{iijjkl} / \gsix$ & $-0.1954(37)$ & $-0.1008(65)$ & $0.160(18)$ & $1.091(32)$ & $4.62(17)$ & $17.4(13)$  \\
$\chitd^{iijjkk} / \gsix$ & $-0.562(37)$ & $-0.349(10)$ & $0.3942(50)$ & $3.010(89)$ & $8.76(43)$ & $23.4(17)$  \\
$\chitd^{ijklmn} / \gsix$ & $-0.0322(20)$ & $-0.0189(11)$ & $-0.0030(13)$ & $0.0986(57)$ & $2.89(28)$ & $15.6(12)$ \\
\hline
\end{tabular}
}
\caption{Results for the different derivatives of the EQCD pressure at $T=25,~100~\mathrm{GeV}$, number of massless quark flavors $\nf$, and chemical potentials $z =\frac{\nf \mu}{3 \pi^2 T}$; in units of the three-dimensional coupling $\gsq$.}
\label{res_tab_25_100}
\end{table}

\FloatBarrier
\bibliographystyle{unsrt}
\bibliography{references}

\end{document}